\documentclass[aps,prb,twocolumn,superscriptaddress,amsmath,showpacs,floatfix]{revtex4-1}

\usepackage{graphicx} 
\usepackage{color}

\definecolor{darkgreen}{rgb}{0,0.4,0}

\usepackage{color}
\usepackage{amsfonts,amssymb,amsmath}
\usepackage{tabularx,hhline}
\usepackage{tikz}

\begin{document}

\title{The multiple symmetry sustaining phase transitions of spin ice}

\author{V. Raban}
\affiliation{Universit\'e de Lyon, ENS de Lyon, Universit\'e Claude Bernard, CNRS, Laboratoire de Physique, F-69342 Lyon, France} 

\affiliation{Laboratoire Charles Coulomb (L2C), Universit\'e de 
Montpellier, CNRS, Montpellier, France}

\author{C. T. Suen}
\affiliation{Universit\'e de Lyon, ENS de Lyon, Universit\'e Claude Bernard, CNRS, Laboratoire de Physique, F-69342 Lyon, France} 

\author{L. Berthier}

\affiliation{Laboratoire Charles Coulomb (L2C), Universit\'e de 
Montpellier, CNRS, Montpellier, France}

\author{P. C. W. Holdsworth}

\affiliation{Universit\'e de Lyon, ENS de Lyon, Universit\'e Claude Bernard, CNRS, Laboratoire de Physique, F-69342 Lyon, France} 

\date{\today}

\begin{abstract}
We present the full phase diagram of the dumbbell model of spin ice as a function of temperature, chemical potential and staggered chemical potential which breaks the translational lattice symmetry in favour of charge crystal ordering. We observe a double winged structure with five possible phases, monopole fluid (spin ice), fragmented single monopole crystal phases and double monopole crystal, the zinc blend structure. Our model provides a skeleton for liquid-liquid phase transitions and for the winged structures observed for itinerant magnets under pressure and external field. We relate our results to recent experiments on Ho$_2$Ir$_2$O$_7$ and propose a wide ranging set of new experiments that exploit the phase diagram, including high pressure protocols, dynamical scaling of Kibble-Zurek form and universal violations of the fluctuation-dissipation theorem.
\end{abstract}

\maketitle

\section{Introduction}

\begin{figure} [b]
\centering{\includegraphics[scale=0.3]{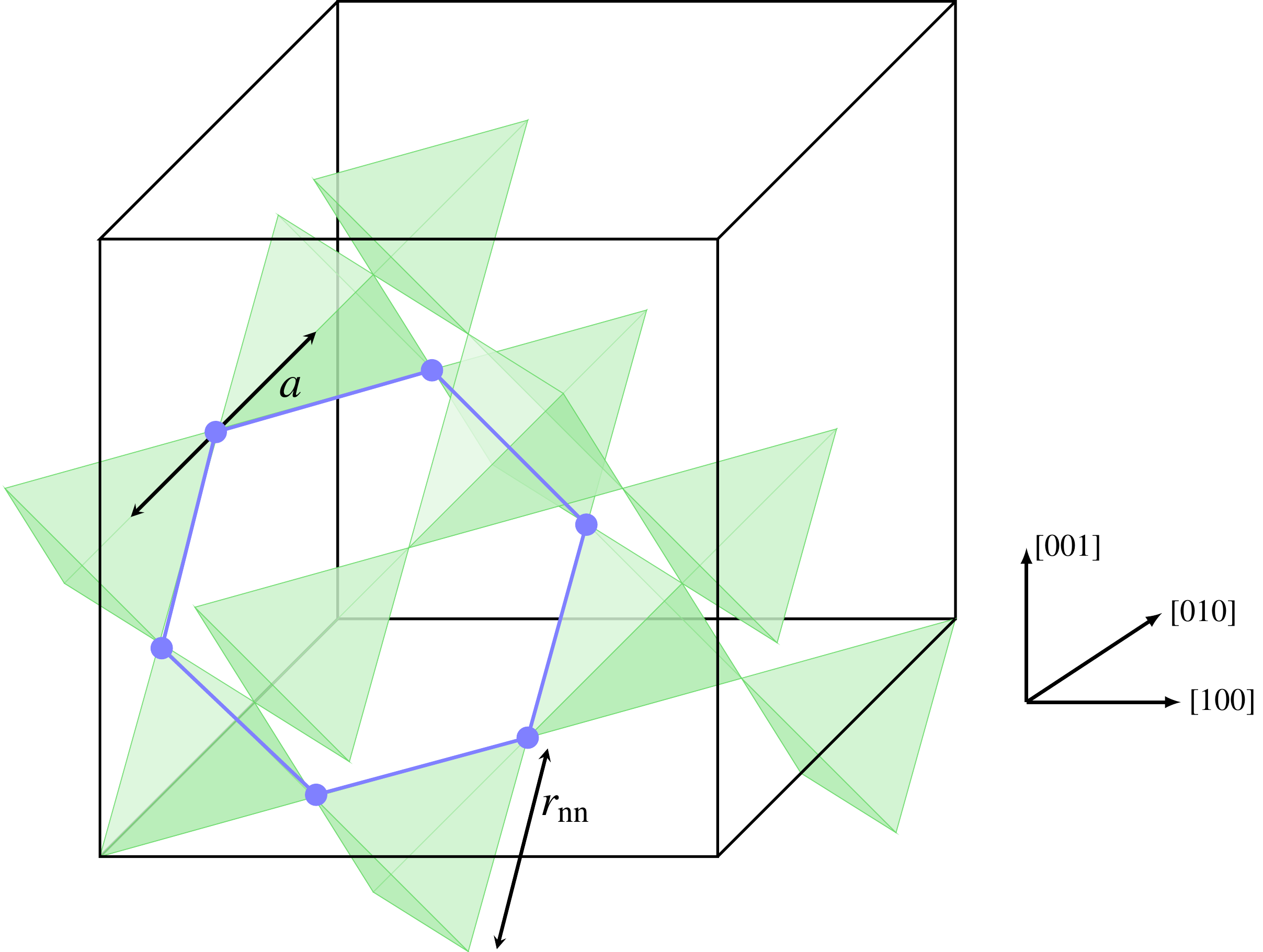}}
\centering{\includegraphics[scale=0.25]{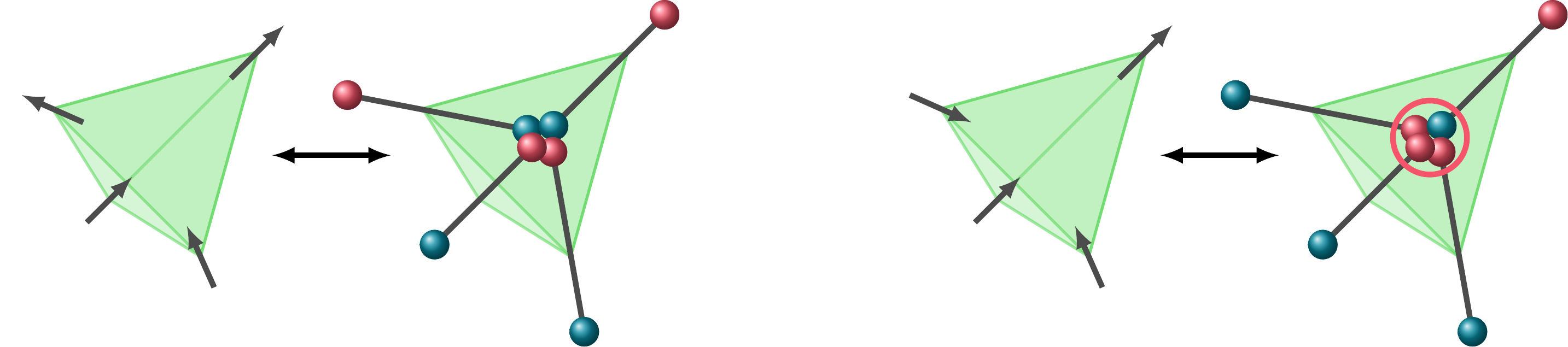}}
\caption[$\;\;$ Pyrochlore and dumbbells]{{\textbf{From spins to dumbbells:} The point dipoles are extended to needles touching at the diamond lattice centres. The needles carry magnetic flux and charge $q=\pm m/a$ at each end. In a 2in-2out configuration (left)  the vertex is charge neutral. A 3in-1out (3out-1in) configuration carries a monopole charge $Q=2m/a$ ($-Q=-2m/a$) (left). A 4in (4out) configuration carries a double monopole charge $2Q=4m/a$ ($-2Q=-4m/a$).}}
\label{Dumbbell}
\end{figure}

Over the last decade, spin ice models and materials \cite{Harris1997,Bramwell2001} have emerged as model systems for the study of generalized electrostatics on a lattice \cite{Isakov2004,Castelnovo2008,Ryzhkin2005,Jaubert2009,Castelnovo2011,Brooks2014,Kaiser2015}.
The emergence of the electrostatics can best be seen by replacing  the point dipole moments of spin ice by infinitesimally thin magnetic needles, lying along the axes linking the centres of adjoining tetrahedra \cite{Moller2006}(see Fig.~\ref{Dumbbell}). Within this dumbbell approximation \cite{Castelnovo2008}, the pyrochlore lattice of magnetic moments transforms \cite{denHertog2000,Isakov2005} into a diamond lattice of vertices for magnetic charge. The  needles carry magnetic flux and dumbbells of effective magnetic charge which touch at the vertices. By construction the ensemble of low energy ``Pauling states'' \cite{Pauling1935} with two spins into and two out of each tetrahedron are degenerate in this approximation, with charge neutrality imposed at each vertex. 
These ground
states form a vacuum from which magnetic monopole quasi-particles are excited by reversing
the orientation of a needle, breaking the ice rules on a pair of neighbouring
sites \cite{Castelnovo2008}. Double monopoles can also be created by reversing a second needle, for a vertex with all needles in or all out. The emerging Coulomb fluid of magnetic origin is often referred to as a magnetolyte \cite{Jaubert2019} in analogy with its electrical counterpart.

In this paper we study the full phase diagram of the dumbbell model, including a staggered chemical potential, $\Delta$, which breaks a $Z_2$ translational symmetry of the diamond lattice, favouring monopole and double monopole crystallisation into bi-partite ionic cristals. The staggered chemical potential lifts the degeneracy between single and double monopoles at the crystallisation transition in a manner compatible with the staggered internal magnetic field offered by iridium ions in the spin ice material Ho$_2$Ir$_2$O$_7$ \cite{Lefrancois2017}. 

\begin{figure} 
\centering{\includegraphics[scale=0.7]{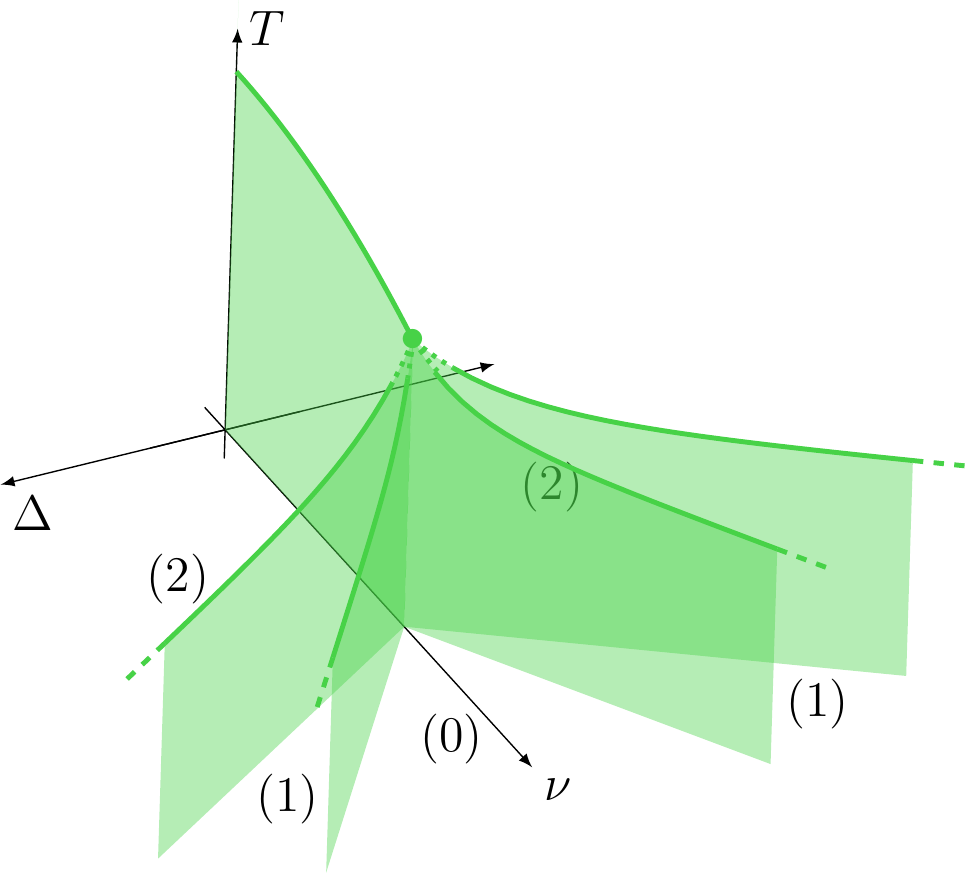}}
\caption[$\;\;$ Phase diagram]{{\textbf{Dumbbell model phase diagram:} {\bf (0)} monopole fluid ($\phi_0$), {\bf (1)} monopole crystal ($\phi_1$), {\bf (2)} double monopole crystal ($\phi_2$). Surfaces show $1^{st}$ order and solid lines $2^{nd}$ order transitions and dotted lines show the multi-critical region. The long dashed lines show the extension of the $2^{nd}$ order lines to infinity. Chemical potentials for single and double monopoles are $\mu=-\nu$, $\mu_2=-4\nu$ (see eqn. (\ref{eq2})). For the phase diagram of the $S=2$ Blume Capel model $\bar{\Delta}$ replaces $\Delta$ and $\bar{\nu}$ replaces $\nu$.}
}
\label{Full-phase}
\end{figure}

As shown in Fig. \ref{Full-phase}, the dumbbell model offers a rich phase diagram in the three dimensional space of parameters $\Delta$, energy scale $\nu$ fixing the monopole and double monopole chemical potentials: $\mu=-\nu$, $\mu_2=-4\nu$, and temperature $T$. The central plane with $\Delta=0$ corresponds to the standard spin ice phase diagram within this approximation \cite{Melko2004,Borzi2014}, with a transition from spin ice to ``all-in-all-out'' (AIAO) order that changes from first  to second order in a multi-critical region. In the monopole language AIAO order corresponds to an ionic crystal of double monopoles with the zinc blend structure. Emerging from this region, there is a double winged structure of phase boundaries that terminate in continuous lines of critical end points. The five phases separated by the boundaries are the Coulomb fluid (spin ice) phase, a fragmented monopole crystal \cite{Brooks2014,Borzi2013} in which the magnetic moments appear to break up into independent divergence full and divergence free parts and the double monopole crystal AIAO phase. 

As $\Delta$ breaks the translational symmetry all transitions, away from the central plane, are symmetry sustaining. In this sense the transition from monopole fluid to single monopole crystal is thermodynamically equivalent to the liquid-gas transition and that from single to double monopole crystal is equivalent to liquid-liquid transitions observed experimentally in supercooled liquids \cite{Katayama2000,Sastry2003,Brovchenko2005}. Entirely analogous sets of phase transitions also occur in itinerant magnetic compounds under pressure and in the presence of an external field  \cite{Taufour2017,Taufour2011}. A consequence of our work is that we are able to offer a generic framework and minimal model to generate such seemingly exotic behaviour,  occurring in diverse domains of physics and chemistry.

Inspired by the $S=2$ Blume-Capel model \cite{Plascak1993}, in the next section we will provide concrete and quantitative evidence for the existence of the double winged phase diagram shown in Fig.~\ref{Full-phase}, introducing general Blume-Capel models, providing a detailed explanation of the multi-critical region and investigating one of the continuous set of critical end points that takes the model from the spin ice monopole fluid to fragmented monopole crystal.  In section {\bf III} we present dynamical finite size scaling results in the region of the critical point and show that it exihibits dynamical Kibble-Zurek scaling in the three dimensional Ising universality class. We also present results showing the universal violation of the fluctuation-dissipation relation consistent with this universality class. In section {\bf IV} we relate our results to the observed monopole driven phase transition for spin ice materials in a magnetic field $\vec H$ in the $[111]$ direction showing that monopole crystallisation thermodynamics leads to a quantitative prediction of the $\vec H, T$ phase diagram.
In section {\bf V} we give some discussion, putting our results in the wider context of liquid-liquid phase transitions and the temperature-field-pressure phase diagram of itinerant magnets. We conclude this section,  returning to frustrated magnets, in particular Ho$_2$Ir$_2$O$_7$ and the possibility of observing such a rich phase diagram and its consequences in future experiments. 

The Kelvin energy scale is used throughout, fixing Boltzmann's constant to unity. We also set the permeability of free space $\mu_0=1$ so that the field $\vec H$ is measured in Tesla. 
We follow standard notation for spin ice simulations and refer to a dimensionless length $L$, measured in cubic units. Each cubic cell contains 16 spins (dumbbells) so that the number of tetrahedra (monopole sites), $N_0=8L^3$. In this paper quantitative measures refer to the spin ice material Dy$_2$Ti$_2$O$_7$ (DTO) for which diamond lattice constant $a=4.33$ \AA, the nearest neighbour spin distance $r_{nn}=\frac{\sqrt{3}a}{2}=3.74$ \AA,  and cube length $a_c=\frac{4a}{\sqrt{3}}\approx10$ \AA (see Fig.~(\ref{Dumbbell})).

\section{Monopole Crystal Phase Diagram}

The dumbbell model is an excellent approximation to the dipolar spin ice model (DSI) which is characterised by short range exchange interactions and dipole interactions which provide long range forces for the monopole quasi-particles \cite{denHertog2000,Henelius2016}. The dumbbell model captures all features of the DSI except for a low temperature ordering transition which indicates the lifting of the degeneracy of the Pauling states. Above this energy scale, the DSI shows a phase transition on varying the ratio of the exchange terms to dipolar interaction, taking the model from the spin ice phase to the AIAO phase \cite{Melko2001}. The transition appears to change from first to second order via a multi-critical point \cite{Borzi2014}. 

\subsection{Blume-Capel models}

\begin{figure} 
\centering{\includegraphics[scale=0.7]{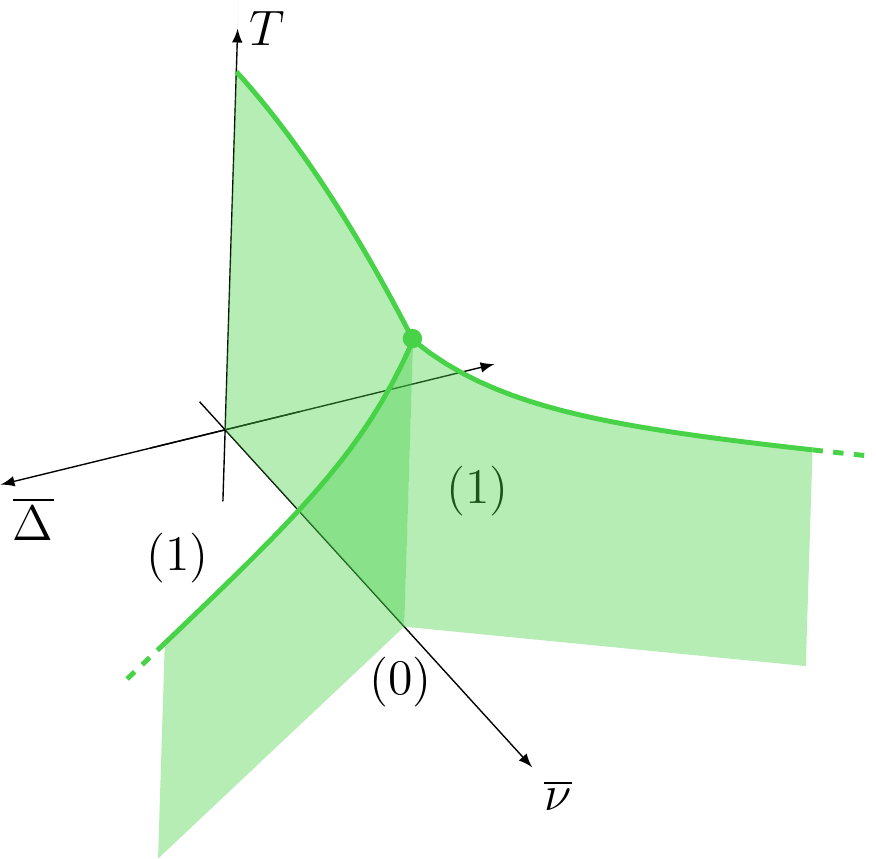}}
\caption[$\;\;$ Phase diagram]{{\textbf{Blume Capel S=1:} {\bf (0)} Paramagnet, {\bf (1)} antiferromagnet. Surfaces show $1^{st}$ and lines $2^{nd}$ order transitions and the point shows a tri-critical point.}}
\label{BC1}
\end{figure}

Such physics is generically provided \cite{Cardy1996} by the Blume-Capel (BC1) model \cite{Blume1966,Capel1966}, developed by Blume, Emery and Griffiths \cite{Blume1971} to study mixtures of $^3$He and $^4$He.  In this model  Ising-like degrees of freedom, which could be spins or occupation numbers for a neutral two component lattice fluid, take on values, $S_i=0,\pm 1$. Contact with spin ice corresponds to the antiferromagnetic case with spins on a bipartite lattice such as square, cubic or diamond with energy function
\begin{equation}
{\cal{H}}_{BC}=J\sum_{ij} S_i S_j +\bar{\nu} \sum_i S_i^2 -\bar{\Delta} \sum_{i=1,N_0}(-1)^i S_i \;,\label{HBC}
\end{equation}
where $J<0$ is a coupling constant, $\bar{\nu}$ is the energy scale for exciting a site $i$, $\bar{\Delta}$ is a staggered field that breaks the $Z_2$ symmetry of the bipartite lattice. Although the function ${\cal{H}}_{BC}$ may generically be referred to as the Hamiltonian, for future reference we take the Hamiltonian to be the many body term only. The parameters $\bar{\nu}$ and $\bar{\Delta}$ can be interpreted as Lagrange multipliers which allow the evolution from the canonical to less constrained ensembles, so that the single site terms contribute to the free energy but not the internal energy \cite{Landau1959}. A suitable order parameter can be defined
\begin{equation}
\phi = \bigg\langle\frac{1}{N_0} \sum_{i=1,N_0}\phi_i\bigg\rangle, 
\label{OP}
 \end{equation}
where $\langle \dots\rangle$ is a thermal average. The term $\phi_i=\epsilon_i S_i$ distinguishes the two sublattices with $\epsilon_i=1$ on an $A$ site and $-1$ on a $B$ site.

For $\bar{\Delta}=0$, on increasing $\bar{\nu}$, the transition changes from $2^{nd}$ order, in the Ising universality class, to $1^{st}$ order via a tri-critical point. The staggered term $\bar{\Delta}$ is conjugate to $\phi$ and therefore guarantees a winged structure, as shown in Fig. (\ref{BC1}). The first order transitions terminate along a line of  critical end points for finite $\bar{\Delta}$ and temperature. The winged phase boundaries and finite temperature critical end points stretch out to $\bar{\Delta} \rightarrow\infty$, as even when the site occupation is  perfectly partitioned with $S_i=1$ on $A$ sites only and $S_i=-1$ on $B$ sites only, the interaction between the sublattices remains, allowing for a singular jump in site occupation at finite temperature. As $\bar{\Delta}$ breaks the lattice symmetry, the transitions at the critical end points are symmetry sustaining. They are characterised by an emergent Ising like order parameter at each point and in this sense are liquid-gas like. 

The Blume-Capel model can be extended \cite{Plascak1993} to higher values of $S$. Of particular interest is $S=2$ (BC2) which greatly resembles the dumbbell model of spin ice.
%
%
The order parameter is now defined on the interval $-2\le \phi\le 2$ and according to mean field \cite{Plascak1993} and pair approximation calculations \cite{Lara1998} the BC2 model allows for two ordered phases corresponding to $|\phi| \sim 1$ (referred to as $\phi_1$) and $|\phi| \sim 2$ ($\phi_2$) as well as the disordered phase with $\phi\approx 0$ ($\phi_0$). As a consequence, adding a finite staggered field, $\bar{\Delta}$ to the BC2 energy function will open out a double winged structure as shown qualitatively in Fig. (\ref{Full-phase}) for the dumbbell model and discussed in detail below. 

\subsection{The dumbbell model}

Returning to the dumbbell model, the charge  on vertex $i$ of the diamond lattice takes values $Q_i=0, \pm Q, \pm 2Q$ with $Q=2m/a$, $m$ the magnetic moment associated with a spin and $a$ the lattice constant (see Fig.~(\ref{Dumbbell})), from which one can define a site occupation  variable $\hat{n}_i=\frac{Q_i}{Q}=0,\pm 1, \pm 2$ in analogy with the BC2 model variables $S_i$. A magnetic north (south) monopole carries charge $+Q \, (-Q)$.
Within the dumbbell approximation, the dipolar spin ice Hamiltonian for excitations above the lowest energy 2in-2out states can be written:
\begin{equation}
{\cal{H}}={u(a)\over{2}}\sum_{i\ne j} \left({a\over{r_{ij}}}\right)\hat{n}_i\hat{n}_j + \nu \sum_i \hat{n}_i^2 -\Delta\sum_{i=1,N_0} (-1)^i \hat{n}_i,
\label{eq2}
\end{equation} 
%
where $u(a)={\mu_0 Q^2\over{4\pi a }}$ is the nearest neighbour Coulomb energy scale for a pair of monopoles.
The mapping  thus re-formulates the spin ice problem as a lattice Coulomb fluid in the grand ensemble \cite{Castelnovo2008,Jaubert2009,Jaubert2011,Jaubert2013,Brooks2014,Kaiser2013,Kaiser2018} with chemical potential for monopole and double monopole creation
$\mu=-\nu$ and $\mu_2=-4\nu$ respectively. The chemical potential  $\mu$ can be calculated for each material from the parameters of the corresponding (DSI) and that for double monopoles is constrained to $\mu_2=4\mu$ by the spin Hamiltonian. Here we add a staggered chemical potential term $\Delta$ which lifts the degeneracy for quasi-particles with charge $\pm Q$ (and with charge $\pm 2Q$) on the sublattices $A$ and $B$, $\mu^A_{\pm}=\mu \pm \Delta$, $\mu^B_{\pm} =\mu \mp \Delta$ and the convention is such that $\Delta>0$ reduces the energy scale for creation of monopoles (double monopoles) with positive charge on $A$ sites and with negative charge on $B$ sites.

The Hamiltonian in eqn.~(\ref{eq2}) is a BC2 type energy function with long range Coulomb interactions, with order parameter $\phi$ given by eqn.~(\ref{OP}) and with $\hat{n}_i$ replacing $S_i$. However, the BC2 and dumbbell models are different as they have different configurational phase spaces and so have different entropies. In the dumbbell model one must take into account the fragmented spin background \cite{Brooks2014}, the so-called Dirac strings \cite{Castelnovo2008,Castelnovo2019}, which emerge in the electrostatics as a divergence free electric field giving Coulomb phase correlations \cite{Isakov2004,Henley2010} at low temperature in the $\phi_0$ phase. These strings possess their own configurational entropy independently of the charges. As a consequence, for zero or finite monopole density and even in the monopole crystal phases, the entropy remains different from that of a lattice Coulomb fluid and hence of the BC2 model. The zero temperature limits for these entropies are well known. The entropy number density of the Coulomb fluid phase is the Pauling entropy, $s_0\approx \ln (3/2)=0.405 $ per tetrahedron. The entropy of the fragmented monopole crystal is that of an ensemble of hard core dimers on a diamond lattice \cite{Nagle1966,Brooks2014}, $s_1\approx \ln(1.3)=0.262$ while that of the double monopole crystal is zero. One can develop an expression for the entropy of both monopoles and strings at the Pauling level of approximation \cite{Pauling1935,Ryzhkin2005} which works well in the monopole fluid phase \cite{Kaiser2018,Castelnovo2019}, but breaks down in the crystal phases. More detailed analysis requires a return to the field theoretic description of the charges and its ensuring lattice Helmholtz decomposition \cite{Brooks2014,Maggs2002}.

\subsection{The double winged phase diagram}

\begin{figure} 
\centering{\includegraphics[scale=0.4]{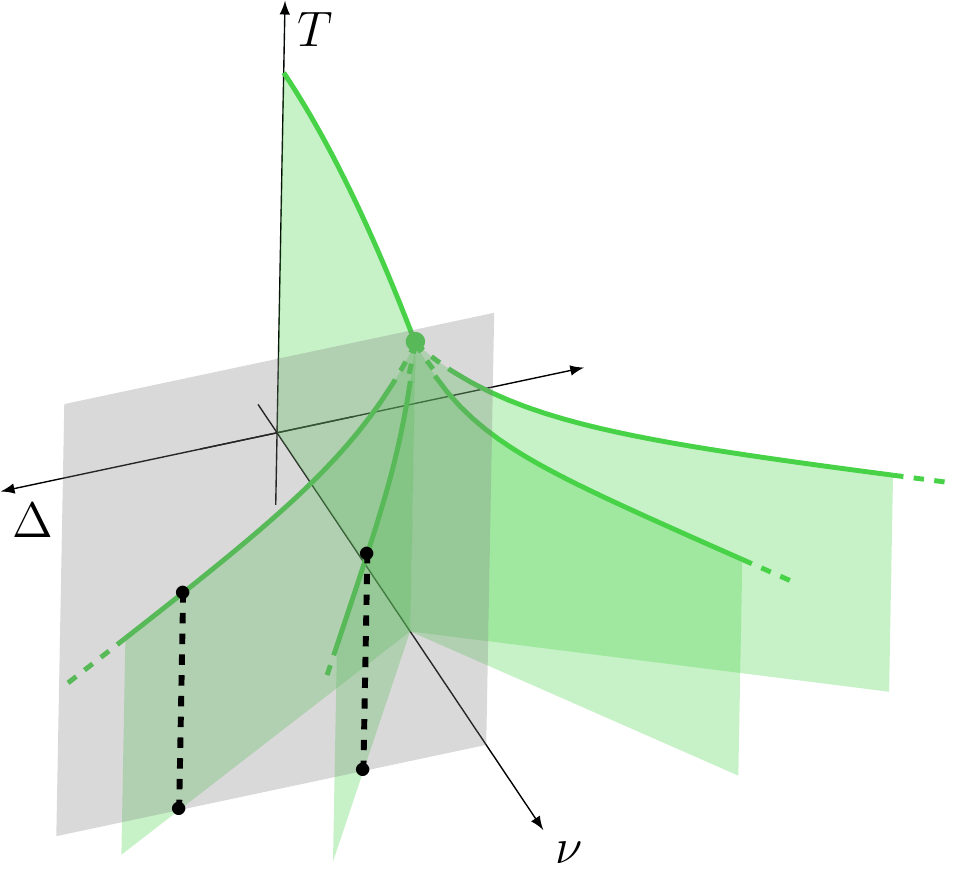}}
\centering{\includegraphics[scale=0.6]{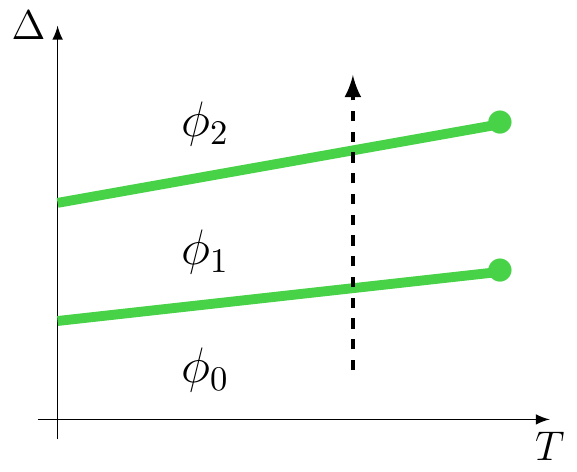}}
\caption[$\;\;$ Phase-space-slice]{{\textbf{$\Delta-T$ plane:} (Left) shaded plane through the full phase diagram at fixed $\nu$ - dotted black lines show the intercepts of the phase boundaries with the plane.
(Right) The fixed $\nu$ plane rotated to give $\Delta$ vs. $T$. Green lines show the phase boundaries, the dotted black line shows an isothermal trajectory in the plane. }}
\label{Phase-space-slice}
\end{figure}

The entropy terms make some quantitative  difference but  similar phase diagrams can be expected for the two models as can be seen from thermodynamic arguments. The monopole free energy can be written
\begin{equation}
\Omega=  N_0 \big ( u_c+\nu n+ 4\nu n_2 - \Delta \phi -sT \big ),
\end{equation}
where $u_c$ and $s$ are the Coulomb energy and entropy number densities.
As we are dealing with ionic crystals, the energy of the three phases are known exactly  at zero temperature  \cite{Brooks2014}: $u_c^{\phi_0}=0$, $u_c^{\phi_1}=-u(a)\alpha/2$, $u_c^{\phi_2}=-2u(a)\alpha$, where $\alpha=1.638$ is the Madelung constant for a diamond lattice. Hence there are zero temperature phase boundaries between the three phases with $\Delta \ge 0$
\begin{eqnarray} 
&\phi_0&:\phi_1, \;\; \Delta=-\frac{u(a)\alpha}{2}+\nu \; K \nonumber \\
&\phi_1&:\phi_2, \;\; \Delta=-\frac{3u(a)\alpha}{2}+3\nu \; K. \label{thermo1}
\end{eqnarray}
Notice that, as both the Coulomb energies and the chemical potentials scale with the square of the charge ($\mu_2=-4\nu$) the five phases intercept the $\Delta=0$ axis at the same point, $\nu^{\ast}=u\alpha/2$. For smaller $\nu$ the Coulomb energy of the double monopole crystal wins out corresponding to spin ice models passing directly into the AIAO phase. However, as $\Delta$ couples linearly to the charge the wings spread out from this point in the $T=0$ plane.

The finite temperature phase boundaries can be estimated from the Clapeyron equation for equilibrium between phases $i$ and $j$:
\begin{equation}
\frac{d\Delta}{dT}= -\frac{s_i - s_j}{\phi_i-\phi_j},
\label{thermo2}
\end{equation}
where $\phi_i$ and $s_i$ are the order parameter and entropy densities of phase $i$. At small temperature we can assume that both order parameter and entropy are constant: $s_0=0.405 $, $s_1=0.262$ and $s_2=0$, $\phi=2, 1$ and $0$,  giving intercepts and slopes for the phase boundaries in a $T-\Delta$ plane for fixed $\nu>\nu^{\ast}$. At higher temperatures this ``fixed entropy approximation'' will break down and the lines should terminate in critical end points as illustrated in Fig. (\ref{Phase-space-slice}).


%
\begin{figure}
\centering{\includegraphics[scale=0.4]{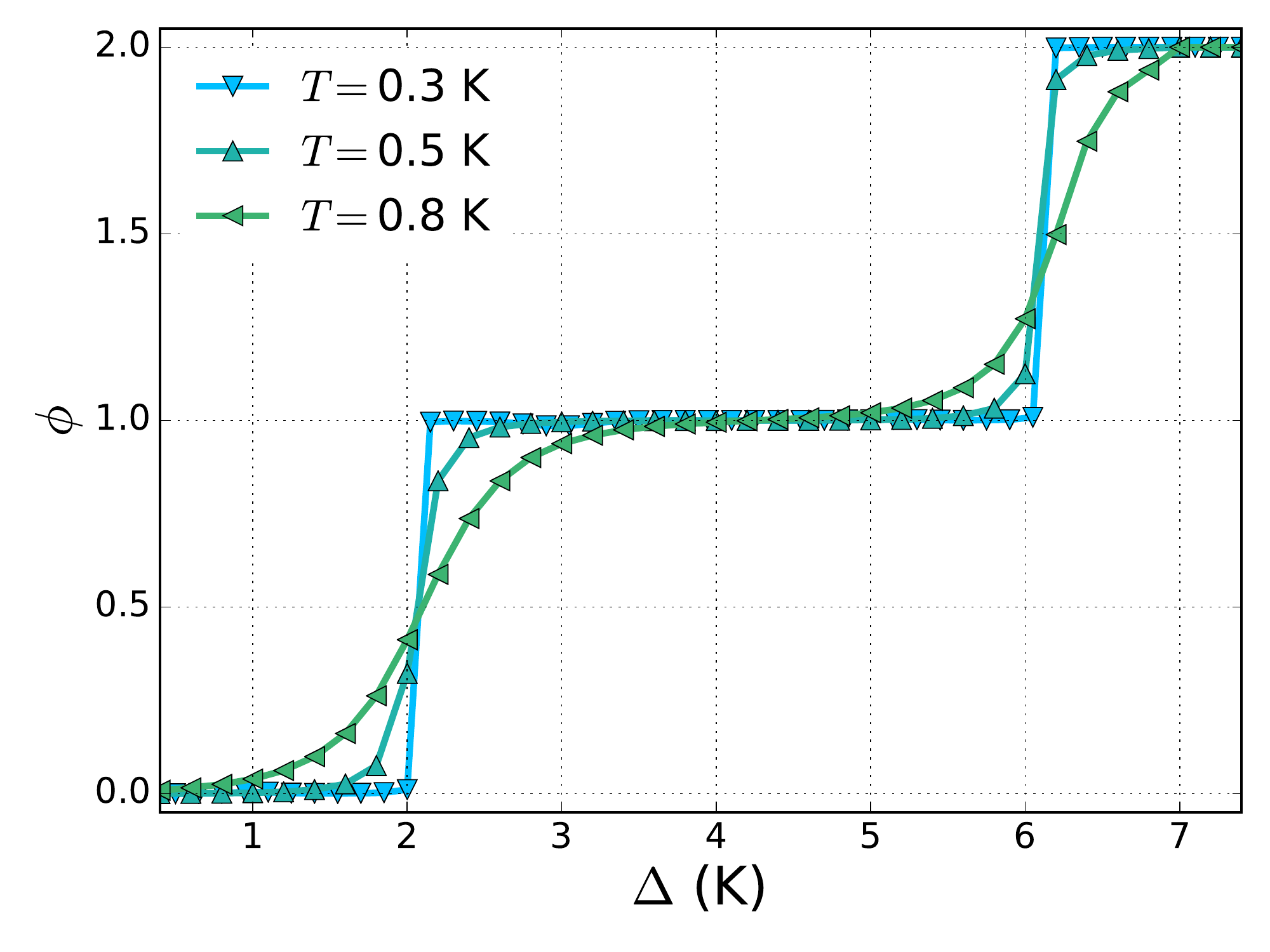}}
\centering{\includegraphics[scale=0.4]{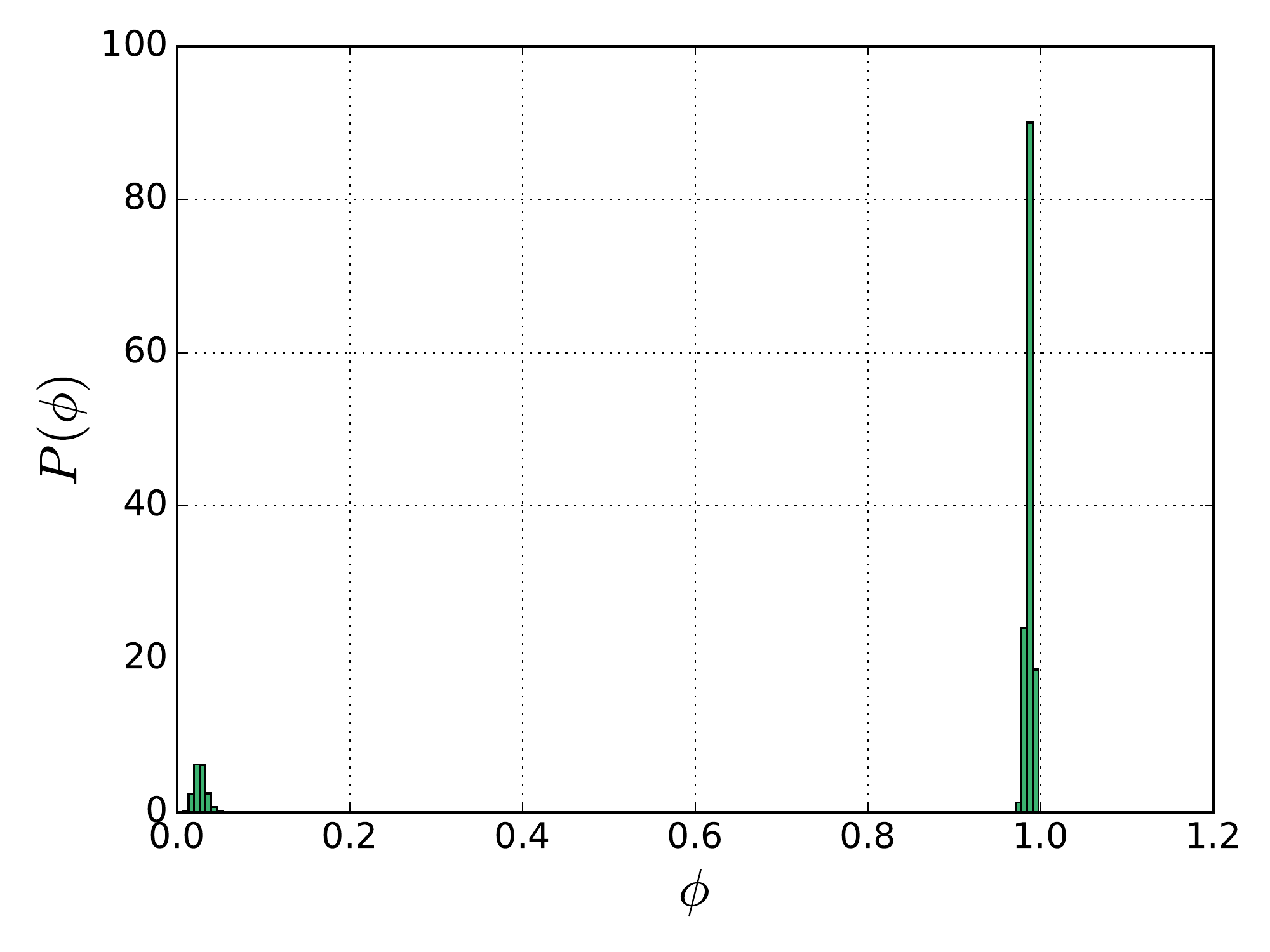}}
\caption[$\;\;$ Cut through]{{\textbf{Multiple monopole crystallization:} (Upper) The order parameter $\phi$ {\it vs.} $\Delta$ simulated from the dumbbell model at fixed $\nu=4.35$ K. Simulations for $N_0=4096$ ($L=8$) and periodic boundaries. All values are in Kelvin. (Lower) Probability density $P(\phi)$ for $\Delta=2.05$ K, $T=0.3$K and $\nu=4.35$ K. 
}}
\label{Cut-through}
\end{figure}

We have  tested this proposition numerically. In Fig.~(\ref{Cut-through}) we show $\phi$ vs $\Delta$ for simulations with $L=8$ for different temperatures for fixed $\nu=4.35$ K and $u(a)=2.88$ K, values estimated for DTO \footnote{The value of $u(a)=2.88$ used here corresponds to a magnetic moment for the spins, $m=9.87\mu_B$, as deduced from crystal field calculations \cite{Yavorskii2008} rather than the $10 \mu_B$ often quoted in the literature, which gives $u(a)=3.07$K and a corresponding difference in the energy scale for the phase diagram} . 
For the lowest temperature, sharp steps are indeed observed in $\phi$ from $\phi \approx 0$ to $\phi \approx 1$ and from $\phi\approx 1$ to $\phi \approx 2$ at a value slightly greater than $2$ K and $6$ K respectively. The data is consistent with two  $1^{st}$ order phase transitions from $\phi_0$ to $\phi_1$ and from $\phi_1$ to $\phi_2$. As the temperature is increased the steps in $\phi$ become rounded, consistent with the model passing through a critical end point with the transitions evolving to crossovers at high temperature. The singular nature of the transition between $\phi_0$ and $\phi_1$ at $T=0.3$ is confirmed in the lower panel where we show the probability density $P(\phi)$ estimated during the simulation. The distribution is sharply peaked near $\phi=1$ but shows a lower peak in probability near $\phi=0$, consistent with fluctuations between metastable states separated by a finite jump in order parameter space. The inequality in the peak heights shows that for these parameters, the system has passed into the ordered phase. The lower peak in distribution occurs at a small but finite value of $\phi$, consistently with $\Delta$ breaking the $Z_2$ symmetry of the lattice even in the $\phi_0$ phase. The five phases confirming the double winged structure are indeed the Coulomb fluid (spin ice) phase ($\phi_0$), the two fragmented monopole crystal phases \cite{Brooks2014,Borzi2013} ($\phi_1$) and  the double monopole crystal AIAO phases ($\phi_2$). 

The position of the $1^{st}$ order transitions in parameter space can be estimated using eqns. (\ref{thermo1}) and (\ref{thermo2}). 
Taking the DTO values for $\nu$, and $u(a)$ and
$\Delta>0$ the zero temperature intercept of the two phase boundaries are
\begin{eqnarray} 
&\phi_0&:\phi_1, \;\; \Delta=1.99 \; K \nonumber \\
&\phi_1&:\phi_2, \;\; \Delta=5.97 \; K.
\end{eqnarray}
Assuming complete jumps in the order parameter at the transition,  one finds for $T=0.3$ K
\begin{eqnarray} 
&\phi_0&:\phi_1, \;\; \Delta=2.03 \; K \nonumber \\
&\phi_1&:\phi_2, \;\; \Delta=6.05 \; K,
\end{eqnarray}
in close agreement with the results of Fig. (\ref{Cut-through}).

\subsection{A critical end point}

\begin{figure}
\centering{\includegraphics[scale=0.38]{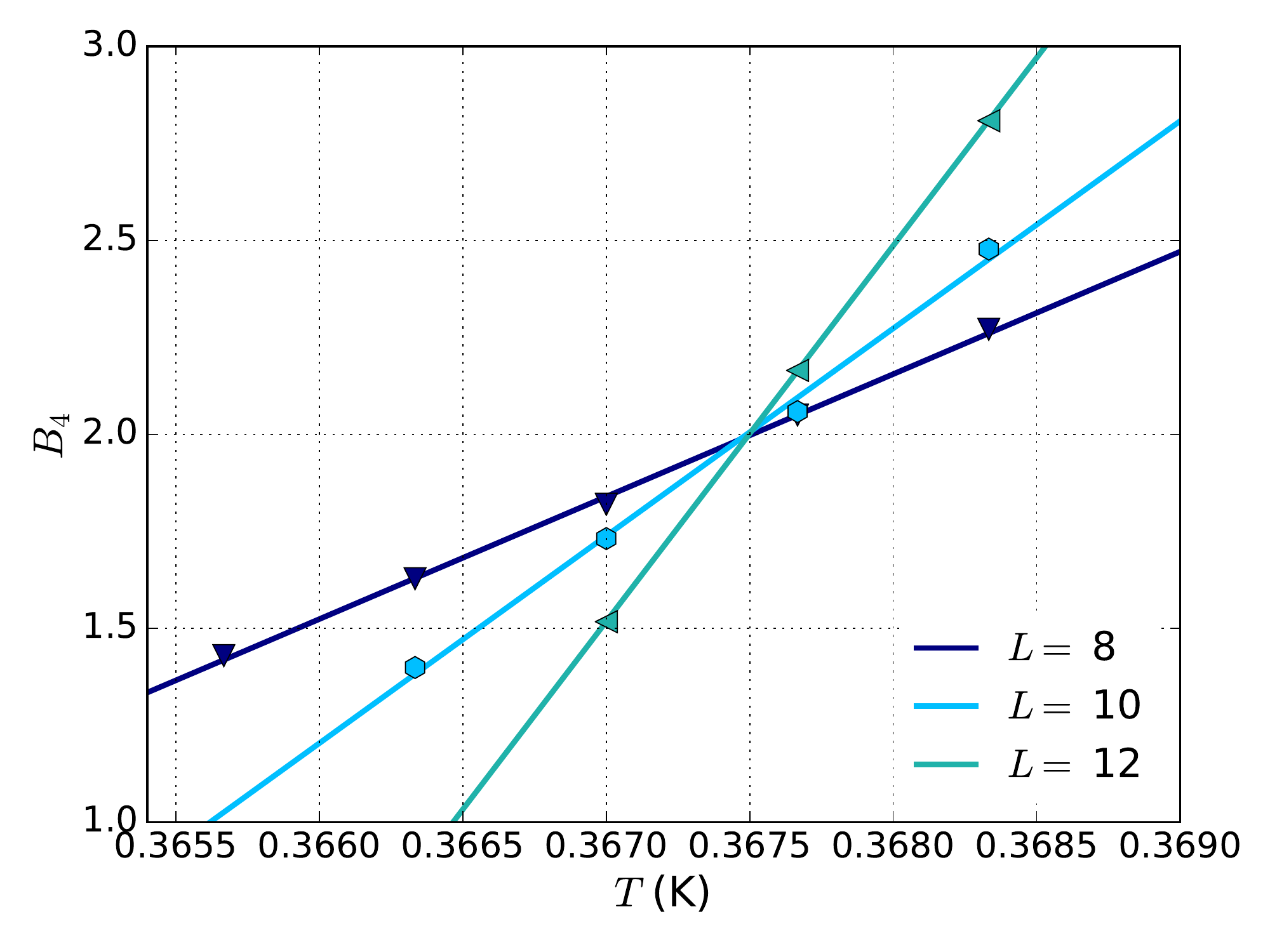}}
\centering{\includegraphics[scale=0.38]{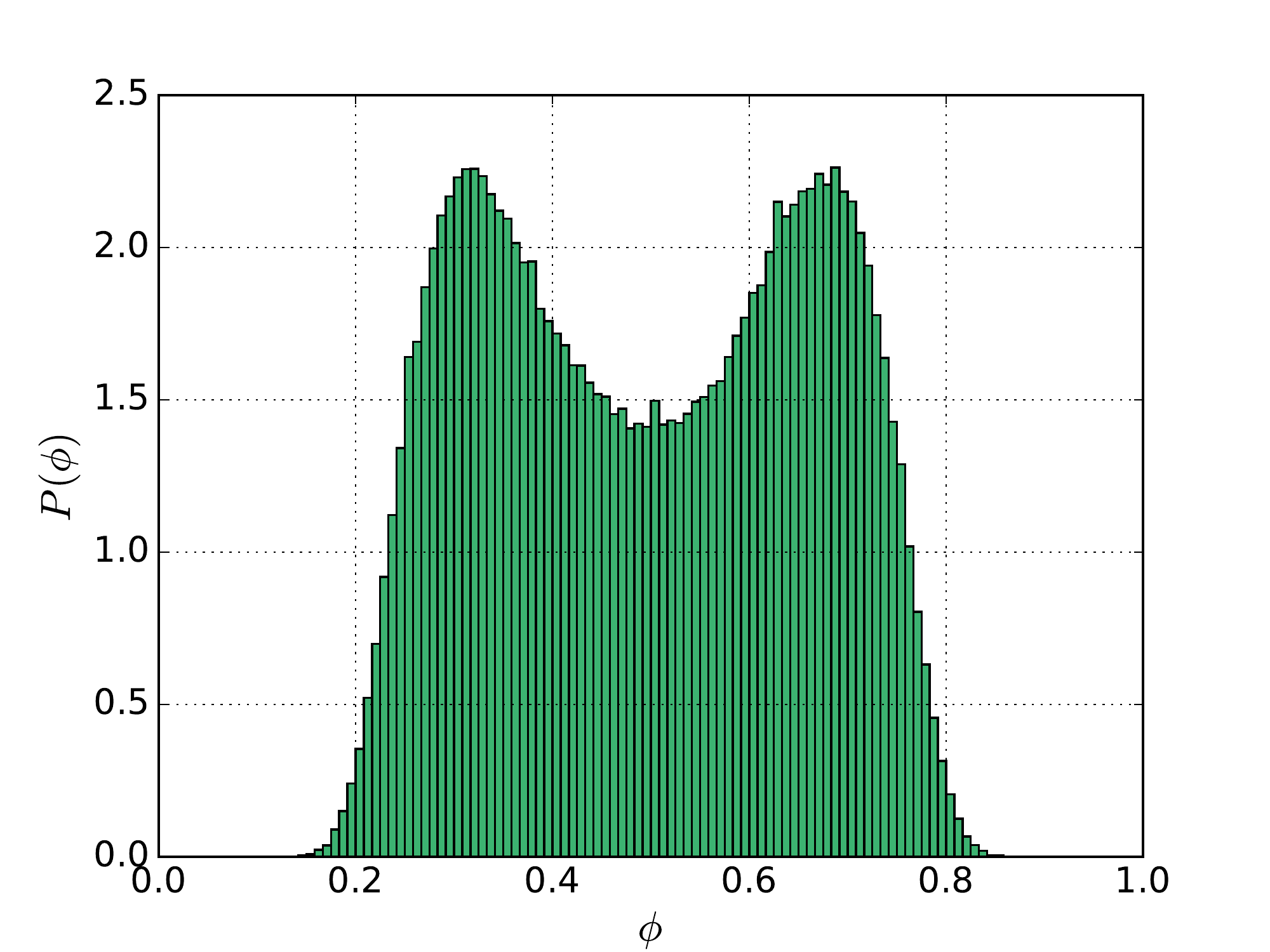}}
\caption[$\;\;$ Binder]{{\textbf{Locating the critical point.} 
(Top) The Binder cumulant $B_4$ for the emergent order parameter $\varphi=\phi-\phi_c$ for fixed $\nu=4.35$ K and $\Delta=\Delta_c=2.03745$ K and $\phi_c=0.42$ at the $\phi_1:\phi_0$ phase boundary (see text). 
(Bottom) Probability density $P(\phi)$ at the critical end point, $\nu=4.35$ K, $\Delta_c=2.03745$, $T_c=0.36752$. 
}}
\label{Binder}
\end{figure}

We have made a quantitative estimate of the position of one critical end point, that for the transition from $\phi_1$ to $\phi_0$ for $\nu=4.35$ K. This can be extracted from the crossings of the Binder cumulant~\cite{binder1981finite}, $B_4$ for the emergent Ising like order parameter at the critical end point, $\varphi=\phi-\phi_c$, where $\phi_c=\langle \phi(T_c)\rangle$:
\begin{equation}
B_4(T,\Delta)=\frac{\langle(\varphi)^4\rangle}{\langle(\varphi)^2\rangle^2}. 
\end{equation}
The parameters $\Delta_c$, $T_c$ and $\phi_c$ were estimated using an iterative procedure. 
A first estimate of $T_c$ and $\Delta_c$ was made by following the evolution of $P(\phi)$ from a double to single peak distribution. From here a more accurate estimate of $\Delta$ was found from the maximum of the susceptibility for $\phi$. This estimate was found to be invariant under small temperature changes and the result can be established with high precision  \cite{Hamp2015}. We find $\Delta_c=2.03745\pm 0.00005$ K. The evolution of $B_4$ with temperature for this $\Delta_c$ is shown in Fig. (\ref{Binder}) for system sizes $L=8,10,12$ and for $\phi_c=0.42$. 
A crossing point is found for $T=0.36752\pm0.00001$ K with $B_4(T,\Delta)=2.03 \pm 0.01$. The crossing value should be compared with other Ising like systems: $B_4(T_c)=1.60$ for the 3D Ising model \cite{Fenz2007} and $B_4(T_c,H_c)=1.86$ for spin ice with field $\vec H$ along the $[111]$ cubic axis \cite{Hamp2015}. We found that the value depends on $\phi_c$, reducing to $\approx 1.60$ for $\phi_c=0.5$, with crossing at $T=0.3672$ K,  but in this case the crossing was not so accurately defined. From this analysis we estimate $T_c=0.3675 \pm 0.0003$ K. In Fig. (\ref{Binder}) we show the probability density function, $P(\phi)$ calculated at $T_c,\Delta_c$ which resembles qualitatively the universal function $P(M)$ for the magnetisation $M$ of the three dimensional Ising model at the critical point \cite{Rummukainen1998,Lopes2016} and is centred on $\phi=0.5$. The universality class of the critical point is discussed further below through a dynamical finite size scaling analysis and the measurement of the fluctuation-dissipation ratio.

%

%
%

\subsection{The multicritical region}\label{multicritical}

How the wings meet in the multicritical region is a rather subtle question. The intersection of  the five phases on the $\Delta=0$ plane at a single penta-critical point is unlikely, as the plane is characterised by two variables $T$ and $\nu$ only. This allows the system to tune to a tri-critical point in which both the quadratic and quartic terms in an expansion of the free energy in $\varphi$ are zero \cite{Cardy1996}. However, a penta-critical point would require the annulation of the sixth order term which, without a third parameter would be accidental. In the model studied here, emergent from the DSI, the monopole and double monopole costs are fixed: $\mu_2=4\nu$. Floating $\mu_2$ away from this value could allow the tuning necessary to establish penta-criticality but the evidence presented below suggests that in our case the wings meet in two stages which indeed maintains the tri-criticality of the BC1 model.

\begin{figure}
\centering{\includegraphics[scale=0.7]{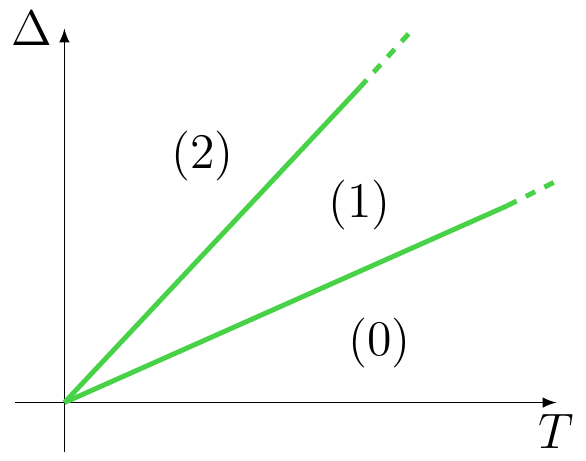}}
\centering{\includegraphics[scale=0.7]{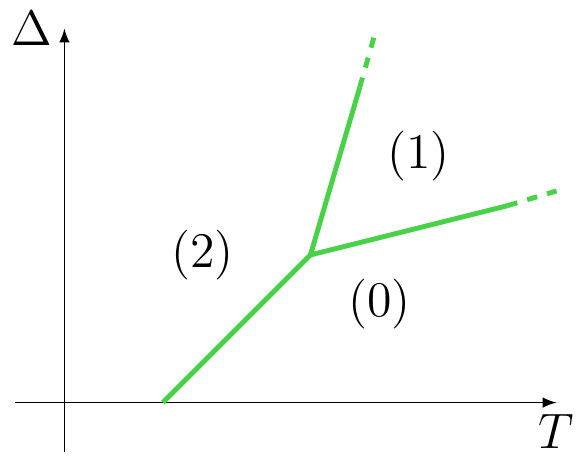}}
\caption[$\;\;$ Binder]{{\textbf{Low temperature ($T\ll \nu$):}  {\bf (Left)} $\nu=\nu^{\ast}$, {\bf (Right)} $\nu < \nu^{\ast}$. {\bf (0)} monopole fluid ($\phi_0$), {\bf (1)} monopole crystal ($\phi_1$), {\bf (2)} double monopole crystal ($\phi_2$)  . }}
\label{Multi}
\end{figure}
%

%
%
The five phases do however meet at $T=0, \Delta=0,\nu=\nu^{\ast}$.
A phase boundary between $\phi_2$ and $\phi_0$ then rises from the five phase intercept 
along
\begin{equation}
 4(\nu^{\ast}-\nu)=s_0T,
 \end{equation}
 where we have again assumed a constant entropy approximation, valid for $T\ll \nu^{\ast}$. Within this approximation the Pauling entropy of the spin ice vacuum gives a finite slope away from $T=0$ which takes the system away from $\phi_1$ and this phase is suppressed everywhere in the $\Delta=0$ plane except the special point at $T=0$. This can be seen in detail by analysis of the three different free energies. 
 As a consequence, $\Delta-T$ planes for $\nu=\nu^{\ast}$ and $\nu\leq\nu^{\ast}$ take the form shown in Fig.~(\ref{Multi}) at low temperature. In the latter case there is a finite temperature order disorder transition between $\phi_2$ and $\phi_0$ along the $\Delta=0$ axis, ensuring that the $\phi_2$, $\phi_1$ and $\phi_0$ phases meet at a triple point for finite $\Delta$. The slopes of the phase boundaries, $\eta_{ij}$ can be estimated from eqn. (\ref{thermo2}): $\eta_{21}=(s_1-s_2)=0.262$, $\eta_{10}=(s_0-s_1)=0.143$ and $\eta_{20}=\frac{1}{2}(s_0-s_2)=0.203$ and the triple point, which is allowed because of the linear dependence between the three boundary curves, occurs at
 \begin{equation}
 T_t=\frac{2(\nu^{\ast}-\nu)}{2s_1-s_0}.
 \end{equation}
 %
As $|\nu-\nu^{\ast}|$ increases, the $\Delta=0$ transition temperature increases until at the tricritical point the transition changes from $1^{st}$ to $2^{nd}$ order, at which point the line structure in Fig.~(\ref{Multi}b) will have evaporated through  critical end points.

Heating up to the critical end points should therefore lead to the wings meeting in two stages with three separate tri-critical points, with all lines meeting tangentially \cite{Taufour2016}. Two of these being the critical termination of the triple points for finite positive and negative $\Delta$ and the third, a classic tri-critical point separating ordered and disordered phases for $\Delta=0$. 

\begin{figure}
\centering{\includegraphics[scale=1.1]{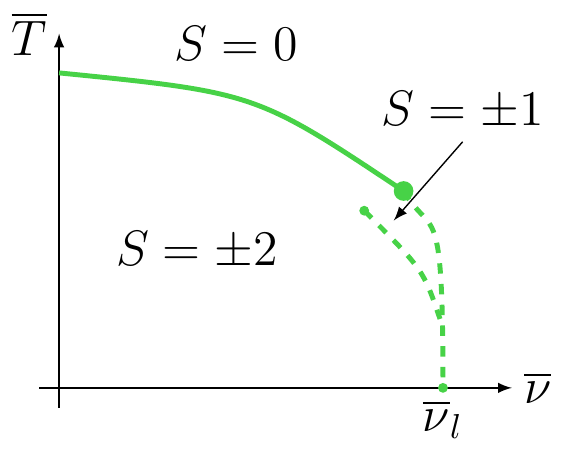}}
\caption[$\;\;$ Blume Capel $S=2$]{{\textbf{Symmetry breaking in the $S=2$ Blume Capel model:} The $\bar{\Delta}=0$ plane of the $S=2$ Blume Capel model \cite{Lara1998}}.}
\label{BC2-phase}
\end{figure}

This scenario can be compared with that of the BC2 model. In this case, the same five phase intercept occurs at $T=0$ but the $\phi_2$ : $\phi_0$ phase boundary now rises vertically as the entropy of both phases approach zero as $T$ goes to zero. 
However, at the level of mean field and pair approximation calculations \cite{Plascak1993,Lara1998} a small sliver of $\phi_1$ appears at higher temperatures, stabilised by the entropy of spin fluctuations. The $\phi_1$ : $\phi_2$ boundary ends at a critical point in the $\bar{\Delta}=0$ plane as shown in Fig. \ref{BC2-phase}. This suggests that the tri-critical point of the BC1 model is again maintained with this time, separate intercepts onto the central plane for the two wings for positive and for negative $\bar{\Delta}$. 

The undershoot and overshoot of the wing interceptions in the dumbbell and BC2 models illustrates the accidental nature of penta-criticality for this set of parameters and strongly suggests that a generalised model with independent $\mu$ and $\mu_2$ could be tuned to include a penta-critical point.

\section{Dynamic Scaling at a Critical End Point}\label{KZ}

\subsection{Critical slowing down}

\begin{figure} 
\centering{\includegraphics[scale=0.4]{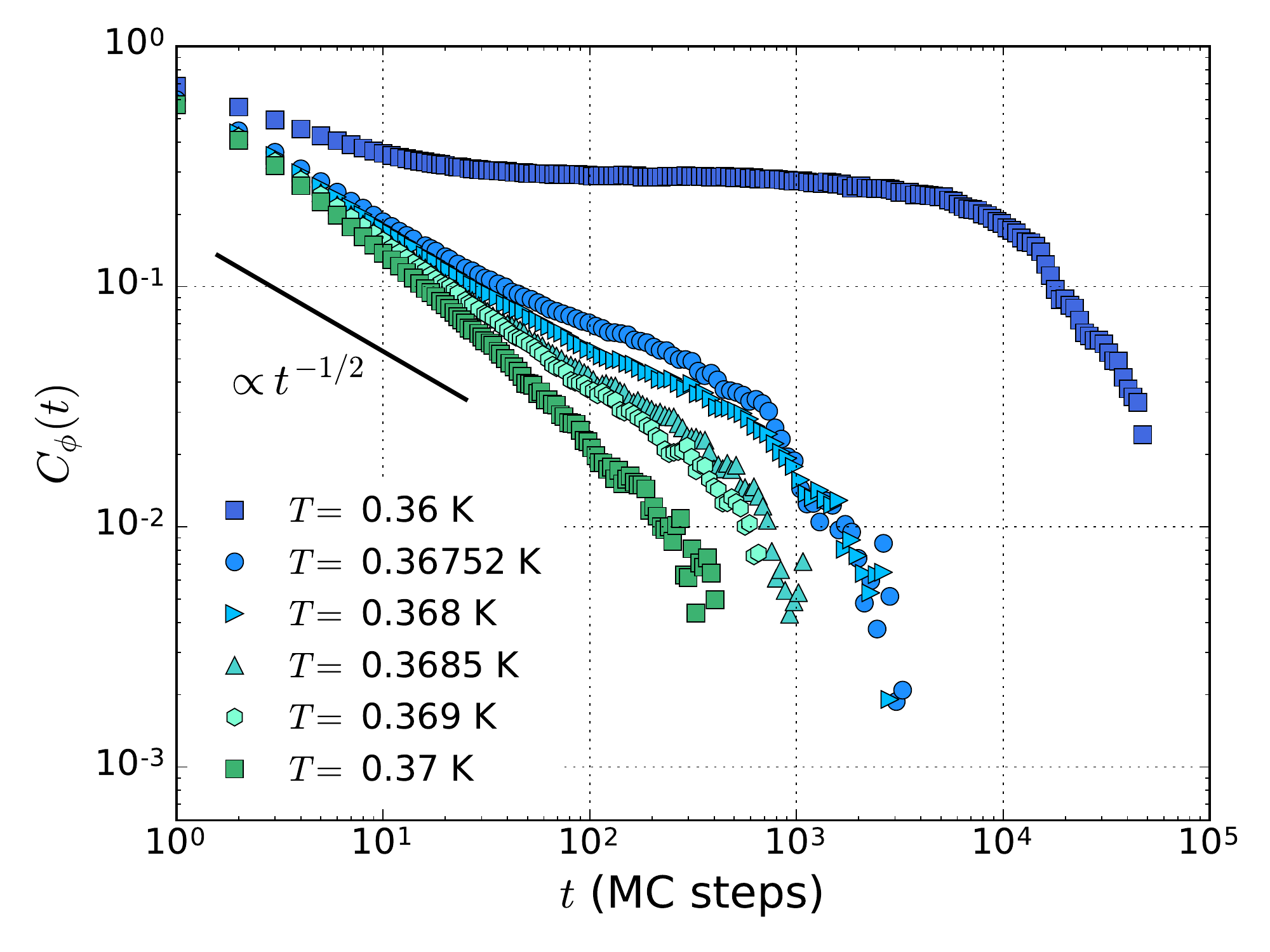}}
\caption[$\;\;$ Autocorrelations]{{\textbf{Critical correlations at equilibrium:} Autocorrelation function for the monopole crystal $C_{\phi}(t)$ {\it vs.} $t$ for temperatures close to the critical temperature. Solid black line shows power law decay with exponent $-1/2$.}}
\label{Dyn}
\end{figure}

Along the lines of critical end points there are divergent time scales associated with  the diverging correlation lengths and critical slowing down. In Fig. (\ref{Dyn}) we show the evolution of the auto-correlation function 
\begin{equation}
C_\phi(t)=\frac{\left<\phi_i(t)\phi_i(0)\right>-\left<\phi_i(t)\right>\left<\phi_i(0)\right>}{
\left<\phi_i(0)^2\right> - \left<\phi_i(0)\right>^2}, 
\label{eq:cphi}
\end{equation}
with  Metropolis Monte Carlo time as the critical end point for $\nu=4.35$ K, $\Delta_c=2.03745$ is approached along the temperature axis. Note that in  
eqn.~(\ref{eq:cphi}) we study the critical dynamics using the local spin-spin autocorrelation function, which is distinct from the autocorrelation function of the global order parameter $\phi$. The spin autocorrelation function is statistically easier to access, but it also captures the critical slowing down. The data shows decay of correlations at equilibrium for a system of size $L=12$.

As the transition is approached from above the correlation time increases and $C_\phi(t)$ develops a powerlaw decay with exponent $\alpha \approx -1/2$,  out to a maximum of the order of $10^3$ Monte Carlo steps per dumbbell. The best power law is observed for a temperature $T\approx 0.369$ K, higher than the $T_c=0.3675$ estimated from analysis of the Binder cumulant. 

Within the critical region time scales and length scales are bridged via the dynamical critical exponent~\cite{Hohenberg1977} $z$. 
The correlation time $\tau$ diverges with the correlation length $\xi$ as 
\begin{equation}
\frac{\tau}{\tau_0}=\left(\frac{\xi}{a}\right)^z. \label{divt}
\end{equation}
Hence, as the spatial correlation function for the local order parameter in dimension $d$ scales with distance in the critical region as $r^{-(d-1-\eta)}$, with $d-\eta$ the anomalous dimension of the universality class, one  expects dynamical scaling of the form $C_\phi(t)\sim t^{-(d-1-\eta)/z}$. Taking $\eta \simeq 0.0363 \ll 1$ and $z \approx 2$, which should be the case for local dynamics in the three dimensional Ising universality class, one finds an exponent $\alpha \approx 1/2$ as observed. The shift in effective transition temperature away from the Binder crossing point is expected and is due to finite size effects. 

The cut off of the power law is compatible with the finite size cut off of $\xi$: $\xi_{max} < L'=\frac{4La}{\sqrt{3}}$. Taking $z\approx 2$, $L=12$ and microscopic time $\tau_0$ equal to one Monte Carlo time step indeed gives a cut off to the critical scaling of the order of $10^3$ Metropolis time steps. Below the critical temperature the time correlation function develops a plateau which decays at longer time scales. This is consistent with a change of regime in the dense crystalline phase where decay of correlations is due to the creation and propagation of monopole holes \cite{Jaubert2015}.



\subsection{Kibble-Zurek scaling}

A more quantitative picture of the emergent universality class of the critical end point can be achieved by following the dynamical Kibble-Zurek \cite{Kibble1980,Zurek1985} scaling protocol proposed in [\onlinecite{Hamp2015}]. In this scenario the field like scaling variable is swept in time through a cycle with characteristic time scale $\tau_Q$:
\begin{equation}
\lambda(t) =\frac{\Delta(t)-\Delta_c}{\Delta_c}=\lambda_0 \sin\left(\frac{\pi t}{2\tau_Q}\right), \label{sweep}
\end{equation}
with temperature fixed at $T_c$. Far from the critical point the equilibrium time scale is small compared with $\tau_Q$ so that the evolution is adiabatic but as the critical point is approached the equilibrium time scale diverges. As a consequence, at a given point in each cycle the system falls out of equilibrium creating hysteresis loops in the thermodynamic observables, whose magnitude depends on sweep time. 

\begin{figure} 
\centering{\includegraphics[scale=0.35]{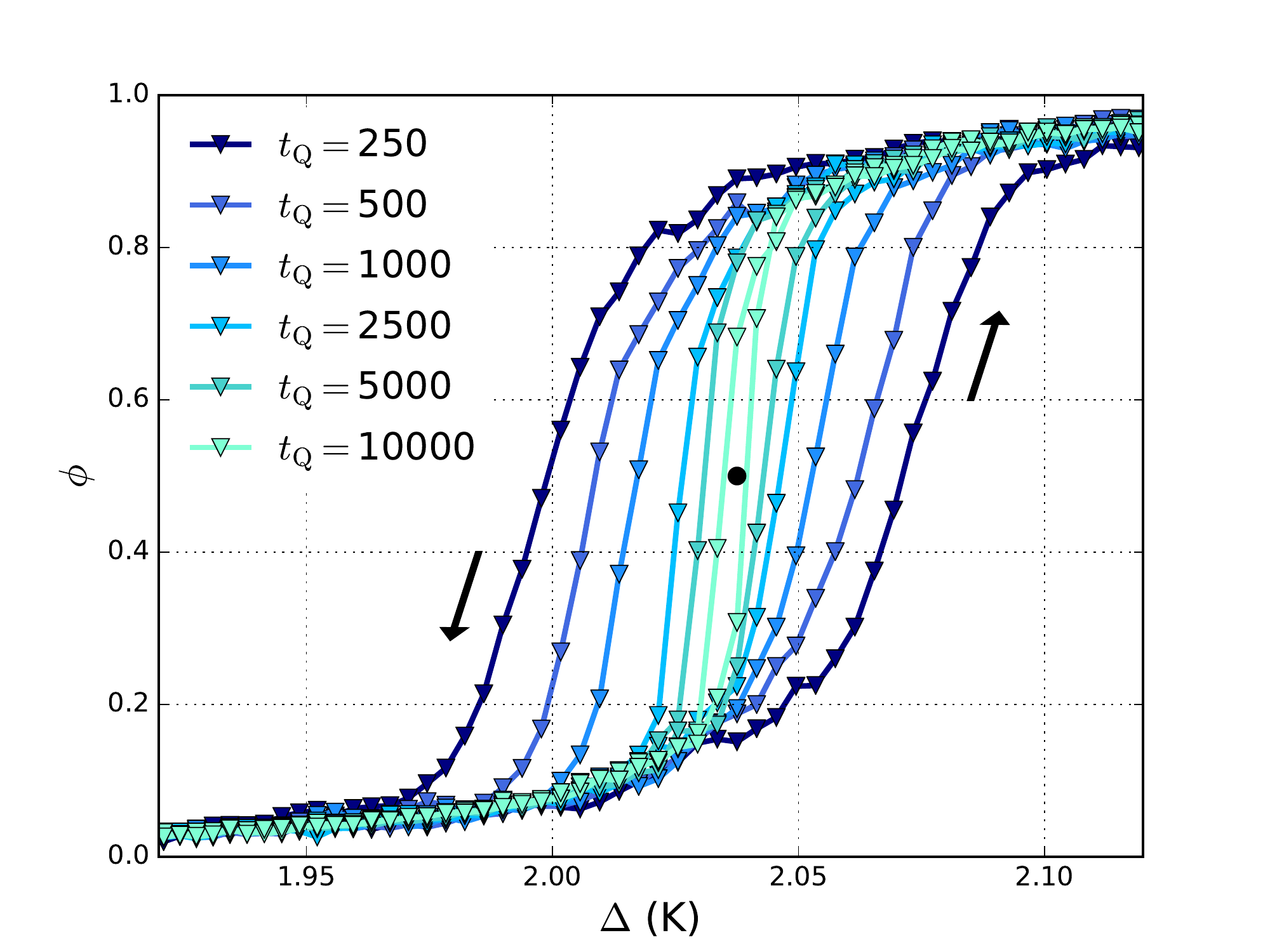}}
\caption[$\;\;$ Hysteresis loops]{{\textbf{Hysteresis loops:} $\phi$ {\it. vs.} $\Delta$ for different sweep rates. System size $L=8$. The $\phi(t)$ is a configurational average of $500$ samples, each starting at equilibrium at $t=0$.}}
\label{Hyst}
\end{figure}

In Fig. \ref{Hyst} we show the evolution of $\phi$ with $\Delta$ at $T_c$ for a system of size $L=8$ and $\lambda_0=0.1$ for different $\tau_Q$. Hysteresis loops centred on $\phi=\phi_c$ indeed appear and their amplitude falls to zero as $\tau_Q$ increases. 

Following eqn.~(\ref{divt}), the correlation time diverges along the field axes as $\tau \sim \tau_0 \lambda^{-z\nu_{\lambda}}$ with $\nu_{\lambda}$ the field driven correlation length exponent, so that $\tau$ and sweep time are related through eqn.~(\ref{sweep}). The crossover from adiabatic to out of equilibrium response occurs around the point $\left|\frac{d\tau}{dt}\right|=1$, which fixes a characteristic Kibble-Zurek time scale, $t_{KZ}=\tau_0\left(\frac{\tau_Q}{\tau_0}\right)^{z\nu_{\lambda}/(z\nu_{\lambda}+1)}$.

 In the critical region the fall from equilibrium of the emergent order parameter $\varphi$  is captured by the dynamical scaling hypothesis \cite{Hohenberg1977}
\begin{equation}
\varphi(t)= \left(\frac{1}{t_{KZ}}\right)^{\frac{D_{\lambda}}{z}} {\cal{G}}\left(\frac{t}{t_{KZ}}\right),
\end{equation}
where $D_{\lambda}$ is the scaling dimension of the field and ${\cal{G}}(x)$ is a scaling function allowing for data collapse for different data sets. For short range systems, up to and including the upper critical dimension $D_{\lambda}= \frac{d\nu_{\lambda} -1}{\nu_{\lambda}}$, while in the Gaussian regime $D_{\lambda}=1$. 

\begin{figure}
\vspace{2em}
\centering{\includegraphics[scale=0.35]{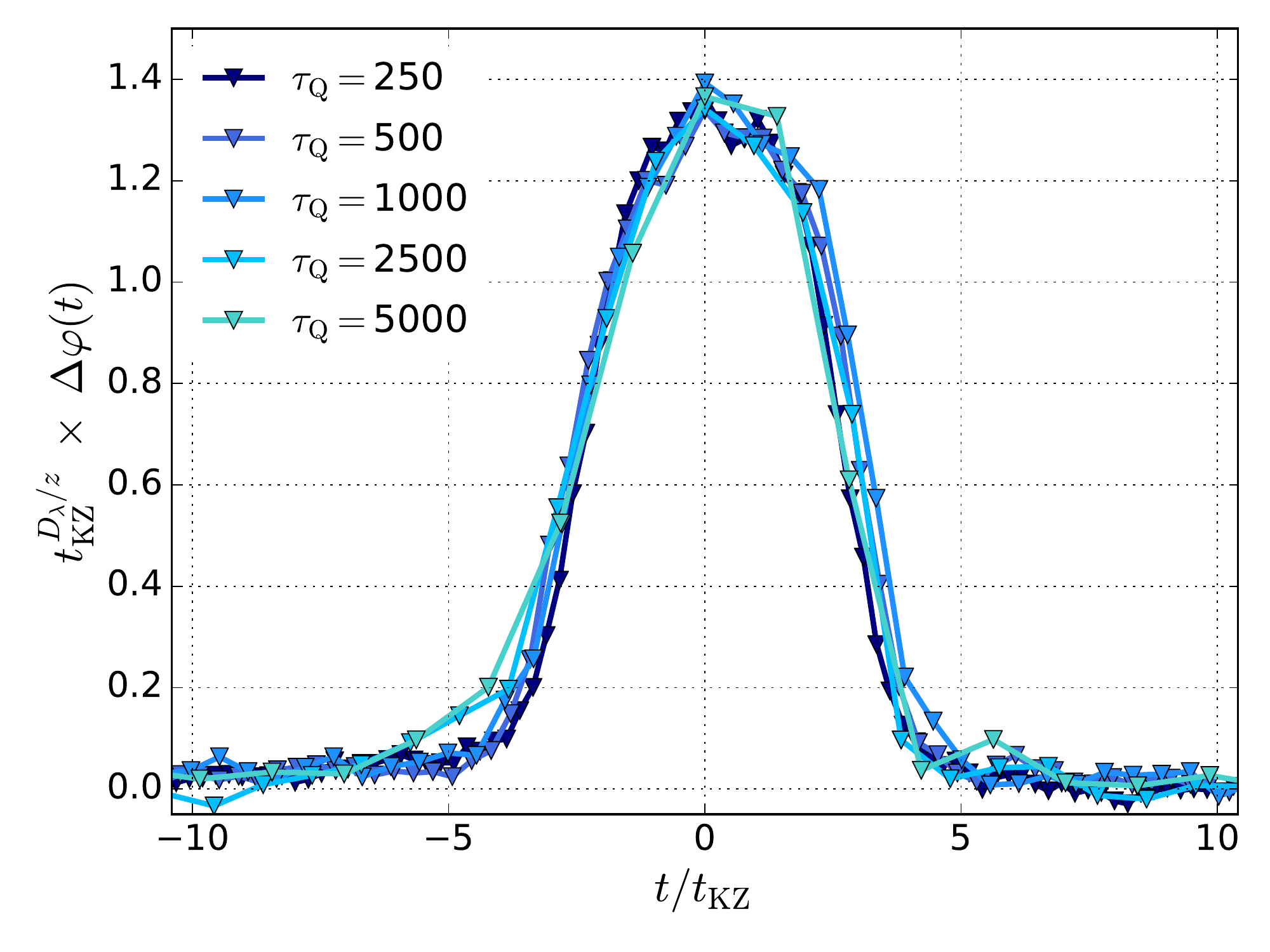}}
\caption[$\;\;$ KZ scaling]{{\textbf{Kibble-Zurek scaling:} $\Delta\varphi$ 
weighted by $t_{KZ}^{D_{\lambda}/z}$ as a function of 
$\left(\frac{t}{t_{KZ}}\right)$ for $\nu_{\lambda}=0.4$ and $z=2$.}}
\label{Dyn-C}
\end{figure}

In Fig. (\ref{Dyn-C}) we show the Kibble-Zurek scaling collapse for $\Delta \varphi=\varphi(\lambda)\uparrow-\varphi(\lambda)\downarrow$, the difference in order parameter values on an up and down swing of the cycle. We find a convincing collapse using known values for the three dimensional Ising universality class \cite{Pelissetto2002} and local stochastic dynamics \cite{Wansleben1991}, $\nu_{\lambda}=0.4$, $z=2$. We do not have access to large enough system sizes or high enough resolution on our data to distinguish between three dimensional XY and Ising universality classes but the collapse shown is superior to that found using Gaussian exponents. Hence, as in [\onlinecite{Hamp2015}] for the critical point observed for spin ice in a $[111]$ field, we can exclude  the possibility of the long range Coulomb interactions influencing the the universal fluctuations. 

It is worth remarking that, however accurate the data, the field scaling Kibble-Zurek  protocol cannot unambiguously establish Ising universality, as the procedure accesses only one of the two static scaling dimensions, $D_{\lambda}$; $\nu_{\lambda}$ being independent of the second dimension $D_{\epsilon}$. This yields  $D_{\lambda}=\frac{1}{\delta\nu_{\lambda}}=\frac{\beta}{\nu}$, where exponents have their usual meaning \cite{Cardy1996}, establishing weak universality only \cite{Suzuki1992}. This in principle allows for variation of $\beta$ and $\nu$ within the weak universality constraint \cite{Taroni2008}. A thermal Kibble-Zurek protocol would fix the two static exponents through the presence of both $D_{\lambda}$ and $\nu$ although one would then have a three parameter fit ($D_{\lambda}, \nu, z)$ for a single expression. 

\subsection{Aging and fluctuation-dissipation ratio}

\label{sec:FDT}


A further remarkable consequence of the diverging time scale at the critical point is that if the system is suddenly quenched from a high temperature to $T_c$, it will not reach equilibrium within the time window offered by  experiments or simulation. As a result, systems quenched to criticality display universal aging properties, reported in an extensive literature~\cite{godreche2000response,godreche2000responseb,berthier2001nonequilibrium,henkel2001aging,PhysRevE.68.065101,calabrese2002two,calabrese2002aging,mayer2003fluctuation,mayer2004reply,calabrese2005ageing,PRUDNIKOV2015774} showing explicitly  that the tools developed in the context of materials with slow glassy dynamics are highly relevant for aging critical dynamics. 

Two important properties emerge from the out-of-equilibrium dynamics. First, the time correlation function in eqn.~(\ref{eq:cphi}) is no longer time translationally invariant, so that one needs to explicitly follow the dependence on the time spent at criticality since the quench. 
As a result the system slowly ages towards equilibrium in a manner reminiscent of disordered glassy systems~\cite{bouchaud1998out}. 
Second, the fluctuation-dissipation theorem (FDT) which, in equilibrium connects linear response functions to time correlations functions, is no longer valid. In glassy materials, violations of the FDT have been found to take simple forms with appealing physical interpretations~\cite{cugliandolo1993analytical,cugliandolo1997energy,bouchaud1998out,crisanti2003violation}. Studies of FDT violations in systems quenched to criticality show that the deviations from the equilibrium relation contains direct information about the universality class of the model~\cite{godreche2000response,calabrese2002two}.


Inspired by these studies, we consider a numerical protocol in which the temperature is instantaneously varied from $T=1000~K$ to $T=T_c=0.3685~K$, and denote $t_w$ the ``waiting time'' spent at $T_c$ since the quench. We then define 
\begin{equation}
\tilde{C}_\phi(t,t_w)=\frac{\left<\phi_i(t_w)\phi_i(t)\right>-\left<\phi_i(t_w)\right>\left<\phi_i(t)\right>}{
\left<\phi_i(t)^2\right> - \left<\phi_i(t)\right>^2}, 
\label{eq:cphiaging}
\end{equation} 
where the waiting time dependance is now made explicit. Indeed, as with other critical systems, we find that the time decay of the spin auto-correlation function is not just a function of $t-t_w$ but now depends explicitly of both times. 
We also define the linear response function associated with the time correlation function in eqn.~(\ref{eq:cphiaging}) as 
\begin{equation}
\chi_\phi (t,t_w) = \frac{\partial \langle \phi_i (t) \rangle}{\partial \Delta_i(t_w)},
\end{equation}
where $\Delta_i$ is the field conjugate to the local order parameter $\phi_i$. We introduce the normalised response function $\tilde{\chi}_\phi = \chi_\phi / (\left<\phi_i(t)^2\right> - \left<\phi_i(t)\right>^2)$, such that the equilibrium FDT reads $\tilde{\chi} = (1-\tilde{C}_\phi)/T$. 

In the aging regime following a quench, the FDT is not expected to be satisfied, and it can generically be rewritten as 
\begin{equation}
\tilde{\chi}_\phi(t,tw) = \frac{X(t,t_w)}{T} \left( 1-\tilde{C}_\phi(t,t_w) \right),
\label{eq:X}
\end{equation}
which defines the fluctuation-dissipation ratio $X(t,t_w)$~\cite{cugliandolo1993analytical}. Physically, eqn.~(\ref{eq:X}) is appealing as it has the same mathematical form as in equilibrium, with the difference that the thermal bath temperature is replaced by an effective temperature $T / X(t,t_w)$~\cite{cugliandolo1997energy}. 

\begin{figure}
\centering{\includegraphics[width=0.99\columnwidth]{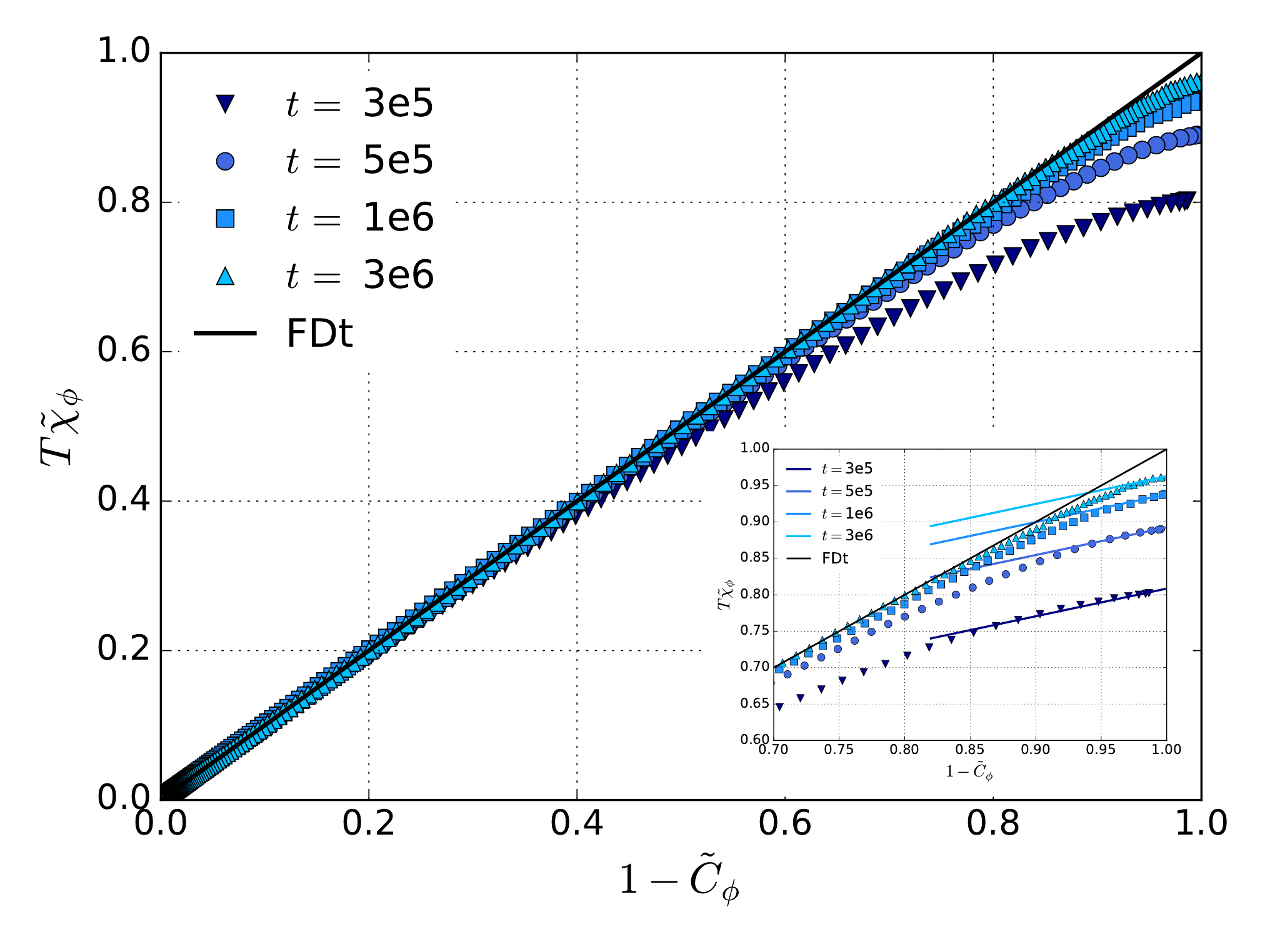}}
\caption{ \textbf{FDT violations} Parametric plot of the response function against the correlation function, for various fixed times $t$ and using $t_w$ as a running parameter after a quench at the critical point $T_c,\Delta_c$. The system size is $N_0=13824$ ($L=12$), and data are averaged over 500 independent quenches. The inset show limiting values for long times. The solid lines show the expected universal value $X_\infty = 0.38$ for the 3D Ising universality class. }
\label{fig:FDT}
\end{figure}

In Fig.~(\ref{fig:FDT}) we display FDT violations by representing $T \tilde{\chi}_\phi(t,t_w)$ as a function of 
$1-\tilde{C}_\phi(t,t_w)$, for a fixed time $t$ and using $t_w$ as a running parameter in the plot~\cite{mayer2004reply}. We repeat these measurements for a series of $t$ values. In order to achieve statistical accuracy, we adapt the most efficient Monte Carlo tools presented in refs. \cite{chatelain2003far,ricci2003measuring,berthier2007efficient} to the dumbell model.   

The relevance of this representation is obvious as the slope of these curves is a direct measure of the fluctuation-dissipation ratio, by virtue of eqn.~(\ref{fig:FDT}). Close to the origin, corresponding to short time differences $t-t_w$, the equilibrium FDT is obeyed and the parametric response-correlation plot is linear with slope given by the temperature $T$. In contrast, clear deviations from the FDT are observed in the opposite limit of large time differences $t-t_w$, with a fluctuation-dissipation ratio $X(t,t_w)<1$. The physical interpretation is that small-scale (and thus fast) fluctuations rapidly reach thermal equilibrium and display equilibrium FDT, whereas large and slow critical fluctuations retain their non-equilibrium nature and display FDT violations, as seen in other critical systems~\cite{godreche2000response,berthier2001nonequilibrium}. 

The limiting value of the fluctuation-dissipation ratio defined as 
\begin{equation}
X_\infty = \lim_{t_w \to \infty} \lim_{t \to \infty} X(t,t_w) 
\end{equation}  
takes a finite value, specific to a particular universality class.
In the inset of Fig.~\ref{fig:FDT}, we compare the limiting value of the fluctuation-dissipation ratio measured in our simulations to the known value, $X_\infty = 0.38$ measured for the three dimensional Ising model~\cite{godreche2000responseb,PRUDNIKOV2015774}. We find an excellent agreement with our data, which again supports the idea that the critical end point is in this universality class. 

\section{Comparison with spin ice in a $[111]$ field}

\begin{figure}
\centering{\includegraphics[scale=0.35]{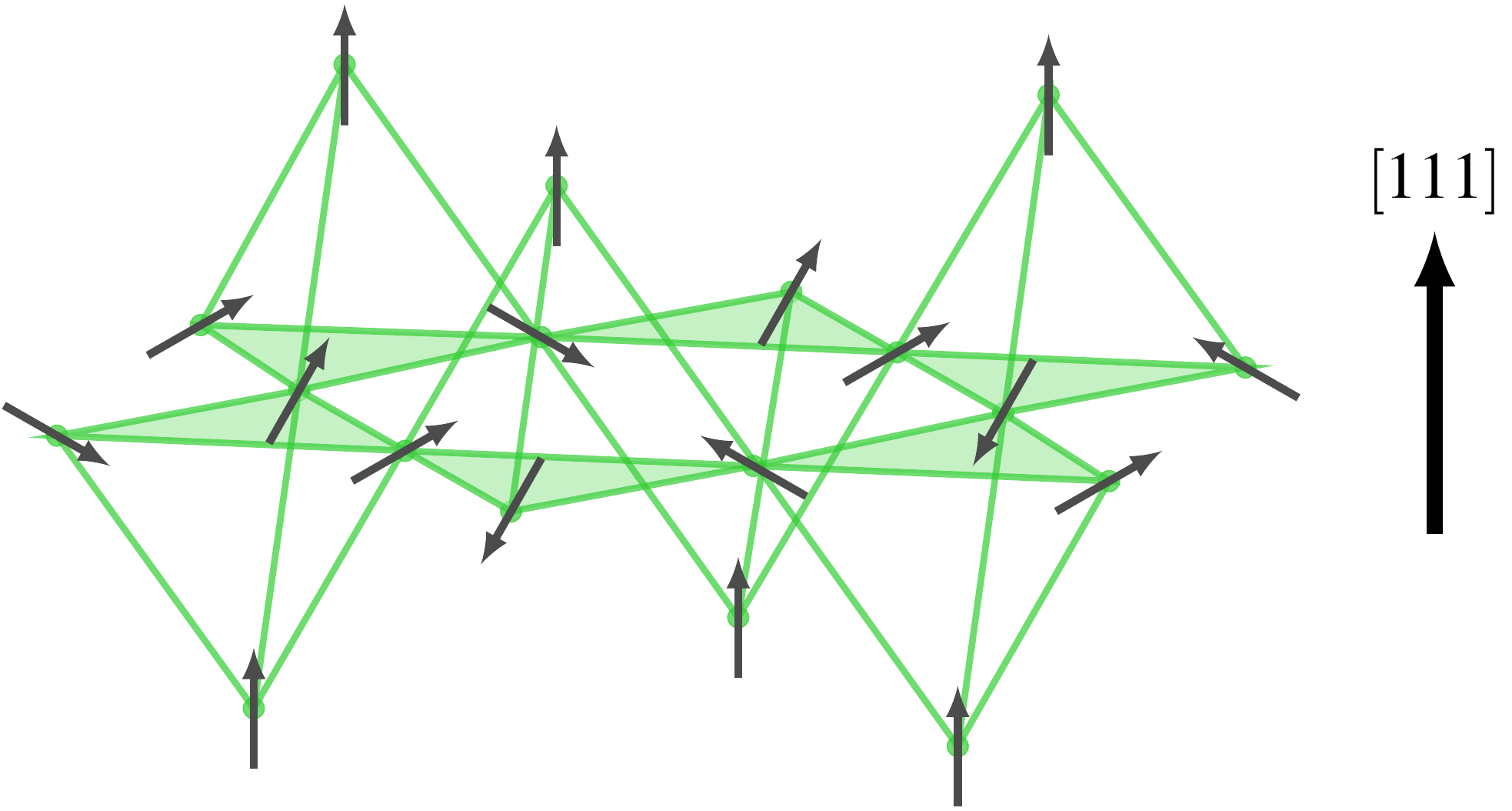}}
\centering{\includegraphics[scale=0.25]{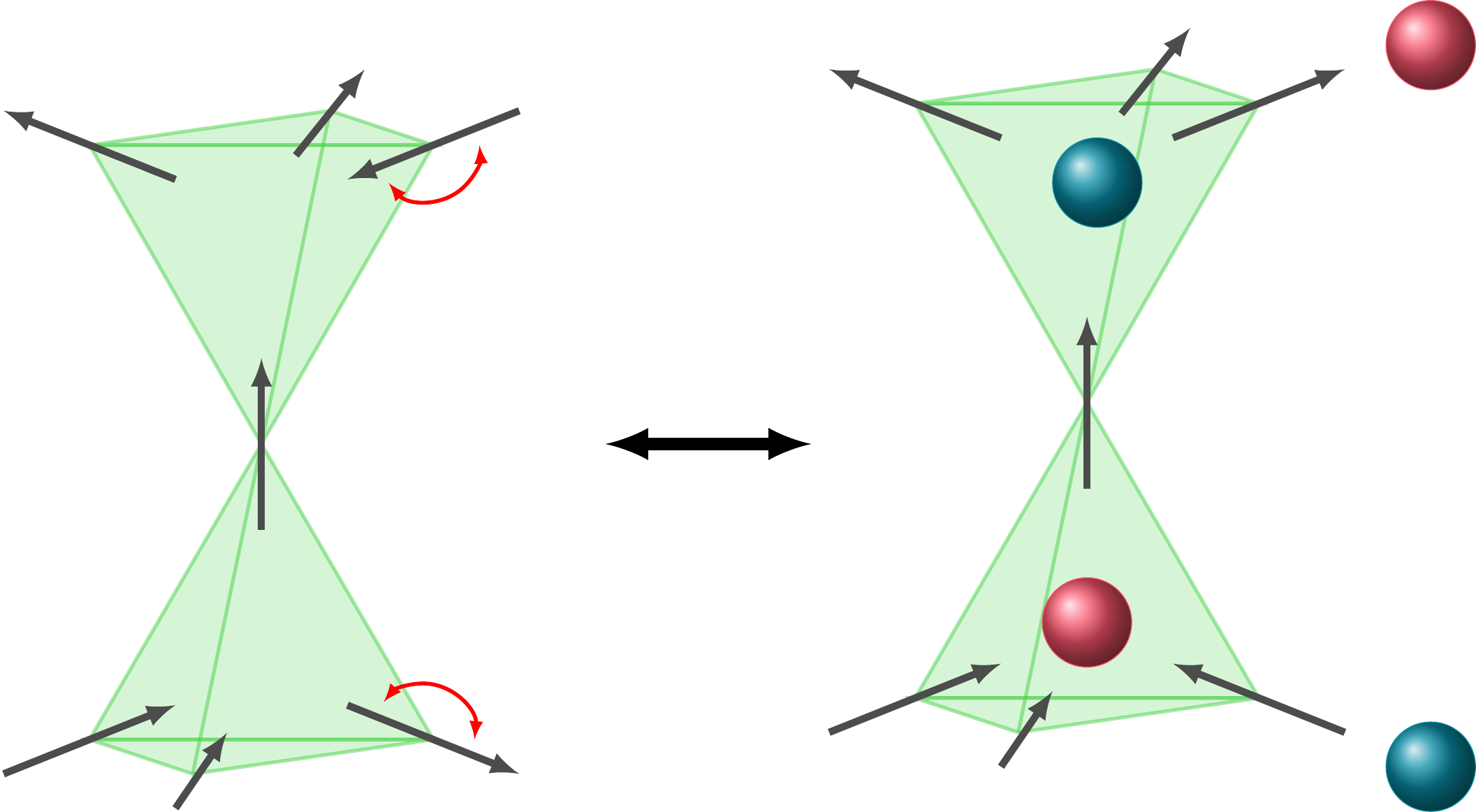}}
\caption[$\;\;$ Monopole creation]{{\textbf{Monopole creation in a $[111]$ field:} \\\noindent {\bf(Upper panel)} connected tetrahedra perpendicular to the $[111]$ axis form kagom\'e planes of spins which are bases for alternating up and down tetetrahedra. The arrow shows the direction of an applied field.\\ \noindent {\bf (Lower panel)}
Flipping the spins indicated (left) creates monopole paires with broken translational symmetry. North pole (+) -red disc, South pole (-) blue disc.}}
\label{Mono-flip}
\end{figure}

\begin{figure}
\centering{\includegraphics[scale=0.95]{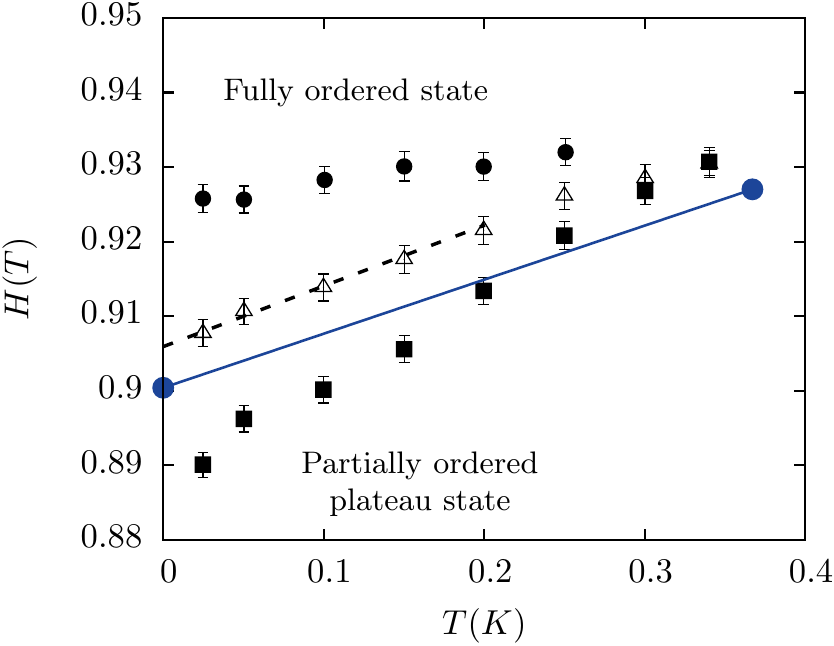}}
\caption[$\;\;$ III field]{{\textbf{Monopole phase transition in a $[111]$ field:}  We reproduce data from Figure 4 of [\onlinecite{Sakakibara2003}], which shows the thermal variation of the transition field of
Dy$_2$Ti$_2$O$_7$ for $H$ parallel to $[111]$.  Solid  circles  (squares)  denote  the  data  points
obtained for  increasing (decreasing) field sweeps. Open triangles  are  the  averaged  critical  field $H_c$, which  show  a  nearly linear temperature variation with $\frac{dH_c}{dT}=0.08$ Tesla K$^{-1}$
(dashed line) at low temperature.  The added blue circles are our estimates of the field strength at $T=0$ and at $T_c$ with blue line the phase boundary given by the constant entropy approximation. }}
\label{111-comp}
\end{figure}

At present, the only experimentally observable phase transition driven by monopole ordering is that observed with magnetic field placed along the $[111]$ crystal axis \cite{Sakakibara2003,Aoki2004,Higashinaka2004}, $\vec H = \frac{H}{\sqrt{3}}(1,1,1)$.  A field of modest strength selects a subset of Pauling states with the moments of the spins lying parallel to the field axis aligned in the field direction. The system maintains a finite entropy related to configurations of the three spins of each tetrahedron with components lying in the kagome planes perpendicular to the field direction  \cite{Isakov2004a}  (see Fig. (\ref{Mono-flip})). On increasing the field at low temperature, a first order transition is observed to a fully ordered state of 3in-1out/3out-1in tetrahedra.  As the temperature increases the transition line terminates in a critical end point. 
 In Fig. (\ref{111-comp}) we reproduce data from Figure 4 of [\onlinecite{Sakakibara2003}], which reports experiments on DTO. The figure shows the estimated phase diagram. The order of magnitude of the applied field is $0.9$ Tesla and the critical temperature is around $0.35$ K. 

The transition has previously been successfully interpreted as a liquid-gas like critical end point of a monopole crystalisation transtion \cite{Castelnovo2008} and in this sense is a close cousin of the transition separating $\phi_0$ and $\phi_1$ discussed above. The main difference is that the external field couples to both of the fragmented components of the magnetic moments \cite{Brooks2014}, providing a staggered chemical potential for the monopoles and introducing a preference for Dirac strings oriented with the field. The field therefore breaks both the magnetic symmetry and the monopole translational symmetry. 

At the transition monopole pairs are created in abondance by flipping spins in the kagome planes as illustrated in Fig. (\ref{Mono-flip}). 
The direct action of the field on the charges is to provide a chemical potential gradient, $\vec \nabla \mu_{\pm}= \pm Q \vec H$ so that, in addition to the energy scale for monopole creation in zero field, there is also a contribution depending on the direction of movement in the field.
The chemical potential gradient alone does not therefore provide a staggered energy profile. If  one of the north monopoles of Fig. (\ref{Mono-flip}) were to continue moving along the $\hat{z}$ axis it would pick up energy at each step in the same manner. However, the constraints of spin ice forbid this: movement between the kagome planes is blocked as, on the magnetic plateau the spins joining the planes point in the wrong direction to allow monopole movement between planes via a single spin flip. Preparing the ground for this move requires flips of loops of spins at high energy cost \cite{Castelnovo2010} so that the monopoles are essentially confined to two-dimensional strips perpendicular to the field axis. 

The chemical potential gradient does provide a staggered energy landscape within this confined space. The difference in potential energy  for a (north) monopole on an $A$ or a $B$ site of a kagome plane is $2\Delta=Q\vec a.\vec H$ where $\vec a = \frac{a}{\sqrt{3}}(-1,-1,1)$ is a lattice vector spanning the two sites. This yields $\Delta=\left(\frac{m H}{3}\right)$, which is just the Zeeman energy of the spin flip in the presence of the field.

Given the similarities, we can repeat the thermodynamic arguments of section II C for modified phases $\phi_0'$, the plateau phase with entropy per tetrahedron $s_o'=0.1616$ \cite{Masafumi2002} and ordered monopole crystal phase $\phi_1'$ with entropy zero. From this, using $m=9.87\mu_B$ \cite{Yavorskii2008} we predict a field for the transition at zero temperature $H(T=0)=0.90$ Tesla and an initial slope for the phase boundary $\frac{d H}{dT}= 0.073$ Tesla K$^{-1}$. We note that the observed critical temperature for DTO \cite{Sakakibara2003} is very close to  our calculated value $T_c=0.367$ K for the $\phi_0:\phi_1$ critical end point. Taking this value  and using the constant entropy approximation we find  $H_c=0.927$ T.

Our predicted phase diagram, shown in Fig.~(\ref{111-comp}) is in quite remarkable quantitative agreement with reference [\onlinecite{Sakakibara2003}]. However, a word of caution is probably in order. As the entropy of the phases $\phi_0, \phi_0'$ and $\phi_1,\phi_1'$ are different there is no reason to expect such quantitative agreement between the two critical temperatures. 
Indeed
simulations of the transition using the dipolar spin ice model \cite{denHertog2000}, while still in excellent qualitative agreement  with the experimental data show a significantly higher critical temperature, $T_c=0.587$ K \cite{Castelnovo2008,Hamp2015}. However, quantitative modelling of experiments with the DSI at such low temperatures requires corrections in the form of further neighbour exchange terms \cite{Henelius2016}, which could also have significant effects on the critical end point \cite{Castelnovo2008}. In general these extra terms reduce the ordering temperature for symmetry breaking among the Pauling states, compared with the original DSI model \cite{Melko2001}. As the dumbbell model has no such ordering transition these corrections may play in its favour, but one could be forced to concede an element of good fortune in this remarkable agreement.  It would clearly be of interest to pursue this subject in future research.

\section{Discussion}

We have shown that the dumbbell model of spin ice has a rich phase diagram with the double winged structure shown in Fig. (\ref{Full-phase}). A key to its existence is the presence of a first order line for the spin ice - AIAO transition in the $\Delta=0$ plane \cite{Borzi2014}. The first order nature of the transition ensures that the singularity survives application of a symmetry breaking field giving symmetry sustaining transitions and the emergence of the wings. The first order transition becomes second order via a tri-critical point as discussed in detail in section \ref{multicritical}. Tri-critical behaviour with first and second order sectors is common in frustrated magnetic systems \cite{Champion2002,Zhitomirsky2012,Shahbazi2008,Sadeghi2015} and is related to the entropy of fluctuations provided by the frustrated geometry. In the case of spin ice one must go beyond the nearest neighbour spin ice model to generate a first order transition as within this approximation the monopoles are non-interacting. Ordering in this case is due uniquely to entropic considerations \cite{Borzi2014} and can only be second order \cite{Castelnovo2008}. Including the dipolar interactions in the spin model provides the emergent monopoles with an energy versus entropy trade off which drives the transition first order. However, truncating the Coulomb interaction beyond nearest neighbour monopoles would not give a quantitative change to the phase diagram. 

We have studied both the dumbbell model and the related Blume-Capel model, the BC2. The five phases of the wings meet either in two stages for dumbbell, or not at all for BC2, entering the $\Delta=0$ plane at two different values of $\nu$ and $T$. This undershoot or overshoot is consistent with there being only two independent variables on the plane. This could however be changed by freeing the double monopole chemical potential from the fixed value, $\mu_2=-4\nu$ of the present model. By tuning $\mu_2$ it should be possible to find a parameter set $\nu$, $T$, $\mu_2$ for which the five phases meet at a single penta-critical point. This  corresponds, at the mean field level to all terms up to and including order $\phi^6$ being zero in an expansion of the free energy. 

\subsection{Liquid-gas, liquid-liquid and symmetry sustaining transitions}

Liquid gas phase transitions have two defining characteristics.
 
Firstly they correspond to crossing lines of phase equilibria in a temperature like - field like phase diagram, between two phases with the same symmetry. If the line of transitions terminates at a critical end point, it is then possible to move analytically from one phase to another by contouring this special point. As a consequence, the only thing that defines the two separate phases is the transition itself. There is broken symmetry at the transition, but it is emergent, separating phase space into high and low density sectors with the same microscopic symmetry. 

The second is that the sustained symmetry is the highest allowed by the Hamiltonian.  The generic case is that of a fluid that changes from low to high density through the control of temperature and pressure, or chemical potential while maintaining continuous translational symmetry. In the quantum case, temperature could be replaced by a coupling constant and thermal fluctuations by quantum fluctuations, allowing for the transition from a long range entangled quantum liquid, such as a quantum spin liquid \cite{Balents2010} to a classical paramagnet, or spin gas phase \cite{Savary2013}. 

The first criterion is ubiquitous thermodynamics and can be generated for any first order transition by the application of a field conjugate to the order parameter characterising the transition. The second is a non-universal property of strongly correlated systems, with the existence of the liquid-gas singularity dependent on the microscopic properties of the model \cite{Barrat2003}.

The transitions discussed in this paper satisfy the first criterion but not the second. In general Coulomb fluids on a bi-partite lattice do not offer a liquid gas transition with the full discrete symmetry of the $T-\nu, \Delta=0$ plane \cite{Kobelev2002}. Rather, such a transition is usurped by sublimation from low density fluid to crystal with broken Z$_2$ translational symmetry, as we have seen here in detail for the diamond lattice. The transitions are therefore liquid-gas like in a weak sense: they are symmetry sustaining but do not maintain the highest translational symmetry offered by the diamond lattice. However, as the two criteria are equivalent from a thermodynamic point of view, Blume-Capel type models and therefore spin ice can be considered as generic systems for studying symmetry sustaining phenomena, often occurring in liquids.

In particular, there has been much work on systems showing liquid-liquid phase transitions. In molten phosphorous \cite{Katayama2000}, silicon \cite{Sastry2003} or water \cite{Brovchenko2005} for example pressure takes the fluid from a low to high density liquid state via a first order transition that terminates in a critical end point. The high density transition often appears in a supercooled state as it is again usurped by crystallisation in thermodynamic equilibrium.
A characteristic of these systems is the capacity to accommodate two kinds of local packing, open (tetrahedral) and close packed. This can be modelled using two hard core repulsion length scales \cite{Franzese2001} but it is also proposed as an emergent phenomenon due to frustration and inhomogeneities in simple fluids \cite{Tanaka2000}. The BC2 model and hence spin ice clearly provides a generic skeleton for this science. If passage from the $\phi_0$ to $\phi_1$ is equivalent a liquid-gas phase transition then that from $\phi_1$ to $\phi_2$ on one side of the double winged phase diagram of Fig.~(\ref{Full-phase})
 is thermodynamically equivalent to a liquid-liquid transition. Detailed comparison with the models presented here could therefore provide new insight into the necessary conditions for liquid-liquid transitions including the possibility of liquid-liquid tri-criticality.

Slightly nearer to home, similar physics is observed in magnetic  itinerant electron systems under pressure. Both LaCrGe$_3$ \cite{Taufour2017} and UGe$_2$ \cite{Taufour2011} show double winged phase diagrams as a function of temperature, pressure and applied field with two ferromagnetic phases extending out to finite field values. The ferromagnetic phase transitions are symmetry sustaining in exact analogy with the transitions presented in this paper, so that the BC2 type models again provide a skeleton for this structure. Interestingly these materials provide experimental examples of the two possible multi-critical regions discussed in section \ref{multicritical}, confirming the accidental nature of the wing connections in the phase diagram. In LaCrGe$_3$ the wings meet in two stages for each field direction, as proposed for spin ice, giving three distinct tri-critical points for positive and negative characteristic fields and for $H=0$. In UGe$_2$ on the other hand, the wings meet the central plane separately, as is apparently the case for the BC2 model.

 In LaCrGe$_3$ the winged phase transitions are extrapolated to terminate at zero temperature and finite field at a series of quantum critical points. This prediction should be contrasted with antiferromagnetic BC2 type models for which the lines of finite temperature critical end points extend out to $\Delta=\pm \infty$. Here, as $\Delta$ becomes large the partitioning of north and south poles on $A$ and $B$ sublattices becomes perfect, but the collective interaction between charges of opposite sign still drives a liquid-gas like discontinuity in the sublattice monopole density at finite temperature. It would certainly be interesting to do more studies for the ferromagnetic case including transverse spin fluctuations, the quantum case being accessible via quantum Monte Carlo simulation.

\subsection{Future experiments in frustrated magnetism}

Motivation for this work has come in large part from experiments on the spin ice material Ho$_2$Ir$_2$O$_7$ (HIO)\cite{Lefrancois2017}. In this material both the Ho$^{3+}$ ions and the Ir$^{4+}$ ions carry a magnetic moment and they sit on interpenetrating pyrochlore structures. The moments of the Ir$^{4+}$ ions order on the scale of $100$ K into an AIAO structure which provides internal magnetic fields which act in turn on the Ho$^{3+}$ magnetic moments. In the monopole picture the internal fields translate into the staggered chemical potential studied here and proposed in [\onlinecite{Brooks2014}] as a mechanism for separating the $\phi_1$ and $\phi_2$ monopole crystal phases and accessing the  $\phi_1$ phase. As temperature is lowered through the 1K range the Ho sublattice continuously develops AIAO order with the ordered moment saturating at 50 $\%$ of the total moment. The leftover moment gives correlated diffuse scattering consistent with a Coulomb phase and the measured characteristics of the powder sample are indeed consistent with the fragmented $\phi_1$ phase.

As the Ho sublattice shows no phase transition none of the winged structure is, as yet observable directly in experiment. However, it is worth noting that a different material in this series, Tb$_2$Ir$_2$O$_7$ (TIO) settles into a ground state with full AIAO order, that is into the $\phi_2$ phase as defined above \cite{Lefrancois2015}. Its sister material Tb$_2$Ti$_2$O$_7$ (TTO) is in some sense spin ice like, falling close to the spin ice AIAO ($\Delta=0$) phase boundary \cite{denHertog2000}. Hence, although TTO remains an enigma \cite{Rau2019}, the fact that TIO fully orders is completely consistent with our logic. If one could chemically tune the values of $\nu$ and $\Delta$ from TIO to HIO one would pass through the $\phi_2:\phi_1$ phase boundary on the way. Once at the values corresponding to HIO, heating up one could hit a further phase boundary but no further transition is required from symmetry arguments as the transitions are symmetry sustaining. Although the planes of first order transitions do not lie perpendicular to the $\nu-\Delta$ planes they do fall with a very steep slope, the inverse of eqn. (\ref{thermo2}). Hence for an accidental value of $\Delta$ it is quite likely that a thermal trajectory would maintain the system well away from the phase boundaries. We propose that this is the case for HIO. 

The above conclusion immediately begs the question of if it is possible to shift the value of $\Delta$ experimentally. One possibility would be to put materials such as HIO or its dysprosium counterpart under pressure. High pressure would presumably change both the strength of the internal fields and the monopole chemical potentials $\mu$ and $\mu_2$ which are combinations of exchange and dipole interactions \cite{Castelnovo2008}. One might expect that increasing the pressure would have the effect of increasing the scale of the antiferromagnetic exchange, therefore reducing the scale of $\mu$, while at the same time increasing the scale of $\Delta$, moving the system towards the $\phi_1:\phi_2$ boundary, but the evolution could equally well be counterintuitive and go in the opposite direction. One could also consider the effects of chemical pressure through the chemical substitution of Ir$^{4+}$ ions with non-magnetic species such as Ti$^{4+}$, or Ge$^{4+}$ which has a smaller ionic radius than its counterparts \cite{Zhou2012}. In order to hit one of the phase boundaries, starting from HIO one would need to shift $\Delta$ and/or $\nu$ on the Kelvin scale, that is on the scale of the exchange constants themselves. These are challenging experiments  that open the door to rich theoretical and numerical problems and the present results provide a motivating framework in which to work.

Given the steepness of the slope of the phase boundaries in Fig. (\ref{Full-phase}), if one did cross a first order plane by altering $\Delta$, further tuning to find the critical end point to the plane should be straightforward, at least  in comparison, giving access to Kibble-Zurek scaling experiments as outlined in section \ref{KZ} and proposed for the critical point in a $[111]$ field \cite{Hamp2015}. The prospect of doing Kibble-Zurek scaling experiments is particularly appealing as critical slowing down gives very weakly diverging time scales and so is difficult to access experimentally. For example, if the microscopic time scale is a nanosecond, getting the divergence into the millisecond range requires a correlation length of 1000 times the microscopic length and a reduced temperature or field of order $10^{-6}$. Such high precision can be avoided by finding systems with either long microscopic length or time scales. Long microscopic length scales occur naturally in cold atom systems, which has recently led to  successful Kibble-Zurek type experiments \cite{Corman2014,Labeyrie2016}. Spin ice, on the other hand is ideally suited because of its naturally long microscopic time scales, for example around a millisecond for DTO \cite{Jaubert2009} so that the plethora of critical points presented here could open the door to many such dynamical experiments. Once accessed, both field like and temperature like protocols are envisageable. 

Our work also suggests that it could be interesting to extend to spin ice materials the type of noise measurements that were previously performed in spin glasses~\cite{herisson2002fluctuation,herisson2004off} to simultaneously detect linear susceptibilities and time correlation functions, in order to experimentally access the fluctuation-dissipation ratio introduced in section \ref{sec:FDT}. 

\section{Conclusion}

Spin ice materials and models have proven to be the source of rich emergent science \cite{Bramwell2001,denHertog2000,Isakov2004,Castelnovo2008,Ryzhkin2005,Fennell2009,Brooks2014}, widening the scope and interest of frustrated magnetism and offering multiple avenues for novel research. In particular the monopole picture, which simplifies a complex and strongly interacting frustrated system to a level in which it can be  addressed in incomparable detail, has provided an unexpected controlled environment in which to study Coulomb fluids both from a field theoretic and charge perspective. We have exploited the full phase diagram of the emergent, on lattice magnetolyte in which both monopoles and double charged monopoles play important roles. In doing so, we have exposed a model system for multiple phase transitions with wide ranging applications. These include fluids showing liquid-liquid phase transitions \cite{Katayama2000,Sastry2003,Brovchenko2005} and itinerant magnetic systems under pressure  \cite{Taufour2017,Taufour2011} as well as extensive new applications within the field of frustrated magnetism. 

\section*{Acknowledgments}

It is a pleasure to thank Claudio Castelnovo, Laurent de Forges de Parny, Ludovic Jaubert, Vojtech Kaiser, Elsa Lhotel, Sylvain Petit, Lucile Savary and Valentin Taufour for useful discussions. We also thank Vojtech Kaiser for sharing numerical codes with us, Camille Scalliet for help with the figures and Zenji Hiroi for authorising the reproduction of our Figure 14. This project was supported in part (VR and PCWH) by ANR grant Listen Monopoles. THS thanks ENS de Lyon for financial support during a research project.
This work was supported by a grant from the Simons Foundation (\# 454933, L. Berthier).


\bibliography{BC-SI-biblio-Sub}

\begin{thebibliography}{98}%
\makeatletter
\providecommand \@ifxundefined [1]{%
 \@ifx{#1\undefined}
}%
\providecommand \@ifnum [1]{%
 \ifnum #1\expandafter \@firstoftwo
 \else \expandafter \@secondoftwo
 \fi
}%
\providecommand \@ifx [1]{%
 \ifx #1\expandafter \@firstoftwo
 \else \expandafter \@secondoftwo
 \fi
}%
\providecommand \natexlab [1]{#1}%
\providecommand \enquote  [1]{``#1''}%
\providecommand \bibnamefont  [1]{#1}%
\providecommand \bibfnamefont [1]{#1}%
\providecommand \citenamefont [1]{#1}%
\providecommand \href@noop [0]{\@secondoftwo}%
\providecommand \href [0]{\begingroup \@sanitize@url \@href}%
\providecommand \@href[1]{\@@startlink{#1}\@@href}%
\providecommand \@@href[1]{\endgroup#1\@@endlink}%
\providecommand \@sanitize@url [0]{\catcode `\\12\catcode `\$12\catcode
  `\&12\catcode `\#12\catcode `\^12\catcode `\_12\catcode `\%12\relax}%
\providecommand \@@startlink[1]{}%
\providecommand \@@endlink[0]{}%
\providecommand \url  [0]{\begingroup\@sanitize@url \@url }%
\providecommand \@url [1]{\endgroup\@href {#1}{\urlprefix }}%
\providecommand \urlprefix  [0]{URL }%
\providecommand \Eprint [0]{\href }%
\providecommand \doibase [0]{http://dx.doi.org/}%
\providecommand \selectlanguage [0]{\@gobble}%
\providecommand \bibinfo  [0]{\@secondoftwo}%
\providecommand \bibfield  [0]{\@secondoftwo}%
\providecommand \translation [1]{[#1]}%
\providecommand \BibitemOpen [0]{}%
\providecommand \bibitemStop [0]{}%
\providecommand \bibitemNoStop [0]{.\EOS\space}%
\providecommand \EOS [0]{\spacefactor3000\relax}%
\providecommand \BibitemShut  [1]{\csname bibitem#1\endcsname}%
\let\auto@bib@innerbib\@empty
\bibitem [{\citenamefont {Harris}\ \emph {et~al.}(1997)\citenamefont {Harris},
  \citenamefont {Bramwell}, \citenamefont {{McMorrow}}, \citenamefont
  {Zeiske},\ and\ \citenamefont {Godfrey}}]{Harris1997}%
  \BibitemOpen
  \bibfield  {author} {\bibinfo {author} {\bibfnamefont {M.~J.}\ \bibnamefont
  {Harris}}, \bibinfo {author} {\bibfnamefont {S.~T.}\ \bibnamefont
  {Bramwell}}, \bibinfo {author} {\bibfnamefont {D.~F.}\ \bibnamefont
  {{McMorrow}}}, \bibinfo {author} {\bibfnamefont {T.}~\bibnamefont {Zeiske}},
  \ and\ \bibinfo {author} {\bibfnamefont {K.~W.}\ \bibnamefont {Godfrey}},\
  }\href {\doibase 10.1103/PhysRevLett.79.2554} {\bibfield  {journal} {\bibinfo
   {journal} {Physical Review Letters}\ }\textbf {\bibinfo {volume} {79}},\
  \bibinfo {pages} {2554} (\bibinfo {year} {1997})}\BibitemShut {NoStop}%
\bibitem [{\citenamefont {Bramwell}\ and\ \citenamefont
  {Gingras}(2001)}]{Bramwell2001}%
  \BibitemOpen
  \bibfield  {author} {\bibinfo {author} {\bibfnamefont {S.~T.}\ \bibnamefont
  {Bramwell}}\ and\ \bibinfo {author} {\bibfnamefont {M.~J.~P.}\ \bibnamefont
  {Gingras}},\ }\href {\doibase 10.1126/science.1064761} {\bibfield  {journal}
  {\bibinfo  {journal} {Science}\ }\textbf {\bibinfo {volume} {294}},\ \bibinfo
  {pages} {1495} (\bibinfo {year} {2001})}\BibitemShut {NoStop}%
\bibitem [{\citenamefont {Isakov}\ \emph
  {et~al.}(2004{\natexlab{a}})\citenamefont {Isakov}, \citenamefont {Gregor},
  \citenamefont {Moessner},\ and\ \citenamefont {Sondhi}}]{Isakov2004}%
  \BibitemOpen
  \bibfield  {author} {\bibinfo {author} {\bibfnamefont {S.~V.}\ \bibnamefont
  {Isakov}}, \bibinfo {author} {\bibfnamefont {K.}~\bibnamefont {Gregor}},
  \bibinfo {author} {\bibfnamefont {R.}~\bibnamefont {Moessner}}, \ and\
  \bibinfo {author} {\bibfnamefont {S.~L.}\ \bibnamefont {Sondhi}},\ }\href
  {\doibase ARTN 167204} {\bibfield  {journal} {\bibinfo  {journal} {Physical
  Review Letters}\ }\textbf {\bibinfo {volume} {93}},\ \bibinfo {pages}
  {167204} (\bibinfo {year} {2004}{\natexlab{a}})}\BibitemShut {NoStop}%
\bibitem [{\citenamefont {Castelnovo}\ \emph {et~al.}(2008)\citenamefont
  {Castelnovo}, \citenamefont {Moessner},\ and\ \citenamefont
  {Sondhi}}]{Castelnovo2008}%
  \BibitemOpen
  \bibfield  {author} {\bibinfo {author} {\bibfnamefont {C.}~\bibnamefont
  {Castelnovo}}, \bibinfo {author} {\bibfnamefont {R.}~\bibnamefont
  {Moessner}}, \ and\ \bibinfo {author} {\bibfnamefont {S.~L.}\ \bibnamefont
  {Sondhi}},\ }\href {\doibase 10.1038/nature06433} {\bibfield  {journal}
  {\bibinfo  {journal} {Nature}\ }\textbf {\bibinfo {volume} {451}},\ \bibinfo
  {pages} {42} (\bibinfo {year} {2008})}\BibitemShut {NoStop}%
\bibitem [{\citenamefont {Ryzhkin}(2005)}]{Ryzhkin2005}%
  \BibitemOpen
  \bibfield  {author} {\bibinfo {author} {\bibfnamefont {I.~A.}\ \bibnamefont
  {Ryzhkin}},\ }\href {http://link.springer.com/article/10.1134/1.2103216}
  {\bibfield  {journal} {\bibinfo  {journal} {Journal of Experimental and
  Theoretical Physics}\ }\textbf {\bibinfo {volume} {101}},\ \bibinfo {pages}
  {481{\textendash}486} (\bibinfo {year} {2005})}\BibitemShut {NoStop}%
\bibitem [{\citenamefont {Jaubert}\ and\ \citenamefont
  {Holdsworth}(2009)}]{Jaubert2009}%
  \BibitemOpen
  \bibfield  {author} {\bibinfo {author} {\bibfnamefont {L.~D.~C.}\
  \bibnamefont {Jaubert}}\ and\ \bibinfo {author} {\bibfnamefont {P.~C.~W.}\
  \bibnamefont {Holdsworth}},\ }\href {\doibase 10.1038/nphys1227} {\bibfield
  {journal} {\bibinfo  {journal} {Nature Physics}\ }\textbf {\bibinfo {volume}
  {5}},\ \bibinfo {pages} {258} (\bibinfo {year} {2009})}\BibitemShut {NoStop}%
\bibitem [{\citenamefont {Castelnovo}\ \emph {et~al.}(2011)\citenamefont
  {Castelnovo}, \citenamefont {Moessner},\ and\ \citenamefont
  {Sondhi}}]{Castelnovo2011}%
  \BibitemOpen
  \bibfield  {author} {\bibinfo {author} {\bibfnamefont {C.}~\bibnamefont
  {Castelnovo}}, \bibinfo {author} {\bibfnamefont {R.}~\bibnamefont
  {Moessner}}, \ and\ \bibinfo {author} {\bibfnamefont {S.~L.}\ \bibnamefont
  {Sondhi}},\ }\href {\doibase 10.1103/PhysRevB.84.144435} {\bibfield
  {journal} {\bibinfo  {journal} {Physical Review B}\ }\textbf {\bibinfo
  {volume} {84}},\ \bibinfo {pages} {144435} (\bibinfo {year}
  {2011})}\BibitemShut {NoStop}%
\bibitem [{\citenamefont {Brooks-Bartlett}\ \emph {et~al.}(2014)\citenamefont
  {Brooks-Bartlett}, \citenamefont {Banks}, \citenamefont {Jaubert},
  \citenamefont {Harman-Clarke},\ and\ \citenamefont
  {Holdsworth}}]{Brooks2014}%
  \BibitemOpen
  \bibfield  {author} {\bibinfo {author} {\bibfnamefont {M.~E.}\ \bibnamefont
  {Brooks-Bartlett}}, \bibinfo {author} {\bibfnamefont {S.~T.}\ \bibnamefont
  {Banks}}, \bibinfo {author} {\bibfnamefont {L.~D.~C.}\ \bibnamefont
  {Jaubert}}, \bibinfo {author} {\bibfnamefont {A.}~\bibnamefont
  {Harman-Clarke}}, \ and\ \bibinfo {author} {\bibfnamefont {P.~C.~W.}\
  \bibnamefont {Holdsworth}},\ }\href {\doibase 10.1103/PhysRevX.4.011007}
  {\bibfield  {journal} {\bibinfo  {journal} {Phys. Rev. X}\ }\textbf {\bibinfo
  {volume} {4}},\ \bibinfo {pages} {011007} (\bibinfo {year}
  {2014})}\BibitemShut {NoStop}%
\bibitem [{\citenamefont {Kaiser}\ \emph {et~al.}(2015)\citenamefont {Kaiser},
  \citenamefont {Bramwell}, \citenamefont {Holdsworth},\ and\ \citenamefont
  {Moessner}}]{Kaiser2015}%
  \BibitemOpen
  \bibfield  {author} {\bibinfo {author} {\bibfnamefont {V.}~\bibnamefont
  {Kaiser}}, \bibinfo {author} {\bibfnamefont {S.~T.}\ \bibnamefont
  {Bramwell}}, \bibinfo {author} {\bibfnamefont {P.~C.~W.}\ \bibnamefont
  {Holdsworth}}, \ and\ \bibinfo {author} {\bibfnamefont {R.}~\bibnamefont
  {Moessner}},\ }\href {\doibase 10.1103/PhysRevLett.115.037201} {\bibfield
  {journal} {\bibinfo  {journal} {Phys. Rev. Lett.}\ }\textbf {\bibinfo
  {volume} {115}},\ \bibinfo {pages} {037201} (\bibinfo {year}
  {2015})}\BibitemShut {NoStop}%
\bibitem [{\citenamefont {M\"oller}\ and\ \citenamefont
  {Moessner}(2006)}]{Moller2006}%
  \BibitemOpen
  \bibfield  {author} {\bibinfo {author} {\bibfnamefont {G.}~\bibnamefont
  {M\"oller}}\ and\ \bibinfo {author} {\bibfnamefont {R.}~\bibnamefont
  {Moessner}},\ }\href {\doibase 10.1103/PhysRevLett.96.237202} {\bibfield
  {journal} {\bibinfo  {journal} {Phys. Rev. Lett.}\ }\textbf {\bibinfo
  {volume} {96}},\ \bibinfo {pages} {237202} (\bibinfo {year}
  {2006})}\BibitemShut {NoStop}%
\bibitem [{\citenamefont {den Hertog}\ and\ \citenamefont
  {Gingras}(2000)}]{denHertog2000}%
  \BibitemOpen
  \bibfield  {author} {\bibinfo {author} {\bibfnamefont {B.~C.}\ \bibnamefont
  {den Hertog}}\ and\ \bibinfo {author} {\bibfnamefont {M.~J.~P.}\ \bibnamefont
  {Gingras}},\ }\href {\doibase 10.1103/PhysRevLett.84.3430} {\bibfield
  {journal} {\bibinfo  {journal} {Phys. Rev. Lett.}\ }\textbf {\bibinfo
  {volume} {84}},\ \bibinfo {pages} {3430} (\bibinfo {year}
  {2000})}\BibitemShut {NoStop}%
\bibitem [{\citenamefont {Isakov}\ \emph {et~al.}(2005)\citenamefont {Isakov},
  \citenamefont {Moessner},\ and\ \citenamefont {Sondhi}}]{Isakov2005}%
  \BibitemOpen
  \bibfield  {author} {\bibinfo {author} {\bibfnamefont {S.~V.}\ \bibnamefont
  {Isakov}}, \bibinfo {author} {\bibfnamefont {R.}~\bibnamefont {Moessner}}, \
  and\ \bibinfo {author} {\bibfnamefont {S.~L.}\ \bibnamefont {Sondhi}},\
  }\href {\doibase ARTN 217201} {\bibfield  {journal} {\bibinfo  {journal}
  {Phys. Rev. Lett.}\ }\textbf {\bibinfo {volume} {95}},\ \bibinfo {pages}
  {217201} (\bibinfo {year} {2005})}\BibitemShut {NoStop}%
\bibitem [{\citenamefont {Pauling}(1935)}]{Pauling1935}%
  \BibitemOpen
  \bibfield  {author} {\bibinfo {author} {\bibfnamefont {L.}~\bibnamefont
  {Pauling}},\ }\href@noop {} {\bibfield  {journal} {\bibinfo  {journal}
  {Journal of the American Chemical Society}\ }\textbf {\bibinfo {volume} {57}}
  (\bibinfo {year} {1935})}\BibitemShut {NoStop}%
\bibitem [{\citenamefont {Jaubert}\ and\ \citenamefont
  {Udagawa}(2019)}]{Jaubert2019}%
  \BibitemOpen
  \bibinfo {editor} {\bibfnamefont {L.~D.~C.}\ \bibnamefont {Jaubert}}\ and\
  \bibinfo {editor} {\bibfnamefont {M.}~\bibnamefont {Udagawa}},\ eds.,\
  \href@noop {} {\emph {\bibinfo {title} {Spin Ice}}}\ (\bibinfo  {publisher}
  {Springer},\ \bibinfo {year} {2019})\BibitemShut {NoStop}%
\bibitem [{\citenamefont {Lefran{\c c}ois}\ \emph {et~al.}(2017)\citenamefont
  {Lefran{\c c}ois}, \citenamefont {Cathelin}, \citenamefont {Lhotel},
  \citenamefont {Robert}, \citenamefont {Lejay}, \citenamefont {Colin},
  \citenamefont {Canals}, \citenamefont {Damay}, \citenamefont {Ollivier},
  \citenamefont {F{\aa}k}, \citenamefont {Chapon}, \citenamefont {Ballou},\
  and\ \citenamefont {Simonet}}]{Lefrancois2017}%
  \BibitemOpen
  \bibfield  {author} {\bibinfo {author} {\bibfnamefont {E.}~\bibnamefont
  {Lefran{\c c}ois}}, \bibinfo {author} {\bibfnamefont {V.}~\bibnamefont
  {Cathelin}}, \bibinfo {author} {\bibfnamefont {E.}~\bibnamefont {Lhotel}},
  \bibinfo {author} {\bibfnamefont {J.}~\bibnamefont {Robert}}, \bibinfo
  {author} {\bibfnamefont {P.}~\bibnamefont {Lejay}}, \bibinfo {author}
  {\bibfnamefont {C.~V.}\ \bibnamefont {Colin}}, \bibinfo {author}
  {\bibfnamefont {B.}~\bibnamefont {Canals}}, \bibinfo {author} {\bibfnamefont
  {F.}~\bibnamefont {Damay}}, \bibinfo {author} {\bibfnamefont
  {J.}~\bibnamefont {Ollivier}}, \bibinfo {author} {\bibfnamefont
  {B.}~\bibnamefont {F{\aa}k}}, \bibinfo {author} {\bibfnamefont {L.~C.}\
  \bibnamefont {Chapon}}, \bibinfo {author} {\bibfnamefont {R.}~\bibnamefont
  {Ballou}}, \ and\ \bibinfo {author} {\bibfnamefont {V.}~\bibnamefont
  {Simonet}},\ }\href {\doibase 10.1038/s41467-017-00277-1} {\bibfield
  {journal} {\bibinfo  {journal} {Nature Communications}\ }\textbf {\bibinfo
  {volume} {8}},\ \bibinfo {pages} {209} (\bibinfo {year} {2017})}\BibitemShut
  {NoStop}%
\bibitem [{\citenamefont {Melko}\ and\ \citenamefont
  {Gingras}(2004)}]{Melko2004}%
  \BibitemOpen
  \bibfield  {author} {\bibinfo {author} {\bibfnamefont {R.~G.}\ \bibnamefont
  {Melko}}\ and\ \bibinfo {author} {\bibfnamefont {M.~J.~P.}\ \bibnamefont
  {Gingras}},\ }\href {\doibase 10.1088/0953-8984/16/43/R02} {\bibfield
  {journal} {\bibinfo  {journal} {Journal of Physics: Condensed Matter}\
  }\textbf {\bibinfo {volume} {16}},\ \bibinfo {pages} {R1277} (\bibinfo {year}
  {2004})}\BibitemShut {NoStop}%
\bibitem [{\citenamefont {Guruciaga}\ \emph {et~al.}(2014)\citenamefont
  {Guruciaga}, \citenamefont {Grigera},\ and\ \citenamefont
  {Borzi}}]{Borzi2014}%
  \BibitemOpen
  \bibfield  {author} {\bibinfo {author} {\bibfnamefont {P.~C.}\ \bibnamefont
  {Guruciaga}}, \bibinfo {author} {\bibfnamefont {S.~A.}\ \bibnamefont
  {Grigera}}, \ and\ \bibinfo {author} {\bibfnamefont {R.~A.}\ \bibnamefont
  {Borzi}},\ }\href {\doibase 10.1103/PhysRevB.90.184423} {\bibfield  {journal}
  {\bibinfo  {journal} {Phys. Rev. B}\ }\textbf {\bibinfo {volume} {90}},\
  \bibinfo {pages} {184423} (\bibinfo {year} {2014})}\BibitemShut {NoStop}%
\bibitem [{\citenamefont {Borzi}\ \emph {et~al.}(2013)\citenamefont {Borzi},
  \citenamefont {Slobinsky},\ and\ \citenamefont {Grigera}}]{Borzi2013}%
  \BibitemOpen
  \bibfield  {author} {\bibinfo {author} {\bibfnamefont {R.~A.}\ \bibnamefont
  {Borzi}}, \bibinfo {author} {\bibfnamefont {D.}~\bibnamefont {Slobinsky}}, \
  and\ \bibinfo {author} {\bibfnamefont {S.~A.}\ \bibnamefont {Grigera}},\
  }\href {\doibase 10.1103/PhysRevLett.111.147204} {\bibfield  {journal}
  {\bibinfo  {journal} {Phys. Rev. Lett.}\ }\textbf {\bibinfo {volume} {111}},\
  \bibinfo {pages} {147204} (\bibinfo {year} {2013})}\BibitemShut {NoStop}%
\bibitem [{\citenamefont {Katayama}\ \emph {et~al.}(2000)\citenamefont
  {Katayama}, \citenamefont {Mizutani}, \citenamefont {Utsumi}, \citenamefont
  {Shimomura}, \citenamefont {Yamakata},\ and\ \citenamefont
  {Funakoshi}}]{Katayama2000}%
  \BibitemOpen
  \bibfield  {author} {\bibinfo {author} {\bibfnamefont {Y.}~\bibnamefont
  {Katayama}}, \bibinfo {author} {\bibfnamefont {T.}~\bibnamefont {Mizutani}},
  \bibinfo {author} {\bibfnamefont {W.}~\bibnamefont {Utsumi}}, \bibinfo
  {author} {\bibfnamefont {O.}~\bibnamefont {Shimomura}}, \bibinfo {author}
  {\bibfnamefont {M.}~\bibnamefont {Yamakata}}, \ and\ \bibinfo {author}
  {\bibfnamefont {K.-i.}\ \bibnamefont {Funakoshi}},\ }\href
  {https://doi.org/10.1038/35003143} {\bibfield  {journal} {\bibinfo  {journal}
  {Nature}\ }\textbf {\bibinfo {volume} {403}},\ \bibinfo {pages} {170 EP }
  (\bibinfo {year} {2000})}\BibitemShut {NoStop}%
\bibitem [{\citenamefont {Sastry}\ and\ \citenamefont
  {Austen~Angell}(2003)}]{Sastry2003}%
  \BibitemOpen
  \bibfield  {author} {\bibinfo {author} {\bibfnamefont {S.}~\bibnamefont
  {Sastry}}\ and\ \bibinfo {author} {\bibfnamefont {C.}~\bibnamefont
  {Austen~Angell}},\ }\href {https://doi.org/10.1038/nmat994} {\bibfield
  {journal} {\bibinfo  {journal} {Nature Materials}\ }\textbf {\bibinfo
  {volume} {2}},\ \bibinfo {pages} {739 EP } (\bibinfo {year}
  {2003})}\BibitemShut {NoStop}%
\bibitem [{\citenamefont {Brovchenko}\ \emph {et~al.}(2005)\citenamefont
  {Brovchenko}, \citenamefont {Geiger},\ and\ \citenamefont
  {Oleinikova}}]{Brovchenko2005}%
  \BibitemOpen
  \bibfield  {author} {\bibinfo {author} {\bibfnamefont {I.}~\bibnamefont
  {Brovchenko}}, \bibinfo {author} {\bibfnamefont {A.}~\bibnamefont {Geiger}},
  \ and\ \bibinfo {author} {\bibfnamefont {A.}~\bibnamefont {Oleinikova}},\
  }\bibfield  {booktitle} {\emph {\bibinfo {booktitle} {The Journal of Chemical
  Physics}},\ }\href {\doibase 10.1063/1.1992481} {\bibfield  {journal}
  {\bibinfo  {journal} {The Journal of Chemical Physics}\ }\textbf {\bibinfo
  {volume} {123}},\ \bibinfo {pages} {044515} (\bibinfo {year}
  {2005})}\BibitemShut {NoStop}%
\bibitem [{\citenamefont {Kaluarachchi}\ \emph {et~al.}(2017)\citenamefont
  {Kaluarachchi}, \citenamefont {Bud'ko}, \citenamefont {Canfield},\ and\
  \citenamefont {Taufour}}]{Taufour2017}%
  \BibitemOpen
  \bibfield  {author} {\bibinfo {author} {\bibfnamefont {U.~S.}\ \bibnamefont
  {Kaluarachchi}}, \bibinfo {author} {\bibfnamefont {S.~L.}\ \bibnamefont
  {Bud'ko}}, \bibinfo {author} {\bibfnamefont {P.~C.}\ \bibnamefont
  {Canfield}}, \ and\ \bibinfo {author} {\bibfnamefont {V.}~\bibnamefont
  {Taufour}},\ }\href {\doibase 10.1038/s41467-017-00699-x} {\bibfield
  {journal} {\bibinfo  {journal} {Nature Communications}\ }\textbf {\bibinfo
  {volume} {8}},\ \bibinfo {pages} {546} (\bibinfo {year} {2017})}\BibitemShut
  {NoStop}%
\bibitem [{\citenamefont {Kotegawa}\ \emph {et~al.}(2011)\citenamefont
  {Kotegawa}, \citenamefont {Taufour}, \citenamefont {Aoki}, \citenamefont
  {Knebel},\ and\ \citenamefont {Flouquet}}]{Taufour2011}%
  \BibitemOpen
  \bibfield  {author} {\bibinfo {author} {\bibfnamefont {H.}~\bibnamefont
  {Kotegawa}}, \bibinfo {author} {\bibfnamefont {V.}~\bibnamefont {Taufour}},
  \bibinfo {author} {\bibfnamefont {D.}~\bibnamefont {Aoki}}, \bibinfo {author}
  {\bibfnamefont {G.}~\bibnamefont {Knebel}}, \ and\ \bibinfo {author}
  {\bibfnamefont {J.}~\bibnamefont {Flouquet}},\ }\bibfield  {booktitle} {\emph
  {\bibinfo {booktitle} {Journal of the Physical Society of Japan}},\ }\href
  {\doibase 10.1143/JPSJ.80.083703} {\bibfield  {journal} {\bibinfo  {journal}
  {Journal of the Physical Society of Japan}\ }\textbf {\bibinfo {volume}
  {80}},\ \bibinfo {pages} {083703} (\bibinfo {year} {2011})}\BibitemShut
  {NoStop}%
\bibitem [{\citenamefont {Plascak}\ \emph {et~al.}(1993)\citenamefont
  {Plascak}, \citenamefont {Moreira},\ and\ \citenamefont
  {sáBarreto}}]{Plascak1993}%
  \BibitemOpen
  \bibfield  {author} {\bibinfo {author} {\bibfnamefont {J.}~\bibnamefont
  {Plascak}}, \bibinfo {author} {\bibfnamefont {J.}~\bibnamefont {Moreira}}, \
  and\ \bibinfo {author} {\bibfnamefont {F.}~\bibnamefont {sáBarreto}},\ }\href
  {\doibase https://doi.org/10.1016/0375-9601(93)90250-4} {\bibfield  {journal}
  {\bibinfo  {journal} {Physics Letters A}\ }\textbf {\bibinfo {volume}
  {173}},\ \bibinfo {pages} {360 } (\bibinfo {year} {1993})}\BibitemShut
  {NoStop}%
\bibitem [{\citenamefont {Henelius}\ \emph {et~al.}(2016)\citenamefont
  {Henelius}, \citenamefont {Lin}, \citenamefont {Enjalran}, \citenamefont
  {Hao}, \citenamefont {Rau}, \citenamefont {Altosaar}, \citenamefont
  {Flicker}, \citenamefont {Yavors'kii},\ and\ \citenamefont
  {Gingras}}]{Henelius2016}%
  \BibitemOpen
  \bibfield  {author} {\bibinfo {author} {\bibfnamefont {P.}~\bibnamefont
  {Henelius}}, \bibinfo {author} {\bibfnamefont {T.}~\bibnamefont {Lin}},
  \bibinfo {author} {\bibfnamefont {M.}~\bibnamefont {Enjalran}}, \bibinfo
  {author} {\bibfnamefont {Z.}~\bibnamefont {Hao}}, \bibinfo {author}
  {\bibfnamefont {J.~G.}\ \bibnamefont {Rau}}, \bibinfo {author} {\bibfnamefont
  {J.}~\bibnamefont {Altosaar}}, \bibinfo {author} {\bibfnamefont
  {F.}~\bibnamefont {Flicker}}, \bibinfo {author} {\bibfnamefont
  {T.}~\bibnamefont {Yavors'kii}}, \ and\ \bibinfo {author} {\bibfnamefont
  {M.~J.~P.}\ \bibnamefont {Gingras}},\ }\href {\doibase
  10.1103/PhysRevB.93.024402} {\bibfield  {journal} {\bibinfo  {journal} {Phys.
  Rev. B}\ }\textbf {\bibinfo {volume} {93}},\ \bibinfo {pages} {024402}
  (\bibinfo {year} {2016})}\BibitemShut {NoStop}%
\bibitem [{\citenamefont {Melko}\ \emph {et~al.}(2001)\citenamefont {Melko},
  \citenamefont {den Hertog},\ and\ \citenamefont {Gingras}}]{Melko2001}%
  \BibitemOpen
  \bibfield  {author} {\bibinfo {author} {\bibfnamefont {R.~G.}\ \bibnamefont
  {Melko}}, \bibinfo {author} {\bibfnamefont {B.~C.}\ \bibnamefont {den
  Hertog}}, \ and\ \bibinfo {author} {\bibfnamefont {M.~J.~P.}\ \bibnamefont
  {Gingras}},\ }\href {\doibase 10.1103/PhysRevLett.87.067203} {\bibfield
  {journal} {\bibinfo  {journal} {Phys. Rev. Lett.}\ }\textbf {\bibinfo
  {volume} {87}},\ \bibinfo {pages} {067203} (\bibinfo {year}
  {2001})}\BibitemShut {NoStop}%
\bibitem [{\citenamefont {Cardy}(1996)}]{Cardy1996}%
  \BibitemOpen
  \bibfield  {author} {\bibinfo {author} {\bibfnamefont {J.}~\bibnamefont
  {Cardy}},\ }\href@noop {} {\emph {\bibinfo {title} {Scaling and
  Renormalization in Statistical Physics}}}\ (\bibinfo  {publisher} {Cambridge
  University Press},\ \bibinfo {year} {1996})\BibitemShut {NoStop}%
\bibitem [{\citenamefont {Blume}(1966)}]{Blume1966}%
  \BibitemOpen
  \bibfield  {author} {\bibinfo {author} {\bibfnamefont {M.}~\bibnamefont
  {Blume}},\ }\href {\doibase 10.1103/PhysRev.141.517} {\bibfield  {journal}
  {\bibinfo  {journal} {Phys. Rev.}\ }\textbf {\bibinfo {volume} {141}},\
  \bibinfo {pages} {517} (\bibinfo {year} {1966})}\BibitemShut {NoStop}%
\bibitem [{\citenamefont {{Capel}}(1966)}]{Capel1966}%
  \BibitemOpen
  \bibfield  {author} {\bibinfo {author} {\bibfnamefont {H.~W.}\ \bibnamefont
  {{Capel}}},\ }\href {\doibase 10.1016/0031-8914(66)90027-9} {\bibfield
  {journal} {\bibinfo  {journal} {Physica}\ }\textbf {\bibinfo {volume} {32}},\
  \bibinfo {pages} {966} (\bibinfo {year} {1966})}\BibitemShut {NoStop}%
\bibitem [{\citenamefont {Blume}\ \emph {et~al.}(1971)\citenamefont {Blume},
  \citenamefont {Emery},\ and\ \citenamefont {Griffiths}}]{Blume1971}%
  \BibitemOpen
  \bibfield  {author} {\bibinfo {author} {\bibfnamefont {M.}~\bibnamefont
  {Blume}}, \bibinfo {author} {\bibfnamefont {V.~J.}\ \bibnamefont {Emery}}, \
  and\ \bibinfo {author} {\bibfnamefont {R.~B.}\ \bibnamefont {Griffiths}},\
  }\href {\doibase 10.1103/PhysRevA.4.1071} {\bibfield  {journal} {\bibinfo
  {journal} {Phys. Rev. A}\ }\textbf {\bibinfo {volume} {4}},\ \bibinfo {pages}
  {1071} (\bibinfo {year} {1971})}\BibitemShut {NoStop}%
\bibitem [{\citenamefont {Landau}\ and\ \citenamefont
  {Lifshitz}(1959)}]{Landau1959}%
  \BibitemOpen
  \bibfield  {author} {\bibinfo {author} {\bibfnamefont {L.~D.}\ \bibnamefont
  {Landau}}\ and\ \bibinfo {author} {\bibfnamefont {E.~M.}\ \bibnamefont
  {Lifshitz}},\ }\href@noop {} {\emph {\bibinfo {title} {Course of Theoretical
  Physics Vol. 5 (Statistical Physics Part 1)}}}\ (\bibinfo  {publisher}
  {Permagon},\ \bibinfo {year} {1959})\BibitemShut {NoStop}%
\bibitem [{\citenamefont {Lara}\ and\ \citenamefont
  {Plascak}(1998)}]{Lara1998}%
  \BibitemOpen
  \bibfield  {author} {\bibinfo {author} {\bibfnamefont {D.~P.}\ \bibnamefont
  {Lara}}\ and\ \bibinfo {author} {\bibfnamefont {J.~A.}\ \bibnamefont
  {Plascak}},\ }\href {\doibase 10.1142/S0217979298001198} {\bibfield
  {journal} {\bibinfo  {journal} {International Journal of Modern Physics B}\
  }\textbf {\bibinfo {volume} {12}},\ \bibinfo {pages} {2045} (\bibinfo {year}
  {1998})},\ \Eprint
  {http://arxiv.org/abs/https://doi.org/10.1142/S0217979298001198}
  {https://doi.org/10.1142/S0217979298001198} \BibitemShut {NoStop}%
\bibitem [{\citenamefont {Jaubert}\ \emph {et~al.}(2011)\citenamefont
  {Jaubert}, \citenamefont {Haque},\ and\ \citenamefont
  {Moessner}}]{Jaubert2011}%
  \BibitemOpen
  \bibfield  {author} {\bibinfo {author} {\bibfnamefont {L.~D.~C.}\
  \bibnamefont {Jaubert}}, \bibinfo {author} {\bibfnamefont {M.}~\bibnamefont
  {Haque}}, \ and\ \bibinfo {author} {\bibfnamefont {R.}~\bibnamefont
  {Moessner}},\ }\href
  {http://journals.aps.org/prl/abstract/10.1103/PhysRevLett.107.177202}
  {\bibfield  {journal} {\bibinfo  {journal} {Physical review letters}\
  }\textbf {\bibinfo {volume} {107}},\ \bibinfo {pages} {177202} (\bibinfo
  {year} {2011})}\BibitemShut {NoStop}%
\bibitem [{\citenamefont {Jaubert}\ \emph {et~al.}(2013)\citenamefont
  {Jaubert}, \citenamefont {Harris}, \citenamefont {Fennell}, \citenamefont
  {Melko}, \citenamefont {Bramwell},\ and\ \citenamefont
  {Holdsworth}}]{Jaubert2013}%
  \BibitemOpen
  \bibfield  {author} {\bibinfo {author} {\bibfnamefont {L.~D.~C.}\
  \bibnamefont {Jaubert}}, \bibinfo {author} {\bibfnamefont {M.~J.}\
  \bibnamefont {Harris}}, \bibinfo {author} {\bibfnamefont {T.}~\bibnamefont
  {Fennell}}, \bibinfo {author} {\bibfnamefont {R.~G.}\ \bibnamefont {Melko}},
  \bibinfo {author} {\bibfnamefont {S.~T.}\ \bibnamefont {Bramwell}}, \ and\
  \bibinfo {author} {\bibfnamefont {P.~C.~W.}\ \bibnamefont {Holdsworth}},\
  }\href {\doibase 10.1103/PhysRevX.3.011014} {\bibfield  {journal} {\bibinfo
  {journal} {Phys. Rev. X}\ }\textbf {\bibinfo {volume} {3}},\ \bibinfo {pages}
  {011014} (\bibinfo {year} {2013})}\BibitemShut {NoStop}%
\bibitem [{\citenamefont {Kaiser}\ \emph {et~al.}(2013)\citenamefont {Kaiser},
  \citenamefont {Bramwell}, \citenamefont {Holdsworth},\ and\ \citenamefont
  {Moessner}}]{Kaiser2013}%
  \BibitemOpen
  \bibfield  {author} {\bibinfo {author} {\bibfnamefont {V.}~\bibnamefont
  {Kaiser}}, \bibinfo {author} {\bibfnamefont {S.~T.}\ \bibnamefont
  {Bramwell}}, \bibinfo {author} {\bibfnamefont {P.~C.~W.}\ \bibnamefont
  {Holdsworth}}, \ and\ \bibinfo {author} {\bibfnamefont {R.}~\bibnamefont
  {Moessner}},\ }\href {\doibase 10.1038/nmat3729} {\bibfield  {journal}
  {\bibinfo  {journal} {Nature Materials}\ }\textbf {\bibinfo {volume} {12}},\
  \bibinfo {pages} {1033} (\bibinfo {year} {2013})}\BibitemShut {NoStop}%
\bibitem [{\citenamefont {Kaiser}\ \emph {et~al.}(2018)\citenamefont {Kaiser},
  \citenamefont {Bloxsom}, \citenamefont {Bovo}, \citenamefont {Bramwell},
  \citenamefont {Holdsworth},\ and\ \citenamefont {Moessner}}]{Kaiser2018}%
  \BibitemOpen
  \bibfield  {author} {\bibinfo {author} {\bibfnamefont {V.}~\bibnamefont
  {Kaiser}}, \bibinfo {author} {\bibfnamefont {J.}~\bibnamefont {Bloxsom}},
  \bibinfo {author} {\bibfnamefont {L.}~\bibnamefont {Bovo}}, \bibinfo {author}
  {\bibfnamefont {S.~T.}\ \bibnamefont {Bramwell}}, \bibinfo {author}
  {\bibfnamefont {P.~C.~W.}\ \bibnamefont {Holdsworth}}, \ and\ \bibinfo
  {author} {\bibfnamefont {R.}~\bibnamefont {Moessner}},\ }\href {\doibase
  10.1103/PhysRevB.98.144413} {\bibfield  {journal} {\bibinfo  {journal} {Phys.
  Rev. B}\ }\textbf {\bibinfo {volume} {98}},\ \bibinfo {pages} {144413}
  (\bibinfo {year} {2018})}\BibitemShut {NoStop}%
\bibitem [{\citenamefont {Castelnovo}\ and\ \citenamefont
  {Holdsworth}(2019)}]{Castelnovo2019}%
  \BibitemOpen
  \bibfield  {author} {\bibinfo {author} {\bibfnamefont {C.}~\bibnamefont
  {Castelnovo}}\ and\ \bibinfo {author} {\bibfnamefont {P.~C.~W.}\ \bibnamefont
  {Holdsworth}},\ }\href@noop {} {\emph {\bibinfo {title} {Spin Ice, Edited by
  L. D. C. Jaubert and M. Udagawa}}}\ (\bibinfo  {publisher} {Springer},\
  \bibinfo {year} {2019})\BibitemShut {NoStop}%
\bibitem [{\citenamefont {Henley}(2010)}]{Henley2010}%
  \BibitemOpen
  \bibfield  {author} {\bibinfo {author} {\bibfnamefont {C.~L.}\ \bibnamefont
  {Henley}},\ }\href@noop {} {\bibfield  {journal} {\bibinfo  {journal} {Annual
  Review of Condensed Matter Physics}\ }\textbf {\bibinfo {volume} {1}},\
  \bibinfo {pages} {179} (\bibinfo {year} {2010})}\BibitemShut {NoStop}%
\bibitem [{\citenamefont {Nagle}(1966)}]{Nagle1966}%
  \BibitemOpen
  \bibfield  {author} {\bibinfo {author} {\bibfnamefont {J.~F.}\ \bibnamefont
  {Nagle}},\ }\href {\doibase 10.1103/PhysRev.152.190} {\bibfield  {journal}
  {\bibinfo  {journal} {Phys. Rev.}\ }\textbf {\bibinfo {volume} {152}},\
  \bibinfo {pages} {190} (\bibinfo {year} {1966})}\BibitemShut {NoStop}%
\bibitem [{\citenamefont {Maggs}\ and\ \citenamefont
  {Rossetto}(2002)}]{Maggs2002}%
  \BibitemOpen
  \bibfield  {author} {\bibinfo {author} {\bibfnamefont {A.~C.}\ \bibnamefont
  {Maggs}}\ and\ \bibinfo {author} {\bibfnamefont {V.}~\bibnamefont
  {Rossetto}},\ }\href@noop {} {\bibfield  {journal} {\bibinfo  {journal}
  {Phys. Rev. Lett.}\ }\textbf {\bibinfo {volume} {88}},\ \bibinfo {pages}
  {196402} (\bibinfo {year} {2002})}\BibitemShut {NoStop}%
\bibitem [{Note1()}]{Note1}%
  \BibitemOpen
  \bibinfo {note} {The value of $u(a)=2.88$ used here corresponds to a magnetic
  moment for the spins, $m=9.87\mu _B$, as deduced from crystal field
  calculations \cite {Yavorskii2008} rather than the $10 \mu _B$ often quoted
  in the literature, which gives $u(a)=3.07$K and a corresponding difference in
  the energy scale for the phase diagram}\BibitemShut {NoStop}%
\bibitem [{\citenamefont {Binder}(1981)}]{binder1981finite}%
  \BibitemOpen
  \bibfield  {author} {\bibinfo {author} {\bibfnamefont {K.}~\bibnamefont
  {Binder}},\ }\href@noop {} {\bibfield  {journal} {\bibinfo  {journal}
  {Zeitschrift f{\"u}r Physik B Condensed Matter}\ }\textbf {\bibinfo {volume}
  {43}},\ \bibinfo {pages} {119} (\bibinfo {year} {1981})}\BibitemShut
  {NoStop}%
\bibitem [{\citenamefont {Hamp}\ \emph {et~al.}(2015)\citenamefont {Hamp},
  \citenamefont {Chandran}, \citenamefont {Moessner},\ and\ \citenamefont
  {Castelnovo}}]{Hamp2015}%
  \BibitemOpen
  \bibfield  {author} {\bibinfo {author} {\bibfnamefont {J.}~\bibnamefont
  {Hamp}}, \bibinfo {author} {\bibfnamefont {A.}~\bibnamefont {Chandran}},
  \bibinfo {author} {\bibfnamefont {R.}~\bibnamefont {Moessner}}, \ and\
  \bibinfo {author} {\bibfnamefont {C.}~\bibnamefont {Castelnovo}},\ }\href
  {\doibase 10.1103/PhysRevB.92.075142} {\bibfield  {journal} {\bibinfo
  {journal} {Phys. Rev. B}\ }\textbf {\bibinfo {volume} {92}},\ \bibinfo
  {pages} {075142} (\bibinfo {year} {2015})}\BibitemShut {NoStop}%
\bibitem [{\citenamefont {Fenz}\ \emph {et~al.}(2007)\citenamefont {Fenz},
  \citenamefont {Folk}, \citenamefont {Mryglod},\ and\ \citenamefont
  {Omelyan}}]{Fenz2007}%
  \BibitemOpen
  \bibfield  {author} {\bibinfo {author} {\bibfnamefont {W.}~\bibnamefont
  {Fenz}}, \bibinfo {author} {\bibfnamefont {R.}~\bibnamefont {Folk}}, \bibinfo
  {author} {\bibfnamefont {I.~M.}\ \bibnamefont {Mryglod}}, \ and\ \bibinfo
  {author} {\bibfnamefont {I.~P.}\ \bibnamefont {Omelyan}},\ }\href {\doibase
  10.1103/PhysRevE.75.061504} {\bibfield  {journal} {\bibinfo  {journal} {Phys.
  Rev. E}\ }\textbf {\bibinfo {volume} {75}},\ \bibinfo {pages} {061504}
  (\bibinfo {year} {2007})}\BibitemShut {NoStop}%
\bibitem [{\citenamefont {Rummukainen}\ \emph {et~al.}(1998)\citenamefont
  {Rummukainen}, \citenamefont {Tsypin}, \citenamefont {Kajantie},
  \citenamefont {Laine},\ and\ \citenamefont {Shaposhnikov}}]{Rummukainen1998}%
  \BibitemOpen
  \bibfield  {author} {\bibinfo {author} {\bibfnamefont {K.}~\bibnamefont
  {Rummukainen}}, \bibinfo {author} {\bibfnamefont {M.}~\bibnamefont {Tsypin}},
  \bibinfo {author} {\bibfnamefont {K.}~\bibnamefont {Kajantie}}, \bibinfo
  {author} {\bibfnamefont {M.}~\bibnamefont {Laine}}, \ and\ \bibinfo {author}
  {\bibfnamefont {M.}~\bibnamefont {Shaposhnikov}},\ }\href {\doibase
  https://doi.org/10.1016/S0550-3213(98)00494-5} {\bibfield  {journal}
  {\bibinfo  {journal} {Nuclear Physics B}\ }\textbf {\bibinfo {volume}
  {532}},\ \bibinfo {pages} {283 } (\bibinfo {year} {1998})}\BibitemShut
  {NoStop}%
\bibitem [{\citenamefont {Cardozo}\ and\ \citenamefont
  {Holdsworth}(2016)}]{Lopes2016}%
  \BibitemOpen
  \bibfield  {author} {\bibinfo {author} {\bibfnamefont {D.~L.}\ \bibnamefont
  {Cardozo}}\ and\ \bibinfo {author} {\bibfnamefont {P.~C.~W.}\ \bibnamefont
  {Holdsworth}},\ }\href {http://stacks.iop.org/0953-8984/28/i=16/a=166007}
  {\bibfield  {journal} {\bibinfo  {journal} {Journal of Physics: Condensed
  Matter}\ }\textbf {\bibinfo {volume} {28}},\ \bibinfo {pages} {166007}
  (\bibinfo {year} {2016})}\BibitemShut {NoStop}%
\bibitem [{\citenamefont {Taufour}\ \emph {et~al.}(2016)\citenamefont
  {Taufour}, \citenamefont {Kaluarachchi},\ and\ \citenamefont
  {Kogan}}]{Taufour2016}%
  \BibitemOpen
  \bibfield  {author} {\bibinfo {author} {\bibfnamefont {V.}~\bibnamefont
  {Taufour}}, \bibinfo {author} {\bibfnamefont {U.~S.}\ \bibnamefont
  {Kaluarachchi}}, \ and\ \bibinfo {author} {\bibfnamefont {V.~G.}\
  \bibnamefont {Kogan}},\ }\href {\doibase 10.1103/PhysRevB.94.060410}
  {\bibfield  {journal} {\bibinfo  {journal} {Phys. Rev. B}\ }\textbf {\bibinfo
  {volume} {94}},\ \bibinfo {pages} {060410} (\bibinfo {year}
  {2016})}\BibitemShut {NoStop}%
\bibitem [{\citenamefont {Hohenberg}\ and\ \citenamefont
  {Halperin}(1977)}]{Hohenberg1977}%
  \BibitemOpen
  \bibfield  {author} {\bibinfo {author} {\bibfnamefont {P.~C.}\ \bibnamefont
  {Hohenberg}}\ and\ \bibinfo {author} {\bibfnamefont {B.~I.}\ \bibnamefont
  {Halperin}},\ }\href {\doibase 10.1103/RevModPhys.49.435} {\bibfield
  {journal} {\bibinfo  {journal} {Rev. Mod. Phys.}\ }\textbf {\bibinfo {volume}
  {49}},\ \bibinfo {pages} {435} (\bibinfo {year} {1977})}\BibitemShut
  {NoStop}%
\bibitem [{\citenamefont {Jaubert}(2015)}]{Jaubert2015}%
  \BibitemOpen
  \bibfield  {author} {\bibinfo {author} {\bibfnamefont {L.~D.~C.}\
  \bibnamefont {Jaubert}},\ }\href {\doibase 10.1142/S2010324715400056}
  {\bibfield  {journal} {\bibinfo  {journal} {SPIN}\ }\textbf {\bibinfo
  {volume} {05}},\ \bibinfo {pages} {1540005} (\bibinfo {year} {2015})},\
  \Eprint {http://arxiv.org/abs/https://doi.org/10.1142/S2010324715400056}
  {https://doi.org/10.1142/S2010324715400056} \BibitemShut {NoStop}%
\bibitem [{\citenamefont {Kibble}(1980)}]{Kibble1980}%
  \BibitemOpen
  \bibfield  {author} {\bibinfo {author} {\bibfnamefont {T.}~\bibnamefont
  {Kibble}},\ }\href {\doibase https://doi.org/10.1016/0370-1573(80)90091-5}
  {\bibfield  {journal} {\bibinfo  {journal} {Physics Reports}\ }\textbf
  {\bibinfo {volume} {67}},\ \bibinfo {pages} {183 } (\bibinfo {year}
  {1980})}\BibitemShut {NoStop}%
\bibitem [{\citenamefont {Zurek}(1985)}]{Zurek1985}%
  \BibitemOpen
  \bibfield  {author} {\bibinfo {author} {\bibfnamefont {W.~H.}\ \bibnamefont
  {Zurek}},\ }\href {https://doi.org/10.1038/317505a0} {\bibfield  {journal}
  {\bibinfo  {journal} {Nature}\ }\textbf {\bibinfo {volume} {317}},\ \bibinfo
  {pages} {505 EP } (\bibinfo {year} {1985})}\BibitemShut {NoStop}%
\bibitem [{\citenamefont {Pelissetto}\ and\ \citenamefont
  {Vicari}(2002)}]{Pelissetto2002}%
  \BibitemOpen
  \bibfield  {author} {\bibinfo {author} {\bibfnamefont {A.}~\bibnamefont
  {Pelissetto}}\ and\ \bibinfo {author} {\bibfnamefont {E.}~\bibnamefont
  {Vicari}},\ }\href {\doibase https://doi.org/10.1016/S0370-1573(02)00219-3}
  {\bibfield  {journal} {\bibinfo  {journal} {Physics Reports}\ }\textbf
  {\bibinfo {volume} {368}},\ \bibinfo {pages} {549 } (\bibinfo {year}
  {2002})}\BibitemShut {NoStop}%
\bibitem [{\citenamefont {Wansleben}\ and\ \citenamefont
  {Landau}(1991)}]{Wansleben1991}%
  \BibitemOpen
  \bibfield  {author} {\bibinfo {author} {\bibfnamefont {S.}~\bibnamefont
  {Wansleben}}\ and\ \bibinfo {author} {\bibfnamefont {D.~P.}\ \bibnamefont
  {Landau}},\ }\href {\doibase 10.1103/PhysRevB.43.6006} {\bibfield  {journal}
  {\bibinfo  {journal} {Phys. Rev. B}\ }\textbf {\bibinfo {volume} {43}},\
  \bibinfo {pages} {6006} (\bibinfo {year} {1991})}\BibitemShut {NoStop}%
\bibitem [{\citenamefont {Suzuki}(1974)}]{Suzuki1992}%
  \BibitemOpen
  \bibfield  {author} {\bibinfo {author} {\bibfnamefont {M.}~\bibnamefont
  {Suzuki}},\ }\href {http://dx.doi.org/10.1143/PTP.51.1992} {\bibfield
  {journal} {\bibinfo  {journal} {Progress of Theoretical Physics}\ }\textbf
  {\bibinfo {volume} {51}},\ \bibinfo {pages} {1992} (\bibinfo {year}
  {1974})}\BibitemShut {NoStop}%
\bibitem [{\citenamefont {Taroni}\ \emph {et~al.}(2008)\citenamefont {Taroni},
  \citenamefont {Bramwell},\ and\ \citenamefont {Holdsworth}}]{Taroni2008}%
  \BibitemOpen
  \bibfield  {author} {\bibinfo {author} {\bibfnamefont {A.}~\bibnamefont
  {Taroni}}, \bibinfo {author} {\bibfnamefont {S.~T.}\ \bibnamefont
  {Bramwell}}, \ and\ \bibinfo {author} {\bibfnamefont {P.~C.~W.}\ \bibnamefont
  {Holdsworth}},\ }\href {http://stacks.iop.org/0953-8984/20/i=27/a=275233}
  {\bibfield  {journal} {\bibinfo  {journal} {Journal of Physics: Condensed
  Matter}\ }\textbf {\bibinfo {volume} {20}},\ \bibinfo {pages} {275233}
  (\bibinfo {year} {2008})}\BibitemShut {NoStop}%
\bibitem [{\citenamefont {Godr{\`e}che}\ and\ \citenamefont
  {Luck}(2000)}]{godreche2000response}%
  \BibitemOpen
  \bibfield  {author} {\bibinfo {author} {\bibfnamefont {C.}~\bibnamefont
  {Godr{\`e}che}}\ and\ \bibinfo {author} {\bibfnamefont {J.}~\bibnamefont
  {Luck}},\ }\href@noop {} {\bibfield  {journal} {\bibinfo  {journal} {Journal
  of Physics A: Mathematical and General}\ }\textbf {\bibinfo {volume} {33}},\
  \bibinfo {pages} {1151} (\bibinfo {year} {2000})}\BibitemShut {NoStop}%
\bibitem [{\citenamefont {Godreche}\ and\ \citenamefont
  {Luck}(2000)}]{godreche2000responseb}%
  \BibitemOpen
  \bibfield  {author} {\bibinfo {author} {\bibfnamefont {C.}~\bibnamefont
  {Godreche}}\ and\ \bibinfo {author} {\bibfnamefont {J.}~\bibnamefont
  {Luck}},\ }\href@noop {} {\bibfield  {journal} {\bibinfo  {journal} {Journal
  of Physics A: Mathematical and General}\ }\textbf {\bibinfo {volume} {33}},\
  \bibinfo {pages} {9141} (\bibinfo {year} {2000})}\BibitemShut {NoStop}%
\bibitem [{\citenamefont {Berthier}\ \emph {et~al.}(2001)\citenamefont
  {Berthier}, \citenamefont {Holdsworth},\ and\ \citenamefont
  {Sellitto}}]{berthier2001nonequilibrium}%
  \BibitemOpen
  \bibfield  {author} {\bibinfo {author} {\bibfnamefont {L.}~\bibnamefont
  {Berthier}}, \bibinfo {author} {\bibfnamefont {P.~C.}\ \bibnamefont
  {Holdsworth}}, \ and\ \bibinfo {author} {\bibfnamefont {M.}~\bibnamefont
  {Sellitto}},\ }\href@noop {} {\bibfield  {journal} {\bibinfo  {journal}
  {Journal of Physics A: Mathematical and General}\ }\textbf {\bibinfo {volume}
  {34}},\ \bibinfo {pages} {1805} (\bibinfo {year} {2001})}\BibitemShut
  {NoStop}%
\bibitem [{\citenamefont {Henkel}\ \emph {et~al.}(2001)\citenamefont {Henkel},
  \citenamefont {Pleimling}, \citenamefont {Godreche},\ and\ \citenamefont
  {Luck}}]{henkel2001aging}%
  \BibitemOpen
  \bibfield  {author} {\bibinfo {author} {\bibfnamefont {M.}~\bibnamefont
  {Henkel}}, \bibinfo {author} {\bibfnamefont {M.}~\bibnamefont {Pleimling}},
  \bibinfo {author} {\bibfnamefont {C.}~\bibnamefont {Godreche}}, \ and\
  \bibinfo {author} {\bibfnamefont {J.-M.}\ \bibnamefont {Luck}},\ }\href@noop
  {} {\bibfield  {journal} {\bibinfo  {journal} {Physical review letters}\
  }\textbf {\bibinfo {volume} {87}},\ \bibinfo {pages} {265701} (\bibinfo
  {year} {2001})}\BibitemShut {NoStop}%
\bibitem [{\citenamefont {Henkel}\ and\ \citenamefont
  {Pleimling}(2003)}]{PhysRevE.68.065101}%
  \BibitemOpen
  \bibfield  {author} {\bibinfo {author} {\bibfnamefont {M.}~\bibnamefont
  {Henkel}}\ and\ \bibinfo {author} {\bibfnamefont {M.}~\bibnamefont
  {Pleimling}},\ }\href {\doibase 10.1103/PhysRevE.68.065101} {\bibfield
  {journal} {\bibinfo  {journal} {Phys. Rev. E}\ }\textbf {\bibinfo {volume}
  {68}},\ \bibinfo {pages} {065101} (\bibinfo {year} {2003})}\BibitemShut
  {NoStop}%
\bibitem [{\citenamefont {Calabrese}\ and\ \citenamefont
  {Gambassi}(2002{\natexlab{a}})}]{calabrese2002two}%
  \BibitemOpen
  \bibfield  {author} {\bibinfo {author} {\bibfnamefont {P.}~\bibnamefont
  {Calabrese}}\ and\ \bibinfo {author} {\bibfnamefont {A.}~\bibnamefont
  {Gambassi}},\ }\href@noop {} {\bibfield  {journal} {\bibinfo  {journal}
  {Physical Review E}\ }\textbf {\bibinfo {volume} {66}},\ \bibinfo {pages}
  {066101} (\bibinfo {year} {2002}{\natexlab{a}})}\BibitemShut {NoStop}%
\bibitem [{\citenamefont {Calabrese}\ and\ \citenamefont
  {Gambassi}(2002{\natexlab{b}})}]{calabrese2002aging}%
  \BibitemOpen
  \bibfield  {author} {\bibinfo {author} {\bibfnamefont {P.}~\bibnamefont
  {Calabrese}}\ and\ \bibinfo {author} {\bibfnamefont {A.}~\bibnamefont
  {Gambassi}},\ }\href@noop {} {\bibfield  {journal} {\bibinfo  {journal}
  {Physical Review E}\ }\textbf {\bibinfo {volume} {65}},\ \bibinfo {pages}
  {066120} (\bibinfo {year} {2002}{\natexlab{b}})}\BibitemShut {NoStop}%
\bibitem [{\citenamefont {Mayer}\ \emph {et~al.}(2003)\citenamefont {Mayer},
  \citenamefont {Berthier}, \citenamefont {Garrahan},\ and\ \citenamefont
  {Sollich}}]{mayer2003fluctuation}%
  \BibitemOpen
  \bibfield  {author} {\bibinfo {author} {\bibfnamefont {P.}~\bibnamefont
  {Mayer}}, \bibinfo {author} {\bibfnamefont {L.}~\bibnamefont {Berthier}},
  \bibinfo {author} {\bibfnamefont {J.~P.}\ \bibnamefont {Garrahan}}, \ and\
  \bibinfo {author} {\bibfnamefont {P.}~\bibnamefont {Sollich}},\ }\href@noop
  {} {\bibfield  {journal} {\bibinfo  {journal} {Physical Review E}\ }\textbf
  {\bibinfo {volume} {68}},\ \bibinfo {pages} {016116} (\bibinfo {year}
  {2003})}\BibitemShut {NoStop}%
\bibitem [{\citenamefont {Mayer}\ \emph {et~al.}(2004)\citenamefont {Mayer},
  \citenamefont {Berthier}, \citenamefont {Garrahan},\ and\ \citenamefont
  {Sollich}}]{mayer2004reply}%
  \BibitemOpen
  \bibfield  {author} {\bibinfo {author} {\bibfnamefont {P.}~\bibnamefont
  {Mayer}}, \bibinfo {author} {\bibfnamefont {L.}~\bibnamefont {Berthier}},
  \bibinfo {author} {\bibfnamefont {J.~P.}\ \bibnamefont {Garrahan}}, \ and\
  \bibinfo {author} {\bibfnamefont {P.}~\bibnamefont {Sollich}},\ }\href@noop
  {} {\bibfield  {journal} {\bibinfo  {journal} {Physical Review E}\ }\textbf
  {\bibinfo {volume} {70}},\ \bibinfo {pages} {018102} (\bibinfo {year}
  {2004})}\BibitemShut {NoStop}%
\bibitem [{\citenamefont {Calabrese}\ and\ \citenamefont
  {Gambassi}(2005)}]{calabrese2005ageing}%
  \BibitemOpen
  \bibfield  {author} {\bibinfo {author} {\bibfnamefont {P.}~\bibnamefont
  {Calabrese}}\ and\ \bibinfo {author} {\bibfnamefont {A.}~\bibnamefont
  {Gambassi}},\ }\href@noop {} {\bibfield  {journal} {\bibinfo  {journal}
  {Journal of Physics A: Mathematical and General}\ }\textbf {\bibinfo {volume}
  {38}},\ \bibinfo {pages} {R133} (\bibinfo {year} {2005})}\BibitemShut
  {NoStop}%
\bibitem [{\citenamefont {Prudnikov}\ \emph {et~al.}(2015)\citenamefont
  {Prudnikov}, \citenamefont {Prudnikov}, \citenamefont {Pospelov},\ and\
  \citenamefont {Vakilov}}]{PRUDNIKOV2015774}%
  \BibitemOpen
  \bibfield  {author} {\bibinfo {author} {\bibfnamefont {V.~V.}\ \bibnamefont
  {Prudnikov}}, \bibinfo {author} {\bibfnamefont {P.~V.}\ \bibnamefont
  {Prudnikov}}, \bibinfo {author} {\bibfnamefont {E.~A.}\ \bibnamefont
  {Pospelov}}, \ and\ \bibinfo {author} {\bibfnamefont {A.~N.}\ \bibnamefont
  {Vakilov}},\ }\href {\doibase https://doi.org/10.1016/j.physleta.2015.01.005}
  {\bibfield  {journal} {\bibinfo  {journal} {Physics Letters A}\ }\textbf
  {\bibinfo {volume} {379}},\ \bibinfo {pages} {774 } (\bibinfo {year}
  {2015})}\BibitemShut {NoStop}%
\bibitem [{\citenamefont {Bouchaud}\ \emph {et~al.}(1998)\citenamefont
  {Bouchaud}, \citenamefont {Cugliandolo}, \citenamefont {Kurchan},\ and\
  \citenamefont {Mezard}}]{bouchaud1998out}%
  \BibitemOpen
  \bibfield  {author} {\bibinfo {author} {\bibfnamefont {J.-P.}\ \bibnamefont
  {Bouchaud}}, \bibinfo {author} {\bibfnamefont {L.~F.}\ \bibnamefont
  {Cugliandolo}}, \bibinfo {author} {\bibfnamefont {J.}~\bibnamefont
  {Kurchan}}, \ and\ \bibinfo {author} {\bibfnamefont {M.}~\bibnamefont
  {Mezard}},\ }\href@noop {} {\bibfield  {journal} {\bibinfo  {journal} {Spin
  glasses and random fields}\ ,\ \bibinfo {pages} {161}} (\bibinfo {year}
  {1998})}\BibitemShut {NoStop}%
\bibitem [{\citenamefont {Cugliandolo}\ and\ \citenamefont
  {Kurchan}(1993)}]{cugliandolo1993analytical}%
  \BibitemOpen
  \bibfield  {author} {\bibinfo {author} {\bibfnamefont {L.~F.}\ \bibnamefont
  {Cugliandolo}}\ and\ \bibinfo {author} {\bibfnamefont {J.}~\bibnamefont
  {Kurchan}},\ }\href@noop {} {\bibfield  {journal} {\bibinfo  {journal}
  {Physical Review Letters}\ }\textbf {\bibinfo {volume} {71}},\ \bibinfo
  {pages} {173} (\bibinfo {year} {1993})}\BibitemShut {NoStop}%
\bibitem [{\citenamefont {Cugliandolo}\ \emph {et~al.}(1997)\citenamefont
  {Cugliandolo}, \citenamefont {Kurchan},\ and\ \citenamefont
  {Peliti}}]{cugliandolo1997energy}%
  \BibitemOpen
  \bibfield  {author} {\bibinfo {author} {\bibfnamefont {L.~F.}\ \bibnamefont
  {Cugliandolo}}, \bibinfo {author} {\bibfnamefont {J.}~\bibnamefont
  {Kurchan}}, \ and\ \bibinfo {author} {\bibfnamefont {L.}~\bibnamefont
  {Peliti}},\ }\href@noop {} {\bibfield  {journal} {\bibinfo  {journal}
  {Physical Review E}\ }\textbf {\bibinfo {volume} {55}},\ \bibinfo {pages}
  {3898} (\bibinfo {year} {1997})}\BibitemShut {NoStop}%
\bibitem [{\citenamefont {Crisanti}\ and\ \citenamefont
  {Ritort}(2003)}]{crisanti2003violation}%
  \BibitemOpen
  \bibfield  {author} {\bibinfo {author} {\bibfnamefont {A.}~\bibnamefont
  {Crisanti}}\ and\ \bibinfo {author} {\bibfnamefont {F.}~\bibnamefont
  {Ritort}},\ }\href@noop {} {\bibfield  {journal} {\bibinfo  {journal}
  {Journal of Physics A: Mathematical and General}\ }\textbf {\bibinfo {volume}
  {36}},\ \bibinfo {pages} {R181} (\bibinfo {year} {2003})}\BibitemShut
  {NoStop}%
\bibitem [{\citenamefont {Chatelain}(2003)}]{chatelain2003far}%
  \BibitemOpen
  \bibfield  {author} {\bibinfo {author} {\bibfnamefont {C.}~\bibnamefont
  {Chatelain}},\ }\href@noop {} {\bibfield  {journal} {\bibinfo  {journal}
  {Journal of Physics A: Mathematical and General}\ }\textbf {\bibinfo {volume}
  {36}},\ \bibinfo {pages} {10739} (\bibinfo {year} {2003})}\BibitemShut
  {NoStop}%
\bibitem [{\citenamefont {Ricci-Tersenghi}(2003)}]{ricci2003measuring}%
  \BibitemOpen
  \bibfield  {author} {\bibinfo {author} {\bibfnamefont {F.}~\bibnamefont
  {Ricci-Tersenghi}},\ }\href@noop {} {\bibfield  {journal} {\bibinfo
  {journal} {Physical Review E}\ }\textbf {\bibinfo {volume} {68}},\ \bibinfo
  {pages} {065104} (\bibinfo {year} {2003})}\BibitemShut {NoStop}%
\bibitem [{\citenamefont {Berthier}(2007)}]{berthier2007efficient}%
  \BibitemOpen
  \bibfield  {author} {\bibinfo {author} {\bibfnamefont {L.}~\bibnamefont
  {Berthier}},\ }\href@noop {} {\bibfield  {journal} {\bibinfo  {journal}
  {Physical review letters}\ }\textbf {\bibinfo {volume} {98}},\ \bibinfo
  {pages} {220601} (\bibinfo {year} {2007})}\BibitemShut {NoStop}%
\bibitem [{\citenamefont {Sakakibara}\ \emph {et~al.}(2003)\citenamefont
  {Sakakibara}, \citenamefont {Tayama}, \citenamefont {Hiroi}, \citenamefont
  {Matsuhira},\ and\ \citenamefont {Takagi}}]{Sakakibara2003}%
  \BibitemOpen
  \bibfield  {author} {\bibinfo {author} {\bibfnamefont {T.}~\bibnamefont
  {Sakakibara}}, \bibinfo {author} {\bibfnamefont {T.}~\bibnamefont {Tayama}},
  \bibinfo {author} {\bibfnamefont {Z.}~\bibnamefont {Hiroi}}, \bibinfo
  {author} {\bibfnamefont {K.}~\bibnamefont {Matsuhira}}, \ and\ \bibinfo
  {author} {\bibfnamefont {S.}~\bibnamefont {Takagi}},\ }\href {\doibase
  10.1103/PhysRevLett.90.207205} {\bibfield  {journal} {\bibinfo  {journal}
  {Phys. Rev. Lett.}\ }\textbf {\bibinfo {volume} {90}},\ \bibinfo {pages}
  {207205} (\bibinfo {year} {2003})}\BibitemShut {NoStop}%
\bibitem [{\citenamefont {Aoki}\ \emph {et~al.}(2004)\citenamefont {Aoki},
  \citenamefont {Sakakibara}, \citenamefont {Matsuhira},\ and\ \citenamefont
  {Hiroi}}]{Aoki2004}%
  \BibitemOpen
  \bibfield  {author} {\bibinfo {author} {\bibfnamefont {H.}~\bibnamefont
  {Aoki}}, \bibinfo {author} {\bibfnamefont {T.}~\bibnamefont {Sakakibara}},
  \bibinfo {author} {\bibfnamefont {K.}~\bibnamefont {Matsuhira}}, \ and\
  \bibinfo {author} {\bibfnamefont {Z.}~\bibnamefont {Hiroi}},\ }\href
  {\doibase 10.1143/JPSJ.73.2851} {\bibfield  {journal} {\bibinfo  {journal}
  {Journal of the Physical Society of Japan}\ }\textbf {\bibinfo {volume}
  {73}},\ \bibinfo {pages} {2851} (\bibinfo {year} {2004})},\ \Eprint
  {http://arxiv.org/abs/http://dx.doi.org/10.1143/JPSJ.73.2851}
  {http://dx.doi.org/10.1143/JPSJ.73.2851} \BibitemShut {NoStop}%
\bibitem [{\citenamefont {Higashinaka}\ \emph {et~al.}(2004)\citenamefont
  {Higashinaka}, \citenamefont {Fukazawa}, \citenamefont {Deguchi},\ and\
  \citenamefont {Maeno}}]{Higashinaka2004}%
  \BibitemOpen
  \bibfield  {author} {\bibinfo {author} {\bibfnamefont {R.}~\bibnamefont
  {Higashinaka}}, \bibinfo {author} {\bibfnamefont {H.}~\bibnamefont
  {Fukazawa}}, \bibinfo {author} {\bibfnamefont {K.}~\bibnamefont {Deguchi}}, \
  and\ \bibinfo {author} {\bibfnamefont {Y.}~\bibnamefont {Maeno}},\ }\bibfield
   {booktitle} {\emph {\bibinfo {booktitle} {Journal of the Physical Society of
  Japan}},\ }\href {\doibase 10.1143/JPSJ.73.2845} {\bibfield  {journal}
  {\bibinfo  {journal} {Journal of the Physical Society of Japan}\ }\textbf
  {\bibinfo {volume} {73}},\ \bibinfo {pages} {2845} (\bibinfo {year}
  {2004})}\BibitemShut {NoStop}%
\bibitem [{\citenamefont {Isakov}\ \emph
  {et~al.}(2004{\natexlab{b}})\citenamefont {Isakov}, \citenamefont {Raman},
  \citenamefont {Moessner},\ and\ \citenamefont {Sondhi}}]{Isakov2004a}%
  \BibitemOpen
  \bibfield  {author} {\bibinfo {author} {\bibfnamefont {S.~V.}\ \bibnamefont
  {Isakov}}, \bibinfo {author} {\bibfnamefont {K.~S.}\ \bibnamefont {Raman}},
  \bibinfo {author} {\bibfnamefont {R.}~\bibnamefont {Moessner}}, \ and\
  \bibinfo {author} {\bibfnamefont {S.~L.}\ \bibnamefont {Sondhi}},\ }\href
  {\doibase 10.1103/PhysRevB.70.104418} {\bibfield  {journal} {\bibinfo
  {journal} {Phys. Rev. B}\ }\textbf {\bibinfo {volume} {70}},\ \bibinfo
  {pages} {104418} (\bibinfo {year} {2004}{\natexlab{b}})}\BibitemShut
  {NoStop}%
\bibitem [{\citenamefont {Castelnovo}\ \emph {et~al.}(2010)\citenamefont
  {Castelnovo}, \citenamefont {Moessner},\ and\ \citenamefont
  {Sondhi}}]{Castelnovo2010}%
  \BibitemOpen
  \bibfield  {author} {\bibinfo {author} {\bibfnamefont {C.}~\bibnamefont
  {Castelnovo}}, \bibinfo {author} {\bibfnamefont {R.}~\bibnamefont
  {Moessner}}, \ and\ \bibinfo {author} {\bibfnamefont {S.~L.}\ \bibnamefont
  {Sondhi}},\ }\href {\doibase 10.1103/PhysRevLett.104.107201} {\bibfield
  {journal} {\bibinfo  {journal} {Phys. Rev. Lett.}\ }\textbf {\bibinfo
  {volume} {104}},\ \bibinfo {pages} {107201} (\bibinfo {year}
  {2010})}\BibitemShut {NoStop}%
\bibitem [{\citenamefont {Udagawa}\ \emph {et~al.}(2002)\citenamefont
  {Udagawa}, \citenamefont {Ogata},\ and\ \citenamefont
  {Hiroi}}]{Masafumi2002}%
  \BibitemOpen
  \bibfield  {author} {\bibinfo {author} {\bibfnamefont {M.}~\bibnamefont
  {Udagawa}}, \bibinfo {author} {\bibfnamefont {M.}~\bibnamefont {Ogata}}, \
  and\ \bibinfo {author} {\bibfnamefont {Z.}~\bibnamefont {Hiroi}},\ }\href
  {\doibase 10.1143/JPSJ.71.2365} {\bibfield  {journal} {\bibinfo  {journal}
  {Journal of the Physical Society of Japan}\ }\textbf {\bibinfo {volume}
  {71}},\ \bibinfo {pages} {2365} (\bibinfo {year} {2002})},\ \Eprint
  {http://arxiv.org/abs/http://dx.doi.org/10.1143/JPSJ.71.2365}
  {http://dx.doi.org/10.1143/JPSJ.71.2365} \BibitemShut {NoStop}%
\bibitem [{\citenamefont {Yavors'kii}\ \emph {et~al.}(2008)\citenamefont
  {Yavors'kii}, \citenamefont {Fennell}, \citenamefont {Gingras},\ and\
  \citenamefont {Bramwell}}]{Yavorskii2008}%
  \BibitemOpen
  \bibfield  {author} {\bibinfo {author} {\bibfnamefont {T.}~\bibnamefont
  {Yavors'kii}}, \bibinfo {author} {\bibfnamefont {T.}~\bibnamefont {Fennell}},
  \bibinfo {author} {\bibfnamefont {M.~J.~P.}\ \bibnamefont {Gingras}}, \ and\
  \bibinfo {author} {\bibfnamefont {S.~T.}\ \bibnamefont {Bramwell}},\ }\href
  {\doibase 10.1103/PhysRevLett.101.037204} {\bibfield  {journal} {\bibinfo
  {journal} {Phys. Rev. Lett.}\ }\textbf {\bibinfo {volume} {101}},\ \bibinfo
  {pages} {037204} (\bibinfo {year} {2008})}\BibitemShut {NoStop}%
\bibitem [{\citenamefont {Champion}\ \emph {et~al.}(2002)\citenamefont
  {Champion}, \citenamefont {Bramwell}, \citenamefont {Holdsworth},\ and\
  \citenamefont {Harris}}]{Champion2002}%
  \BibitemOpen
  \bibfield  {author} {\bibinfo {author} {\bibfnamefont {J.~D.~M.}\
  \bibnamefont {Champion}}, \bibinfo {author} {\bibfnamefont {S.~T.}\
  \bibnamefont {Bramwell}}, \bibinfo {author} {\bibfnamefont {P.~C.~W.}\
  \bibnamefont {Holdsworth}}, \ and\ \bibinfo {author} {\bibfnamefont {M.~J.}\
  \bibnamefont {Harris}},\ }\href {\doibase 10.1209/epl/i2002-00546-1}
  {\bibfield  {journal} {\bibinfo  {journal} {Europhysics Letters}\ }\textbf
  {\bibinfo {volume} {57}},\ \bibinfo {pages} {93} (\bibinfo {year}
  {2002})}\BibitemShut {NoStop}%
\bibitem [{\citenamefont {Zhitomirsky}\ \emph {et~al.}(2012)\citenamefont
  {Zhitomirsky}, \citenamefont {Gvozdikova}, \citenamefont {Holdsworth},\ and\
  \citenamefont {Moessner}}]{Zhitomirsky2012}%
  \BibitemOpen
  \bibfield  {author} {\bibinfo {author} {\bibfnamefont {M.~E.}\ \bibnamefont
  {Zhitomirsky}}, \bibinfo {author} {\bibfnamefont {M.~V.}\ \bibnamefont
  {Gvozdikova}}, \bibinfo {author} {\bibfnamefont {P.~C.~W.}\ \bibnamefont
  {Holdsworth}}, \ and\ \bibinfo {author} {\bibfnamefont {R.}~\bibnamefont
  {Moessner}},\ }\href {\doibase 10.1103/PhysRevLett.109.077204} {\bibfield
  {journal} {\bibinfo  {journal} {Phys. Rev. Lett.}\ }\textbf {\bibinfo
  {volume} {109}},\ \bibinfo {pages} {077204} (\bibinfo {year}
  {2012})}\BibitemShut {NoStop}%
\bibitem [{\citenamefont {Shahbazi}\ and\ \citenamefont
  {Mortezapour}(2008)}]{Shahbazi2008}%
  \BibitemOpen
  \bibfield  {author} {\bibinfo {author} {\bibfnamefont {F.}~\bibnamefont
  {Shahbazi}}\ and\ \bibinfo {author} {\bibfnamefont {S.}~\bibnamefont
  {Mortezapour}},\ }\href {\doibase 10.1103/PhysRevB.77.214420} {\bibfield
  {journal} {\bibinfo  {journal} {Phys. Rev. B}\ }\textbf {\bibinfo {volume}
  {77}},\ \bibinfo {pages} {214420} (\bibinfo {year} {2008})}\BibitemShut
  {NoStop}%
\bibitem [{\citenamefont {Sadeghi}\ \emph {et~al.}(2015)\citenamefont
  {Sadeghi}, \citenamefont {Alaei}, \citenamefont {Shahbazi},\ and\
  \citenamefont {Gingras}}]{Sadeghi2015}%
  \BibitemOpen
  \bibfield  {author} {\bibinfo {author} {\bibfnamefont {A.}~\bibnamefont
  {Sadeghi}}, \bibinfo {author} {\bibfnamefont {M.}~\bibnamefont {Alaei}},
  \bibinfo {author} {\bibfnamefont {F.}~\bibnamefont {Shahbazi}}, \ and\
  \bibinfo {author} {\bibfnamefont {M.~J.~P.}\ \bibnamefont {Gingras}},\ }\href
  {\doibase 10.1103/PhysRevB.91.140407} {\bibfield  {journal} {\bibinfo
  {journal} {Phys. Rev. B}\ }\textbf {\bibinfo {volume} {91}},\ \bibinfo
  {pages} {140407} (\bibinfo {year} {2015})}\BibitemShut {NoStop}%
\bibitem [{\citenamefont {Balents}(2010)}]{Balents2010}%
  \BibitemOpen
  \bibfield  {author} {\bibinfo {author} {\bibfnamefont {L.}~\bibnamefont
  {Balents}},\ }\href {\doibase doi:10.1038/nature08917} {\bibfield  {journal}
  {\bibinfo  {journal} {Nature}\ }\textbf {\bibinfo {volume} {464}},\ \bibinfo
  {pages} {199} (\bibinfo {year} {2010})}\BibitemShut {NoStop}%
\bibitem [{\citenamefont {Savary}\ and\ \citenamefont
  {Balents}(2013)}]{Savary2013}%
  \BibitemOpen
  \bibfield  {author} {\bibinfo {author} {\bibfnamefont {L.}~\bibnamefont
  {Savary}}\ and\ \bibinfo {author} {\bibfnamefont {L.}~\bibnamefont
  {Balents}},\ }\href {\doibase 10.1103/PhysRevB.87.205130} {\bibfield
  {journal} {\bibinfo  {journal} {Phys. Rev. B}\ }\textbf {\bibinfo {volume}
  {87}},\ \bibinfo {pages} {205130} (\bibinfo {year} {2013})}\BibitemShut
  {NoStop}%
\bibitem [{\citenamefont {Barrat}\ and\ \citenamefont
  {Hansen}(2003)}]{Barrat2003}%
  \BibitemOpen
  \bibfield  {author} {\bibinfo {author} {\bibfnamefont {J.-L.}\ \bibnamefont
  {Barrat}}\ and\ \bibinfo {author} {\bibfnamefont {J.-P.}\ \bibnamefont
  {Hansen}},\ }\href@noop {} {\emph {\bibinfo {title} {Basic Concepts for
  Simple and Complex Liquids}}}\ (\bibinfo  {publisher} {Cambridge University
  Press},\ \bibinfo {year} {2003})\BibitemShut {NoStop}%
\bibitem [{\citenamefont {Kobelev}\ \emph {et~al.}(2002)\citenamefont
  {Kobelev}, \citenamefont {Kolomeisky},\ and\ \citenamefont
  {Fisher}}]{Kobelev2002}%
  \BibitemOpen
  \bibfield  {author} {\bibinfo {author} {\bibfnamefont {V.}~\bibnamefont
  {Kobelev}}, \bibinfo {author} {\bibfnamefont {A.~B.}\ \bibnamefont
  {Kolomeisky}}, \ and\ \bibinfo {author} {\bibfnamefont {M.~E.}\ \bibnamefont
  {Fisher}},\ }\href {\doibase 10.1063/1.1464827} {\bibfield  {journal}
  {\bibinfo  {journal} {The Journal of Chemical Physics}\ }\textbf {\bibinfo
  {volume} {116}},\ \bibinfo {pages} {7589} (\bibinfo {year} {2002})},\ \Eprint
  {http://arxiv.org/abs/http://dx.doi.org/10.1063/1.1464827}
  {http://dx.doi.org/10.1063/1.1464827} \BibitemShut {NoStop}%
\bibitem [{\citenamefont {Franzese}\ \emph {et~al.}(2001)\citenamefont
  {Franzese}, \citenamefont {Malescio}, \citenamefont {Skibinsky},
  \citenamefont {Buldyrev},\ and\ \citenamefont {Stanley}}]{Franzese2001}%
  \BibitemOpen
  \bibfield  {author} {\bibinfo {author} {\bibfnamefont {G.}~\bibnamefont
  {Franzese}}, \bibinfo {author} {\bibfnamefont {G.}~\bibnamefont {Malescio}},
  \bibinfo {author} {\bibfnamefont {A.}~\bibnamefont {Skibinsky}}, \bibinfo
  {author} {\bibfnamefont {S.~V.}\ \bibnamefont {Buldyrev}}, \ and\ \bibinfo
  {author} {\bibfnamefont {H.~E.}\ \bibnamefont {Stanley}},\ }\href
  {https://doi.org/10.1038/35055514} {\bibfield  {journal} {\bibinfo  {journal}
  {Nature}\ }\textbf {\bibinfo {volume} {409}},\ \bibinfo {pages} {692 EP }
  (\bibinfo {year} {2001})}\BibitemShut {NoStop}%
\bibitem [{\citenamefont {Tanaka}(2000)}]{Tanaka2000}%
  \BibitemOpen
  \bibfield  {author} {\bibinfo {author} {\bibfnamefont {H.}~\bibnamefont
  {Tanaka}},\ }\href {\doibase 10.1103/PhysRevE.62.6968} {\bibfield  {journal}
  {\bibinfo  {journal} {Phys. Rev. E}\ }\textbf {\bibinfo {volume} {62}},\
  \bibinfo {pages} {6968} (\bibinfo {year} {2000})}\BibitemShut {NoStop}%
\bibitem [{\citenamefont {Lefran\ifmmode~\mbox{\c{c}}\else \c{c}\fi{}ois}\
  \emph {et~al.}(2015)\citenamefont {Lefran\ifmmode~\mbox{\c{c}}\else
  \c{c}\fi{}ois}, \citenamefont {Simonet}, \citenamefont {Ballou},
  \citenamefont {Lhotel}, \citenamefont {Hadj-Azzem}, \citenamefont
  {Kodjikian}, \citenamefont {Lejay}, \citenamefont {Manuel}, \citenamefont
  {Khalyavin},\ and\ \citenamefont {Chapon}}]{Lefrancois2015}%
  \BibitemOpen
  \bibfield  {author} {\bibinfo {author} {\bibfnamefont {E.}~\bibnamefont
  {Lefran\ifmmode~\mbox{\c{c}}\else \c{c}\fi{}ois}}, \bibinfo {author}
  {\bibfnamefont {V.}~\bibnamefont {Simonet}}, \bibinfo {author} {\bibfnamefont
  {R.}~\bibnamefont {Ballou}}, \bibinfo {author} {\bibfnamefont
  {E.}~\bibnamefont {Lhotel}}, \bibinfo {author} {\bibfnamefont
  {A.}~\bibnamefont {Hadj-Azzem}}, \bibinfo {author} {\bibfnamefont
  {S.}~\bibnamefont {Kodjikian}}, \bibinfo {author} {\bibfnamefont
  {P.}~\bibnamefont {Lejay}}, \bibinfo {author} {\bibfnamefont
  {P.}~\bibnamefont {Manuel}}, \bibinfo {author} {\bibfnamefont
  {D.}~\bibnamefont {Khalyavin}}, \ and\ \bibinfo {author} {\bibfnamefont
  {L.~C.}\ \bibnamefont {Chapon}},\ }\href {\doibase
  10.1103/PhysRevLett.114.247202} {\bibfield  {journal} {\bibinfo  {journal}
  {Phys. Rev. Lett.}\ }\textbf {\bibinfo {volume} {114}},\ \bibinfo {pages}
  {247202} (\bibinfo {year} {2015})}\BibitemShut {NoStop}%
\bibitem [{\citenamefont {{Rau}}\ and\ \citenamefont
  {{Gingras}}(2018)}]{Rau2019}%
  \BibitemOpen
  \bibfield  {author} {\bibinfo {author} {\bibfnamefont {J.~G.}\ \bibnamefont
  {{Rau}}}\ and\ \bibinfo {author} {\bibfnamefont {M.~J.~P.}\ \bibnamefont
  {{Gingras}}},\ }\href@noop {} {\bibfield  {journal} {\bibinfo  {journal}
  {arXiv e-prints}\ ,\ \bibinfo {eid} {arXiv:1806.09638}} (\bibinfo {year}
  {2018})},\ \Eprint {http://arxiv.org/abs/1806.09638} {arXiv:1806.09638
  [cond-mat.str-el]} \BibitemShut {NoStop}%
\bibitem [{\citenamefont {Zhou}\ \emph {et~al.}(2012)\citenamefont {Zhou},
  \citenamefont {Cheng}, \citenamefont {Hallas}, \citenamefont {Wiebe},
  \citenamefont {Li}, \citenamefont {Balicas}, \citenamefont {Zhou},
  \citenamefont {Goodenough}, \citenamefont {Gardner},\ and\ \citenamefont
  {Choi}}]{Zhou2012}%
  \BibitemOpen
  \bibfield  {author} {\bibinfo {author} {\bibfnamefont {H.~D.}\ \bibnamefont
  {Zhou}}, \bibinfo {author} {\bibfnamefont {J.~G.}\ \bibnamefont {Cheng}},
  \bibinfo {author} {\bibfnamefont {A.~M.}\ \bibnamefont {Hallas}}, \bibinfo
  {author} {\bibfnamefont {C.~R.}\ \bibnamefont {Wiebe}}, \bibinfo {author}
  {\bibfnamefont {G.}~\bibnamefont {Li}}, \bibinfo {author} {\bibfnamefont
  {L.}~\bibnamefont {Balicas}}, \bibinfo {author} {\bibfnamefont {J.~S.}\
  \bibnamefont {Zhou}}, \bibinfo {author} {\bibfnamefont {J.~B.}\ \bibnamefont
  {Goodenough}}, \bibinfo {author} {\bibfnamefont {J.~S.}\ \bibnamefont
  {Gardner}}, \ and\ \bibinfo {author} {\bibfnamefont {E.~S.}\ \bibnamefont
  {Choi}},\ }\href {\doibase 10.1103/PhysRevLett.108.207206} {\bibfield
  {journal} {\bibinfo  {journal} {Physical Review Letters}\ }\textbf {\bibinfo
  {volume} {108}},\ \bibinfo {pages} {207206} (\bibinfo {year}
  {2012})}\BibitemShut {NoStop}%
\bibitem [{\citenamefont {Corman}\ \emph {et~al.}(2014)\citenamefont {Corman},
  \citenamefont {Chomaz}, \citenamefont {Bienaim\'e}, \citenamefont
  {Desbuquois}, \citenamefont {Weitenberg}, \citenamefont {Nascimb\`ene},
  \citenamefont {Dalibard},\ and\ \citenamefont {Beugnon}}]{Corman2014}%
  \BibitemOpen
  \bibfield  {author} {\bibinfo {author} {\bibfnamefont {L.}~\bibnamefont
  {Corman}}, \bibinfo {author} {\bibfnamefont {L.}~\bibnamefont {Chomaz}},
  \bibinfo {author} {\bibfnamefont {T.}~\bibnamefont {Bienaim\'e}}, \bibinfo
  {author} {\bibfnamefont {R.}~\bibnamefont {Desbuquois}}, \bibinfo {author}
  {\bibfnamefont {C.}~\bibnamefont {Weitenberg}}, \bibinfo {author}
  {\bibfnamefont {S.}~\bibnamefont {Nascimb\`ene}}, \bibinfo {author}
  {\bibfnamefont {J.}~\bibnamefont {Dalibard}}, \ and\ \bibinfo {author}
  {\bibfnamefont {J.}~\bibnamefont {Beugnon}},\ }\href {\doibase
  10.1103/PhysRevLett.113.135302} {\bibfield  {journal} {\bibinfo  {journal}
  {Phys. Rev. Lett.}\ }\textbf {\bibinfo {volume} {113}},\ \bibinfo {pages}
  {135302} (\bibinfo {year} {2014})}\BibitemShut {NoStop}%
\bibitem [{\citenamefont {Labeyrie}\ and\ \citenamefont
  {Kaiser}(2016)}]{Labeyrie2016}%
  \BibitemOpen
  \bibfield  {author} {\bibinfo {author} {\bibfnamefont {G.}~\bibnamefont
  {Labeyrie}}\ and\ \bibinfo {author} {\bibfnamefont {R.}~\bibnamefont
  {Kaiser}},\ }\href {\doibase 10.1103/PhysRevLett.117.275701} {\bibfield
  {journal} {\bibinfo  {journal} {Phys. Rev. Lett.}\ }\textbf {\bibinfo
  {volume} {117}},\ \bibinfo {pages} {275701} (\bibinfo {year}
  {2016})}\BibitemShut {NoStop}%
\bibitem [{\citenamefont {H{\'e}risson}\ and\ \citenamefont
  {Ocio}(2002)}]{herisson2002fluctuation}%
  \BibitemOpen
  \bibfield  {author} {\bibinfo {author} {\bibfnamefont {D.}~\bibnamefont
  {H{\'e}risson}}\ and\ \bibinfo {author} {\bibfnamefont {M.}~\bibnamefont
  {Ocio}},\ }\href@noop {} {\bibfield  {journal} {\bibinfo  {journal} {Physical
  Review Letters}\ }\textbf {\bibinfo {volume} {88}},\ \bibinfo {pages}
  {257202} (\bibinfo {year} {2002})}\BibitemShut {NoStop}%
\bibitem [{\citenamefont {H{\'e}risson}\ and\ \citenamefont
  {Ocio}(2004)}]{herisson2004off}%
  \BibitemOpen
  \bibfield  {author} {\bibinfo {author} {\bibfnamefont {D.}~\bibnamefont
  {H{\'e}risson}}\ and\ \bibinfo {author} {\bibfnamefont {M.}~\bibnamefont
  {Ocio}},\ }\href@noop {} {\bibfield  {journal} {\bibinfo  {journal} {The
  European Physical Journal B-Condensed Matter and Complex Systems}\ }\textbf
  {\bibinfo {volume} {40}},\ \bibinfo {pages} {283} (\bibinfo {year}
  {2004})}\BibitemShut {NoStop}%
\bibitem [{\citenamefont {Fennell}\ \emph {et~al.}(2009)\citenamefont
  {Fennell}, \citenamefont {Deen}, \citenamefont {Wildes}, \citenamefont
  {Schmalzl}, \citenamefont {Prabhakaran}, \citenamefont {Boothroyd},
  \citenamefont {Aldus}, \citenamefont {McMorrow},\ and\ \citenamefont
  {Bramwell}}]{Fennell2009}%
  \BibitemOpen
  \bibfield  {author} {\bibinfo {author} {\bibfnamefont {T.}~\bibnamefont
  {Fennell}}, \bibinfo {author} {\bibfnamefont {P.~P.}\ \bibnamefont {Deen}},
  \bibinfo {author} {\bibfnamefont {A.~R.}\ \bibnamefont {Wildes}}, \bibinfo
  {author} {\bibfnamefont {K.}~\bibnamefont {Schmalzl}}, \bibinfo {author}
  {\bibfnamefont {D.}~\bibnamefont {Prabhakaran}}, \bibinfo {author}
  {\bibfnamefont {A.~T.}\ \bibnamefont {Boothroyd}}, \bibinfo {author}
  {\bibfnamefont {R.~J.}\ \bibnamefont {Aldus}}, \bibinfo {author}
  {\bibfnamefont {D.~F.}\ \bibnamefont {McMorrow}}, \ and\ \bibinfo {author}
  {\bibfnamefont {S.~T.}\ \bibnamefont {Bramwell}},\ }\href
  {http://www.sciencemag.org/content/326/5951/415.short} {\bibfield  {journal}
  {\bibinfo  {journal} {Science}\ }\textbf {\bibinfo {volume} {326}},\ \bibinfo
  {pages} {415} (\bibinfo {year} {2009})}\BibitemShut {NoStop}%
\end{thebibliography}%

\end{document}